\documentclass[aps,prd,reprint,superscriptaddress,nofootinbib,longbibliography,floatfix]{revtex4-2}
\usepackage{amsmath,amssymb,bm,graphicx,booktabs,microtype,array,placeins}
\usepackage[colorlinks=true,linkcolor=blue,citecolor=blue,urlcolor=blue]{hyperref}
\graphicspath{{figures/}}
\newcommand{\GeV}{\mathrm{GeV}}
\newcommand{\MeV}{\mathrm{MeV}}
\newcommand{\keV}{\mathrm{keV}}
\newcommand{\dd}{\mathrm{d}}

\newcommand{\sv}{\langle\sigma v\rangle}

\usepackage{orcidlink}

\providecommand{\aap}{A\&A}

\usepackage[normalem]{ulem}

\newcommand{\bea} {\begin{eqnarray}}
\newcommand{\eea} {\end{eqnarray}}

\newcommand{\be}{\begin{eqnarray}}
\newcommand{\ee}{\end{eqnarray}}

\newcommand{\beq}{\begin{equation}}
\newcommand{\eeq}{\end{equation}}

\begin{document}

\title{ Exothermic Higgs portal:  from LUX-ZEPLIN recoil to Gamma Rays}

\author{Subhojit Roy,}
\email{roysb@ucmail.uc.edu }
\author{Jure Zupan}
\email{zupanje@ucmail.uc.edu}

\affiliation{Department of Physics, University of Cincinnati, Cincinnati, Ohio 45221, USA}
\date{\today}

\begin{abstract}
We show that a minimal extension of the Standard Model (SM) by a single complex scalar $\phi$ provides a viable inelastic exothermic, thermal relic,  interpretation of the 
recently reported 
nuclear-recoil candidate from LUX-ZEPLIN (LZ).
The mass terms and Higgs-portal interactions split $\phi$ into two nearly degenerate real scalars, $\phi_{1,2}$. For mass-splitting of a  few hundred keV, $\phi_2$ is cosmologically long-lived and the two states have comparable relic abundances.
Taking $\phi_{1,2}$ masses to be $\sim100\,$GeV, with off-diagonal Higgs coupling fixed by relic abundance, 
the exothermic down-scattering yields an order-one event expectation for the LZ exposure, 
without relying on the high-velocity tail of the dark matter velocity distribution required in endothermic scenarios.
This scenario can be tested using different nuclear targets and with indirect dark matter searches: $\phi_2\to\phi_1\gamma\gamma$ decays give a broad sub-MeV photon continuum that can be probed by INTEGRAL/SPI and  COSI; $\phi_1\phi_2$ annihilation produces a GeV-scale continuum, extending up to $\sim$100 GeV, that is probed by Fermi-LAT observations of dwarf galaxies; current CMB constraints are satisfied and annihilation among dark matter particles captured in the Sun is strongly suppressed.  Next-generation experiments could probe correlated scattering, decay, and annihilation signatures of this thermal Higgs-portal scenario.
\end{abstract}
\maketitle

\section{Introduction}\label{sec:intro}

The LUX-ZEPLIN (LZ) Collaboration recently reported a nuclear-recoil
candidate with energy
$E_R=248\pm23_{\rm stat}\pm23_{\rm sys}\,\keV$
in a $2.84$ tonne-year exposure~\cite{LZ:2026axp}.
Relative to the background-only hypothesis, the event reaches a maximum local significance of $3.4\sigma$ among the interactions tested by LZ, and a global significance of $2.6\sigma$ once the look-elsewhere effect is included~\cite{LZ:2026axp}.
Such a single high recoil energy is difficult to explain with
elastic scattering of halo-bound non-relativistic dark matter (DM), since this predicts a larger event rate at lower
recoil energies.  
Endothermic inelastic scattering, $\chi_1 A\to\chi_2 A$, on the other hand, 
imposes an additional kinematic threshold which suppresses the low-energy recoils and thus shifts
the recoil spectrum toward higher energies~\cite{Tucker-Smith:2001myb}, in agreement with the observed LZ event.
That is, for $\chi_1 A\to\chi_2 A$ scattering, the lighter DM state $\chi_1$ needs to have enough kinetic energy to produce the heavier state $\chi_2$.
For mass splitting of a few hundred keV this cuts out the bulk of the $\chi_1$ velocity distribution, 
resulting in a visibly harder recoil spectrum.

A particularly well-motivated supersymmetric realization of the endothermic inelastic scenario is
the Higgsino, whose two neutral states can form a nearly degenerate
pseudo-Dirac pair and naturally give rise to inelastic scattering. This
possibility has recently been studied in the context of the observed LZ
high-recoil event and in related electroweak constructions
\cite{Fan:2026kxx, Freese:2026sga,
Wu:2026nhi,
Yin:2026jnn,
Nomura:2026qyq,
DiMauro:2026ldr,
Visinelli:2026kgt,
Smirnov:2026aqk,
Du:2026guj,
Bisal:2026khf,
Cheung:2026byg,
Langhoff:2026ujr,
Frolovsky:2026tvq,
Paul:2026okh,
Khan:2026zuj,
Unwin:2026qep,
Murayama:2026apt, Roy:2026wzb, Lee:2026tro}.
Similar inelastic kinematics has also been explored in singlet--doublet and
inert-doublet sectors, sneutrino models, dark-photon and other
$U(1)$ extensions, axion-like portals, transition-dipole models,
neutrino-mass constructions, freeze-in scenarios, and effective
descriptions~\cite{Su:2026rwz,
Yamashita:2026ump,
Lee:2026wof,
Wang:2026ytg,
Das:2026uyy,
Okada:2026eol,
Ahmed:2026qjg,
Du:2026lpa,
Bandyopadhyay:2026gjw,
Borah:2026zwf,
Yuan:2026djt,
Zhu:2026dag,
Asadi:2026iot,
Lee:2026jxl,
He:2026hqz,
Fan:2026hzw,
Kumar:2026lgi,
Okada:2026upm,
Cabo-Almeida:2026uqw,
Borah:2026ris,
Das:2026buc,
He:2026idw,
Lian:2026hpm,
An:2026pkc,
Nagata:2026pbj,
Okada:2026fef,
Ahmed:2026com,
Tang:2026lgu,
Langhoff:2026qgo,
Choudhury:2026qjz, Capolupo:2026uyi}.

A broad range of alternative interpretations has been proposed in response to the LZ candidate. These include elastic-scattering scenarios with
nonstandard momentum or spin dependence, as well as scalar, axion,
magnetic, and extended-Higgs interactions
\cite{
Unwin:2026rdp,
Arcadi:2026kev,
De:2026win,
Khan:2026osp,
Elahi:2026vlm}.
Other proposals invoked boosted dark matter or nonstandard incident
populations
\cite{
Liang:2026coz,
Kannike:2026qyl,
Alhazmi:2026efz,
Heikinheimo:2026kwp,
Mahapatra:2026glu,
Chauhan:2026udz,
McCabe:2026crm,
OHare:2026nqi},
dark-matter absorption \cite{Lou:2026idn}, neutron-disappearance processes
\cite{Aghaie:2026vsu,
Lee:2026zbr,
Uttayarat:2026isp},
and neutrino-induced nuclear recoils
\cite{Jeesun:2026vzo,
Chattaraj:2026fxn,
Brdar:2026ukx}.
Further possibilities involving composite dark sectors, dark QCD,
extra dimensions, or modified gravity have been explored in Refs.
\cite{Sheng:2026tqt,
Jung:2026otm,
Sannino:2026hkc,
Lee:2026xxh}.

Here we instead consider exothermic inelastic scattering~\cite{Batell:2009vb,Finkbeiner:2009mi,Graham:2010ca}, in which a population of the heavier state that survives to the present day down-scatters as $\chi_2A\to\chi_1A$. Exothermic interpretations of the LZ candidate have been proposed in Refs.~\cite{Dent:2026bji,deLima:2026shq,Baer:2026fpy,Yang:2026wpb,Fan:2026hzw,An:2026pkc,Xing:2026civ,Gemmell:2026yaw,Oliveira:2026jtn,Jho:2026zzj,Sheikh:2026lxx,Deka:2026zeu,Kim:2026euj}.
A particularly simple realization of inelastic DM is obtained by extending 
the Standard Model (SM) 
with a complex scalar singlet that interacts with the visible
sector through the  doublet Higgs ($H$) portal.
Small symmetry-breaking terms split the
complex field into two nearly degenerate real mass eigenstates,
$\phi_1$ and $\phi_2$ (playing the role of $\chi_1$ and $\chi_2$ above), for which the leading couplings to the visible sector can be 
due to off-diagonal Higgs interactions \cite{Silveira:1985rk,
McDonald:1993ex,
Burgess:2000yq, Ghorbani:2014gka, Casas:2017jjg, DiazSaez:2024nrq, Casas:2017rww, Guo:2021vpb, Hooper:2025fda, Roy:2025zvo},
\begin{equation}
\label{eq:off:portal}
    \mathcal{L}_{\rm int}
    =
    -g\,\phi_1\phi_2
    \left(H^\dagger H-\frac{v_h^2}{2}\right).
\end{equation}
After electroweak symmetry breaking, this generates the
$h\phi_1\phi_2$ coupling responsible for inelastic scattering. 
The same off-diagonal interaction also sets the relic abundances of $\phi_1$ and $\phi_2$ via freeze-out of coannihilations, $\phi_1\phi_2\to h^\ast\to{\rm SM}$, and inelastic scatterings, $\phi_1+ {\rm SM} \leftrightarrow \phi_2+ {\rm SM}$.

For small enough mass splitting $\Delta m$,
$\phi_2$ can be stable on cosmological timescales. For instance, for 
$\Delta m<2m_e$, the decay
$\phi_2\to\phi_1 e^+e^-$ is kinematically forbidden, and the dominant decay channel is 
$\phi_2\to\phi_1\gamma\gamma$, induced via emission of an off-shell Higgs boson.
Since the available energy is shared between the two photons, this decay
produces a broad sub-MeV spectrum rather than a monochromatic line.
Such connections between the LZ high-recoil event and sub-MeV photon signals
have been explored in other inelastic dark-matter scenarios
\cite{Yamashita:2026ump, Wu:2026nhi,
Yang:2026wpb, Jho:2026zzj,
Dutra:2026fng}.
In the minimal model, however, the strength of the diphoton signal is fully predicted, since the decay is induced by the same off-diagonal Higgs interaction as the nuclear recoil.
Existing
INTEGRAL/SPI data on photon emissions and future observations with the Compton Spectrometer and
Imager (COSI)~\cite{Roques:2003xg,Tomsick:2023aue} 
therefore provide a complementary probe of this scenario.

The inelastic
interpretations of the LZ event typically face stringent constraints from DM solar capture and subsequent annihilation. For the endothermic scenario, for instance, the larger velocities reached by DM particles inside the Sun make
the inelastic endothermic transitions more easily accessible.
For Higgsino DM, these larger velocities can
allow endothermic scattering on heavy solar nuclei, followed by capture
and annihilation predominantly into $W^+W^-$ and $ZZ$. The resulting
high-energy neutrino flux leads to strong constraints from IceCube and
Super-Kamiokande
\cite{Pospelov:2026ewn,
Bose:2026ndd,
DiMauro:2026dqp,
Nguyen:2026lui,
Ghosh:2026txe} (for extension to
broader classes of
electroweak multiplets,
see Ref. \cite{Luengas:2026tlf}).

The situation is qualitatively different for an off-diagonal scalar
Higgs portal. In regions of parameter space in which the exothermic capture
$\phi_2 A\to\phi_1 A$ is efficient, while the reverse transition
$\phi_1 A\to\phi_2 A$ is suppressed, most of the captured DM solar population is in the form of  
the lighter state. This then reduces the importance of tree-level mixed annihilation
channel $\phi_1\phi_2\to h^\ast\to{\rm SM}$.

Even if at tree level only the off-diagonal portal, Eq. \eqref{eq:off:portal}, is present, the 
 diagonal interactions are generated at one-loop. These then give rise to elastic scattering and
same-state annihilation of the individual dark states. The $\phi_1\phi_1$ annihilation in the Sun 
can be more important than the $\phi_1\phi_2$ one, if $\phi_1$ population dominates. However, as we will show later, 
even in this case the annihilation rate into neutrinos is still many orders of magnitude below the current experimental
solar-neutrino sensitivity, in the region of parameter space relevant
for LZ interpretation.

The paper is organized as follows. 
Section~\ref{sec:model} introduces the inelastic scalar Higgs-portal model
and describes the thermal evolution that determines the relic
abundance and the surviving excited-state population.
In Sec.~\ref{sec:lz}, we calculate the exothermic xenon recoil rate,
map the thermal-relic parameter space predicting order-one
selected events, and then study a representative benchmark and
its finite-momentum nuclear response. Section~\ref{sec:indirect} examines the correlated signals from
radiative decay and present-day annihilation, confronting
them with INTEGRAL/SPI, Fermi-LAT dwarf-galaxy, and CMB
constraints, and discusses complementary collider and
direct-detection probes. In Sec.~\ref{sec:solar}, we study solar capture,
the nonthermal evolution of the captured populations, and
the resulting annihilation rates and their implications
for solar-neutrino searches. We conclude in Sec.~\ref{sec:conclusions}.

\section{The Inelastic Higgs-Portal Dark Matter Model}
\label{sec:model}

The inelastic scalar Higgs-portal model~\cite{Hooper:2025fda, Roy:2025zvo}
extends the SM
by a complex
scalar singlet $\phi$ coupled to the Higgs doublet $H$. The tree-level
scalar potential involving $H$ and $\phi$ is given by
\beq
\begin{split}
\label{Vtreetot}
V(H,\phi)={}&
-\mu_h^2 H^\dagger H
+\lambda_h(H^\dagger H)^2
\\
& +m_0^2|\phi|^2 
+\frac{1}{2}\left(\rho_0^2\phi^2+\rho_0^{2*}\phi^{*2}\right)
\\
&+
\left(
\kappa|\phi|^2
+\frac{1}{2}\left(\eta\phi^2+\eta^*\phi^{*2}\right)
\right)H^\dagger H .
\end{split}
\eeq
The parameters $\rho_0^2$ and $\eta$ may in general be complex. If they are zero, 
the singlet sector possesses a global
$U(1)$ symmetry,  which is otherwise broken
 to a remaining $Z_2$ symmetry $\phi\to-\phi$   (and similarly for the real components, $\phi_{1,2}\to-\phi_{1,2}$). 
The lightest
$Z_2$-odd state is therefore stable.

After electroweak symmetry breaking (EWSB),
\begin{equation}
\label{Higgsdoublet}
H=
\left(
G^+,\,
\frac{v_h+h+iG^0}{\sqrt{2}}
\right)^T,
\qquad
v_h=246~{\rm GeV}.
\end{equation}
Defining
\begin{equation}
m^2\equiv m_0^2+\frac{\kappa v_h^2}{2},
\qquad
\rho^2\equiv\rho_0^2+\frac{\eta v_h^2}{2},
\end{equation}
the complex singlet splits into two real mass eigenstates,
$\phi_1$ and $\phi_2$, with
\begin{equation}
m_{\phi_{1,2}}^2=m^2\mp|\rho^2|,
\end{equation}
where we use the convention $m_{\phi_2}>m_{\phi_1}$ and define
$\Delta m\equiv m_{\phi_2}-m_{\phi_1}>0$.
For $\Delta m\ll m_{\phi_{1,2}}$ the mass splitting is given by
\begin{equation}
\Delta m\simeq \frac{|\rho^2|}{m} .
\end{equation}

Here, $\rho^2$ is a complex parameter since both $\rho_0^2$ and $\eta$  are complex. These parameters can be decomposed into their real and imaginary parts as follows:
\be
\rho_0^2 = \rho_{0,R}^2 + i \rho_{0,I}^2 , ~\quad\eta = \eta_R + i \eta_I, ~\quad\rho^2 = \rho_R^2 + i \rho_I^2.
\ee
The complex field $\phi$ similarly decomposes in terms of the mass eigenstates as
\begin{equation}
\phi = \alpha \phi_1 + \beta \phi_2, \qquad
\phi^* = \alpha^* \phi_1 + \beta^* \phi_2,
\end{equation}
where $\alpha = e^{-i\theta}/\sqrt{2}$ and $\beta = i e^{-i\theta}/\sqrt{2}$. The mixing angle $\theta$ diagonalizes the singlet mass matrix via the rotation
\begin{equation}
\label{eqn:mneut}
{\cal U} = \begin{pmatrix}
\cos\theta & -\sin\theta \\
\sin\theta & \cos\theta
\end{pmatrix},
\end{equation}
with 
\begin{equation}
\cos 2\theta = -\frac{\rho_R^2}{|\rho^2|} \, , \qquad \sin 2\theta = -\frac{\rho_I^2}{|\rho^2|}.
\end{equation}

In the mass basis the Higgs-portal interactions  are
\begin{equation}
{\cal L}_{\phi H}
=
-\left(
f_1\phi_1^2+f_2\phi_2^2+g\phi_1\phi_2
\right)
\left(H^\dagger H-\frac{v_h^2}{2}\right),
\label{eq:Lh-mass}
\end{equation}
where
\begin{subequations}
\label{eq:f1:f2:g:rel}
\begin{align}
f_1&=
\frac{\kappa}{2}
+\frac{1}{2}\eta_R\cos2\theta
+\frac{1}{2}\eta_I\sin2\theta,
\label{f1rel}
\\
f_2&=
\frac{\kappa}{2}
-\frac{1}{2}\eta_R\cos2\theta
-\frac{1}{2}\eta_I\sin2\theta,
\label{f2rel}\\
g&=
\eta_R\sin2\theta-\eta_I\cos2\theta.
\label{grel}
\end{align}
\end{subequations}
The $f_1$ and $f_2$ control diagonal Higgs interactions, and $g$
the off-diagonal ones. 
We focus on the
predominantly off-diagonal regime, 
\beq
|f_{1,2}|\ll|g|,
\eeq
 which differs
qualitatively from the conventional real-singlet Higgs portal~\cite{Silveira:1985rk,McDonald:1993ex,Burgess:2000yq,Cline:2013gha,Arcadi:2017kky, Ghosh:2022fzp,Roy:2022gop}, which corresponds to the $f_1\ne0$, $f_2=g=0$ limit.

To simplify the phenomenology we choose $\sin2\theta=0$,
implying
\begin{equation}
\rho_{0,I}^2=-\frac{v_h^2}{2}\eta_I.
\end{equation}
The scalar potential in terms of ($h, \phi_1, \phi_2$) then takes the form
\beq
\begin{split}
V_0={}&
-\frac{1}{2}\mu_h^2h^2
+\frac{1}{4}\lambda_hh^4
+\frac{1}{2}
\left(m_0^2+\frac{\kappa}{2}h^2\right)
(\phi_1^2+\phi_2^2)
\\
&+
\frac{1}{2}
\left(\rho_{0,R}^2+\frac{\eta_R}{2}h^2\right)
(\phi_1^2-\phi_2^2)
-\frac{\eta_I}{2}(h^2-v_h^2)\phi_1\phi_2
\\
&+
\frac{1}{4}
\left(
\lambda_1\phi_1^4+
\lambda_2\phi_2^4+
\lambda_{12}\phi_1^2\phi_2^2
\right).
\label{Vtree}
\end{split}
\eeq

In the remainder of the manuscript we will impose for simplicity
$f_1=f_2=\lambda_{12}=0$.
For $\sin2\theta=0$, the tree-level portal relations reduce to
\begin{equation}
\eta_I=-g,
\qquad
\kappa=\eta_R=0,
\end{equation}
and the underlying mass parameters are
\begin{align}
m_0^2&=
\frac{1}{2}
\left(m_{\phi_1}^2+m_{\phi_2}^2\right),
\nonumber\\
\rho_{0,R}^2&=
\frac{1}{2}
\left(m_{\phi_1}^2-m_{\phi_2}^2\right).
\label{rho0Rsq}
\end{align}

In this limit, the tree-level interaction Lagrangian is
\begin{equation}
{\cal L}_{\rm int}
=
-g\phi_1\phi_2
\left(H^\dagger H-\frac{v_h^2}{2}\right),
\label{eq:restrictedportal}
\end{equation}
which after EWSB generates the $h\phi_1\phi_2$ and
$hh\phi_1\phi_2$ vertices. The unbroken $Z_2$ acts identically on the two
mass eigenstates, $\phi_{1,2}\to -\phi_{1,2}$. As a result, the lighter state $\phi_1$ is stable, whereas $\phi_2$ is a slightly heavier state. 

Note that the dark-sector self-couplings $\lambda_{1,2}$ do not affect the relic abundance or any of the signals considered here, beyond ensuring that the potential is bounded from below. 
With 
$f_1=f_2=\lambda_{12}=0$, a
sufficient tree-level boundedness condition is
\begin{equation}
\lambda_h>0,
\qquad
\lambda_{1,2}\geq0,
\qquad
g^2\leq2\lambda_h\sqrt{\lambda_1\lambda_2}.
\label{eq:bfb}
\end{equation}
The phenomenology we are interested in is fully specified by
\begin{equation}
\left\{
m_{\phi_1},\,\Delta m,\,g
\right\}.
\label{free-para}
\end{equation}
%

% ---------------------------------------------------------
\subsection{Thermal relic abundances}
\label{sec:thermal-picture}
% ---------------------------------------------------------

A central feature of the off-diagonal Higgs portal is that the same
interaction responsible for inelastic nuclear scattering also governs
the thermal evolution of the dark sector. That is, in the $f_1=f_2=0$ limit, 
 the dominant tree-level annihilation
process is
\begin{equation}
\phi_1\phi_2\rightarrow h^{(\ast)}\rightarrow{\rm SM},
\label{eq:mixedann}
\end{equation}
where ``SM'' denotes any SM final state. Including also the subdominant diagonal annihilation channels,
the two-state system in chemical equilibrium is described
by the coannihilation combination of cross sections
\cite{Griest:1990kh},
\beq
\label{eq:coann}
\langle\sigma v\rangle_{\rm eff}
=
r_1^2\langle\sigma v\rangle_{11}
+2r_1r_2\langle\sigma v\rangle_{12}
+r_2^2\langle\sigma v\rangle_{22},
\eeq
where $\langle \sigma v\rangle_{ij}$ are the thermal averages of $\phi_i \phi_j$ annihilation cross sections, while 
\beq
r_i=\frac{n_i}{n_1+n_2}, 
\eeq
are the number density fractions, with $n_i$ ($i=1,2$), the number densities. The factor of two in \eqref{eq:coann} counts the distinguishable mixed-state initial pairs.

A sub-MeV splitting is smaller than the freeze-out temperature,
$\Delta m\ll T_f$, and thus for $T\sim T_f$ the ratio of the two equilibrium densities is close to one
\begin{equation}
\frac{n_2^{\rm eq}}{n_1^{\rm eq}}
\simeq
\left(\frac{m_{\phi_2}}{m_{\phi_1}}\right)^{3/2}
e^{-\Delta m/T}
\simeq1,
\label{eq:eqratio}
\end{equation}
and the two states participate with nearly equal abundances during
freeze-out.

For fixed $m_{\phi_1}$ and $\Delta m$, requiring the observed
thermal relic abundance determines the off-diagonal portal coupling $g$.
We determine the total abundance and the relative populations of the
two states by solving the coupled Boltzmann equations, including mixed
annihilation, elastic scattering with the thermal bath,
bath-induced $\phi_1\leftrightarrow\phi_2$ conversions,
number-conserving pair conversions, and the corresponding inverse
processes.

\begin{figure}[t]
\centering
\includegraphics[width=\columnwidth]{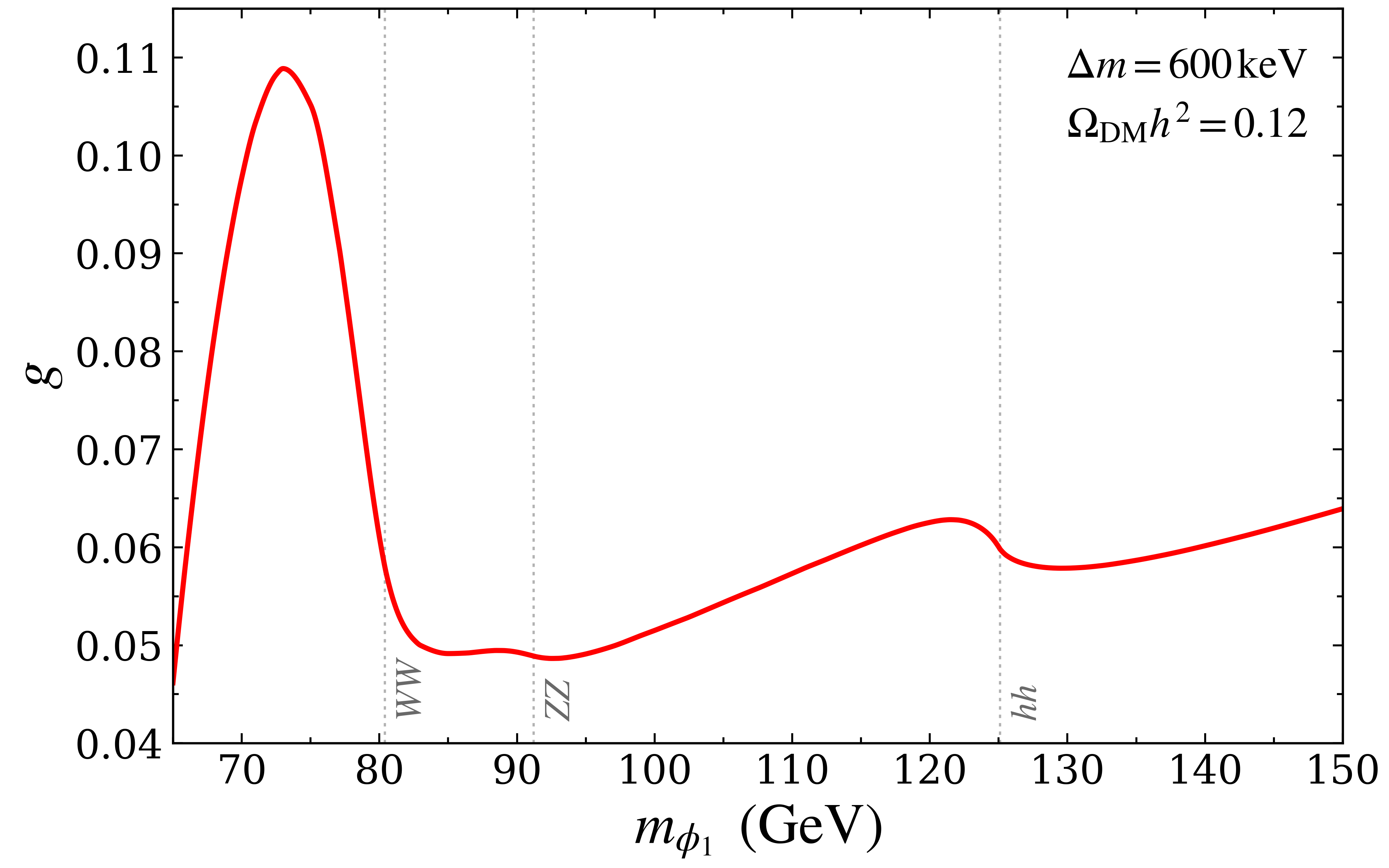}
\caption{Off-diagonal Higgs-portal coupling $g$ required by the observed
thermal relic abundance, $\Omega_{\rm DM}h^2=0.120$, as a function of
$m_{\phi_1}$ for $\Delta m=600\,\keV$. The vertical dotted lines
indicate the $W^+W^-$, $ZZ$, and $hh$ thresholds.}
\label{fig:thermal-coupling}
\end{figure}

Figure~\ref{fig:thermal-coupling} shows the coupling required to reproduce the observed relic abundance as a function of the dark-matter mass, for a representative splitting of $\Delta m=600~\keV$. We consider the mass range $65~\GeV\leq m_{\phi_1}\leq150~\GeV$, extending from just above the Higgs resonance ($m_{\phi_1}\gtrsim m_h/2$) to masses where annihilation into electroweak gauge bosons and Higgs bosons becomes important. Moving away from the Higgs resonance, the
coupling first increases and reaches a maximum near
$m_{\phi_1}\simeq73\,\GeV$. It then decreases as annihilation into
$W$-boson pairs, initially through off-shell bosons and subsequently
through on-shell production, becomes more efficient. The opening of
the $ZZ$ and $hh$ channels produces further changes at higher masses.
Including the precise value of the small mass splitting, $\Delta m\lesssim 1~\MeV$, 
requires a relative adjustment in $g$ of less than $10^{-3}$.

%=========================================================
\section{Exothermic Scattering and the LZ High-Recoil Candidate}
\label{sec:lz}
%=========================================================
%
An exothermic interpretation of the LZ event requires three ingredients: a mass
splitting capable of generating recoils of a few hundred keV,
a substantial heavier-state population surviving until the
present epoch, and an interaction strength sufficient to yield
an observable scattering rate. The relic-density calculation fixes
the portal coupling and predicts the surviving heavier-state fraction.
We now combine these results with the recoil kinematics and nuclear response
to map the expected LZ signal, before analyzing one representative point in detail. 
%
%---------------------------------------------------------
\subsection{Exothermic recoil kinematics}
\label{sec:lz-kinematics}
%---------------------------------------------------------
%
For a target nucleus of mass $m_A$, the minimum incident
velocity required to produce a recoil energy $E_R$ is
\begin{equation}
v_{\min}(E_R)=
\frac{
\left|m_AE_R/\mu_{2A}-\Delta m\right|
}{
\sqrt{2m_AE_R}
},
\qquad
\mu_{2A}=
\frac{m_{\phi_2}m_A}{m_{\phi_2}+m_A}.
\label{eq:vmin}
\end{equation}
A characteristic feature of exothermic scattering is that even for zero minimal velocity,
$v_{\min}=0$, there is a nonzero recoil energy,
\begin{equation}
E_{R,A}^{\star}
=
\frac{\mu_{2A}}{m_A}\Delta m.
\label{eq:ERstar}
\end{equation}
For dark-matter masses of order \(100\,\GeV\) and splittings
of several hundred keV, Eq.~\eqref{eq:ERstar}, the above
characteristic recoil scale is in the few-hundred-keV range. Note that \(E_{R,A}^{\star}\) is comparable but
does
not in general coincide with the maximum of the observed recoil spectrum: the DM halo velocity distribution, nuclear response,
and detector efficiency also enter in this.
%
%---------------------------------------------------------
\subsection{Higgs-mediated nuclear scattering}
\label{sec:lz-rate}
%---------------------------------------------------------

The off-diagonal Higgs interaction mediates coherent scalar
scattering from nuclei. At leading nonrelativistic order,
the differential cross section for DM scattering on xenon isotope with mass number $A$ is given by
\begin{align}
\frac{\dd\sigma_A}{\dd E_R}
={}&
\frac{
g^2m_A
}{
8\pi m_{\phi_2}^2  v^2
}
\frac{
(Af_Nm_N)^2
}{
 m_h^4
}
\frac{W_M^A(q)}{W_M^A(0)},
\label{eq:dsigma-inelastic}
\end{align}
where the recoil four-momentum squared is $q^2=2m_AE_R$. The factor $(Af_Nm_N)^2$ is the squared coherent nuclear charge at $q=0$ and $W_M^A(q)/W_M^A(0) = F^2(q)$ is the normalized momentum dependence of the nuclear response. Here,  $m_N$ is the nucleon mass, $f_N$ parametrizes the
isoscalar Higgs--nucleon scalar coupling, and $W_M^A(q)$
is the isosinglet one-body nuclear response associated with the
coherent scalar interaction (we ignore the subleading isotriplet contribution). We adopt $f_N=0.308$ and
evaluate the recoil spectrum using the natural xenon
isotope mixture and the normalized nuclear responses
implemented in {\tt WIMpy\_NREFT}
\cite{Fitzpatrick:2012ix, Anand:2013yka,Dent:2015zpa, Catena:2019hzw, WIMpyNREFT2024}.
For the rare isotopes without tabulated responses, we
employ the Helm form factor; their contribution to the
total rate is below the percent level. 

The differential recoil rate per unit detector mass is 
\begin{equation}
\frac{\dd R}{\dd E_R}
=
\sum_A
\frac{\xi_A}{m_A}\,
n_2
\int \dd^3v\,
f_{\rm lab}(\mathbf v,t)\,
v\,
\frac{\dd\sigma_A}{\dd E_R}\,
\Theta[v-v_{\min}(E_R)],
\label{eq:recoilrate}
\end{equation}
with $\xi_A$  the isotope mass fraction,
$v=|\mathbf v|$ the incoming DM velocity,
and $f_{\rm lab}(\mathbf v,t)$ the normalized
laboratory-frame velocity distribution. The step function
enforces the minimum-velocity condition for a given recoil
energy.
 We adopt the Standard Halo Model with a truncated
Maxwellian velocity distribution 
and 
$v_0=238~{\rm km\,s^{-1}}$,
$v_{\rm esc}=544~{\rm km\,s^{-1}}$, transformed to the
Earth's frame
\cite{McCabe:2013kea}.
The local number density of $\phi_2$ is
\begin{equation}
n_2=
\frac{r_2\rho_\odot}{\bar m},
\qquad
\bar m=r_1m_{\phi_1}+r_2m_{\phi_2},
\label{eq:n2halo}
\end{equation}
where $\rho_\odot$ is the local dark-matter mass density
and $r_i$ are the present-day number fractions ($r_1=1-r_2$). 
The local DM density is taken to be
$\rho_\odot=0.3~{\rm GeV\,cm^{-3}}$
(nuclear-scattering rate 
scales
linearly with it).
For the parameter scan in Fig.~\ref{fig:lz-thermal-map},
we also include the reverse endothermic process
$\phi_1 A\to\phi_2 A$, with the corresponding
incident density and scattering kinematics.
Its contribution is negligible in the region
predicting an order-one LZ event yield.

%----------------
\subsection{Thermal parameter space and expected LZ yield}
\label{sec:lz-thermal-scan}
%-------------------------
%
For fixed DM mass and splitting, the relic-density
calculation determines the Higgs-portal coupling $g$
and the present-day population of $\phi_2$.
Eq.~\eqref{eq:recoilrate} then predicts the xenon
recoil spectrum and rate without independently
adjusting the scattering strength.

To estimate the number of selected events, we use
the nuclear-recoil efficiency reported by LZ in
Supplemental Fig.~S2 of Ref.~\cite{LZ:2026axp}.
The digitized efficiency $\epsilon(E_R)$ includes
trigger, reconstruction, and analysis-selection effects.
The corresponding event expectation is
\begin{equation}
N_\epsilon
=
\mathcal E f_{\rm sci}
\int_{\mathcal S_\epsilon}
\dd E_R\,
\epsilon(E_R)
\frac{\dd R}{\dd E_R},
\label{eq:nsel}
\end{equation}
where $\mathcal E=2.84$ tonne-year is the LZ exposure
and $\mathcal S_\epsilon=[5.4,299.9]\,\keV$ is the
recoil-energy interval covered by the digitized efficiency.
The LZ analysis reports accidental prompt- and delayed-veto
fractions of $0.01\%$ and $2.86\%$, respectively
~\cite{LZ:2026axp}. We account for these losses
through the survival factor
$f_{\rm sci}=1-0.0001-0.0286=0.9713$,
applied separately from the digitized efficiency.

\begin{figure}[t]
\centering
\includegraphics[width=0.5\textwidth]{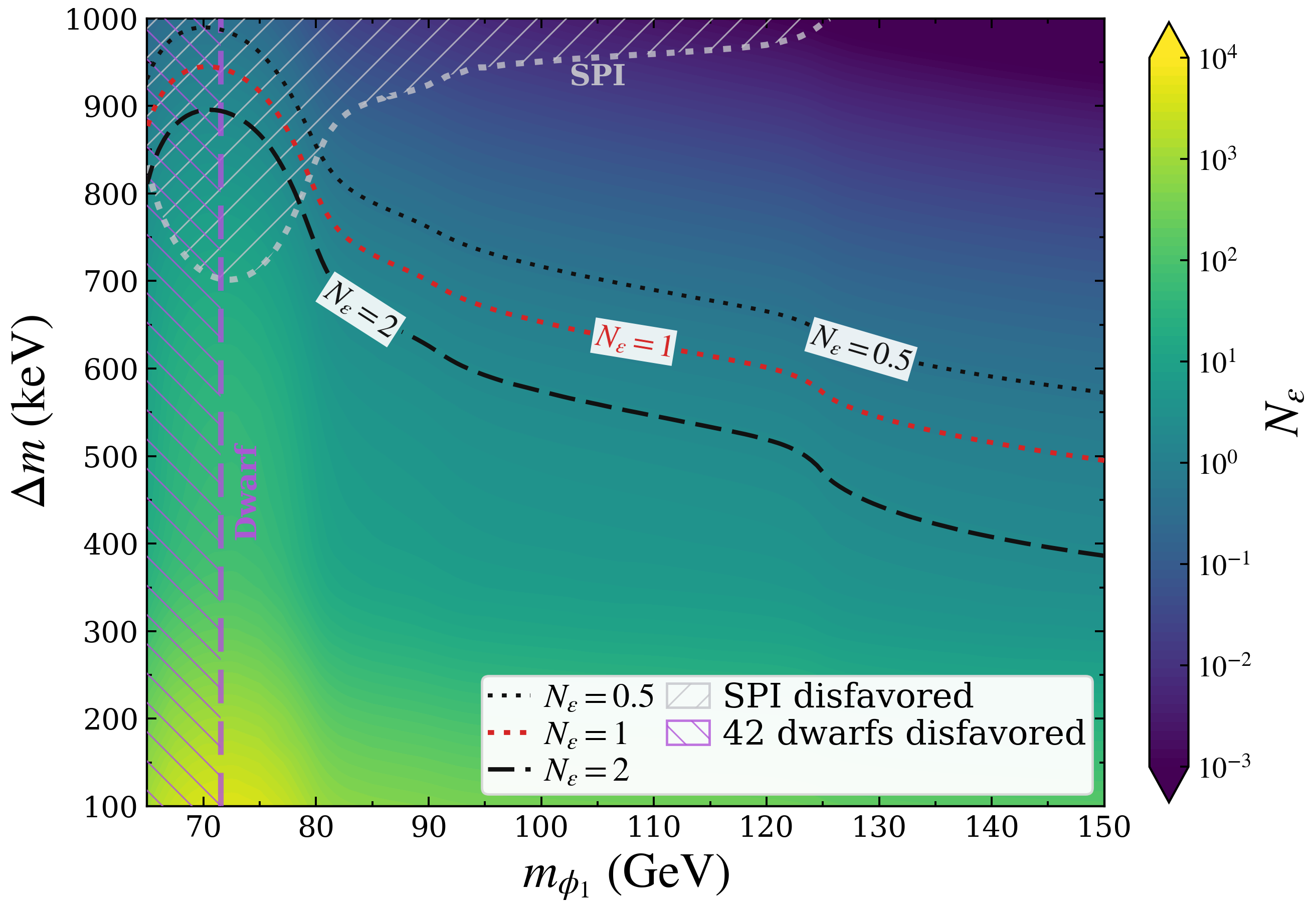}
\caption{
Predicted efficiency-weighted number of xenon recoils,
$N_\epsilon$, in the $(m_{\phi_1},\Delta m)$ plane
for the $2.84$ tonne-year LZ exposure.
The portal coupling $g$ is fixed by the thermal relic
abundance, $\Omega_{\rm DM}h^2=0.120$, while the
present-day state fractions follow from the thermal
evolution. Both exothermic and endothermic scattering
contribute to the recoil yield.
The black dotted, red dotted, and black dashed
contours correspond to $N_\epsilon=0.5$, $1$, and $2$,
respectively. The white hatched region is disfavored by
INTEGRAL/SPI observations of Galactic photon emission,
while the purple hatched region is disfavored by
Fermi-LAT observations of 42 dwarf spheroidal galaxies, see
Sec.~\ref{sec:indirect}.
The recoil rates use the Standard Halo Model,
the finite-momentum one-body scalar nuclear response,
and the LZ nuclear-recoil efficiency.
}
\label{fig:lz-thermal-map}
\end{figure}

Figure~\ref{fig:lz-thermal-map} shows an extended
region in which the predicted LZ event yield is
of order unity, between the $N_\epsilon=0.5$ and
$2$ contours. The shape of this region reflects
the mass dependence of the relic-density coupling
$g$, shown in Fig.~\ref{fig:thermal-coupling},
together with the recoil kinematics and nuclear
response. Near $m_{\phi_1}\sim70$--$75\,\GeV$,
where the required coupling is largest, larger
mass splittings are needed to obtain the same
event yield. At larger dark-matter masses, the
order-one-event region shifts toward smaller
splittings.

The indirect-detection constraints restrict
different parts of this parameter space.
The radiative decay $\phi_2\to\phi_1\gamma\gamma$
produces a sub-MeV photon continuum, which we
constrain using INTEGRAL/SPI measurements of
diffuse hard X-ray and soft gamma-ray emission
from the inner Galaxy~\cite{Calore:2022pks,Siegert:2022jii}.
Complementary constraints arise from Fermi-LAT
observations of 42 dwarf spheroidal galaxies,
which probe gamma rays from present-day dark-matter
annihilation~\cite{McDaniel:2023bju}.
We use the predicted mass-dependent annihilation
spectra in evaluating the dwarf-galaxy constraint.

To define the indirect-detection boundaries
in Fig.~\ref{fig:lz-thermal-map}, we introduce
\begin{equation}
\mathcal R_{\rm SPI}
=
\frac{\tau_{\rm SPI}^{\rm rec}}
{\tau_{\gamma\gamma}},
\qquad
\mathcal R_{\rm dwarf}
=
\frac{\sv_{\rm equiv}}
{\sv_{\rm limit}},
\label{eq:indirect-ratios}
\end{equation}
where $\tau_{\rm SPI}^{\rm rec}$ is the
reconstructed lower lifetime limit from
INTEGRAL/SPI, and $\sv_{\rm limit}$ is the
observed 95\% confidence-level upper limit
on the equivalent annihilation cross section
obtained from the 42-dwarf likelihood.
The SPI constraint is evaluated for the predicted
radiative spectrum, while the dwarf-galaxy
constraint uses the mass-dependent annihilation
spectrum, with only weak dependence on $\Delta m$.
The boundaries correspond to
$\mathcal R_{\rm SPI}=1$ and
$\mathcal R_{\rm dwarf}=1$, respectively,
with $\mathcal R>1$ defining the corresponding
disfavored regions. The SPI constraint disfavors
large mass splittings, while the dwarf-galaxy
bound disfavors $m_{\phi_1}\lesssim72\,\GeV$
over the scanned range.
The calculations underlying these constraints
are discussed in Sec.~\ref{sec:indirect}.
%
%---------------------------------------------------------
\subsection{Representative benchmark and recoil spectrum}
\label{sec:lz-response}
%---------------------------------------------------------
%
For a detailed discussion of the recoil spectrum and the correlated
signals, we consider a benchmark point from the region shown in
Fig.~\ref{fig:lz-thermal-map},
\begin{equation}
m_{\phi_1}=100~{\rm GeV},
~~
\Delta m=600~{\rm keV},~~
g \simeq0.0515.
\label{eq:benchmarkinputs}
\end{equation}
The value of $g$ is from 
the relic-density condition,
$\Omega_{\rm DM}h^2=0.120$,
which for this benchmark 
yields nearly equal present-day $\phi_1$ and $\phi_2$ populations,
\begin{equation}
(r_1,r_2)_{t_0}\simeq(0.501,\,0.499).
\label{eq:latefractions}
\end{equation}
The appreciable population of the heavier state  
provides incident particles for exothermic
scattering. A similar near-equal population is obtained throughout
the mass and splitting range in $0.5 <N_\epsilon < 2$ in Fig. \ref{fig:lz-thermal-map}.

One-loop-induced same-state annihilation does not appreciably modify the
thermal picture, so that these contributions can be safely neglected. For instance, the $\phi_1\phi_1\to W^+W^-$ contribution is
 about $10^{-7}$ times smaller than due to the mixed
annihilation rate.

\begin{table}[t]
\centering
\begin{ruledtabular}
\begin{tabular}{lc}
Quantity & Value\\
\hline
$m_{\phi_1}$ & $100~\GeV$\\
$\Delta m$ & $600~\keV$\\
$g$ & $0.0515 \simeq 0.05$\\
$\Omega_{\rm DM}h^2$ & $0.120$\\
$r_2(t_0)$ & $\simeq0.5$\\
$N_{\rm truth}(5.4\text{--}590~\keV)$ & $2.13$\\
$N_{\epsilon}$ & $1.63$\\
$N_{\epsilon}(100\text{--}200~\keV)$ & $1.18$\\
$N_{\epsilon}(215\text{--}269.9~\keV)$ & $0.216$\\
$\tau_{\gamma\gamma}$ & $4.49\times10^{23}~\mathrm{s}$\\
$\tau_{\gamma\gamma}/\tau_{\rm SPI}^{\rm rec}$ & $41.4$\\
$ \sv_{\rm equiv}$ (${\rm cm^3\,s^{-1}}$) & $2.26\times10^{-26}$\\
\end{tabular}
\end{ruledtabular}
\caption{
Benchmark parameters and predictions for xenon scattering,
radiative decay, and present-day annihilation.
Event counts refer to the $2.84$ tonne-year LZ exposure.
$N_{\rm truth}$ is obtained without detector efficiency, while
$N_\epsilon$ is an efficiency-weighted true-recoil expectation,
not an event-level likelihood result.
$N_\epsilon$ includes the digitized nuclear-recoil efficiency
over its available range, $5.4\leq E_R/\keV\leq299.9$.
The lifetime $\tau_{\gamma\gamma}$ uses leading-order
hadronic matching, and $\tau_{\rm SPI}^{\rm rec}$ is
the reconstructed INTEGRAL/SPI lower limit.
}
\label{tab:benchmark}
\end{table}

\begin{figure}[t]
\centering
\includegraphics[width=0.48\textwidth]{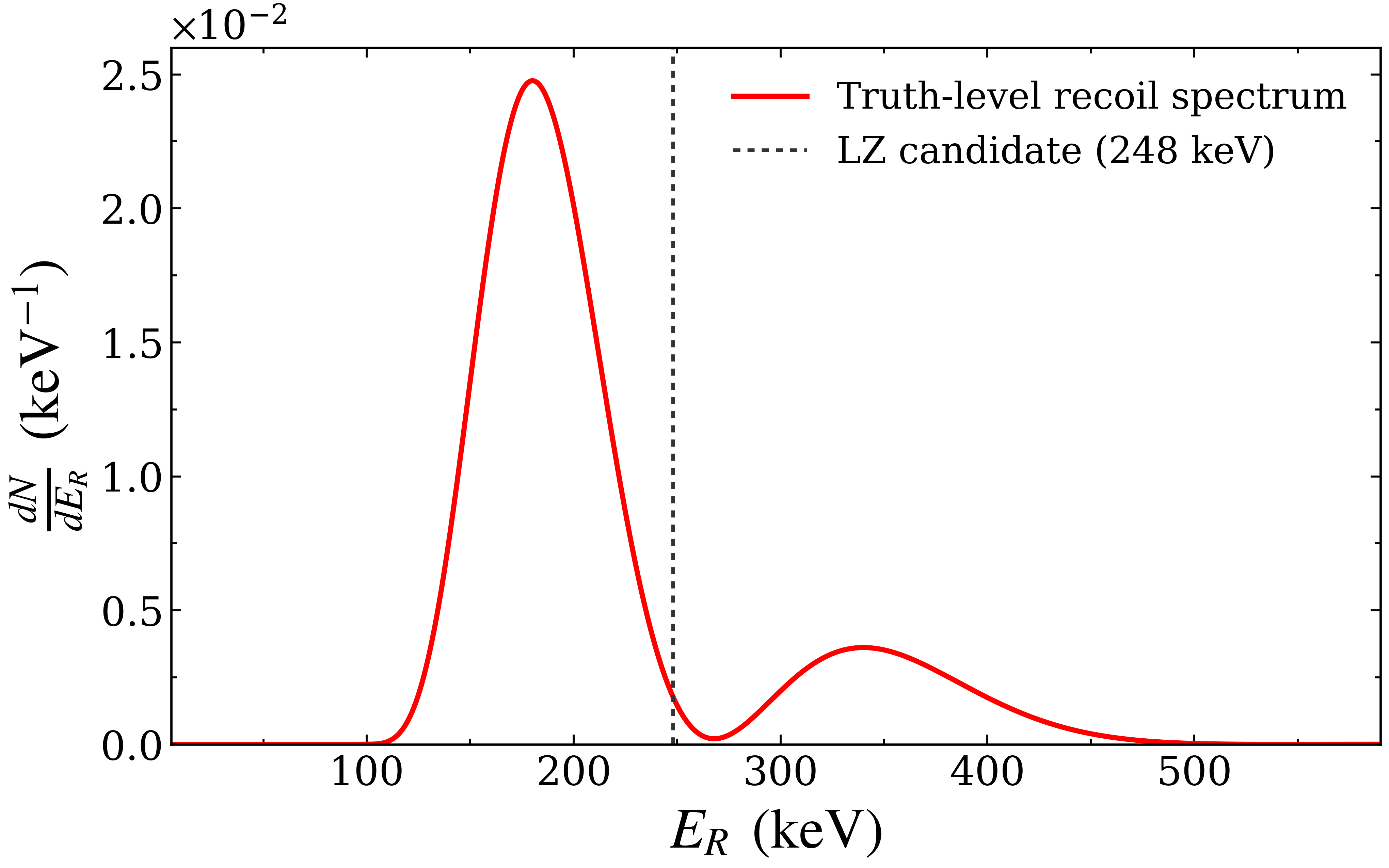}
\caption{
Predicted xenon recoil spectrum $dN_{\rm truth}/dE_R$
for the $2.84$ tonne-year LZ exposure, with
$m_{\phi_1}=100~\mathrm{GeV}$,
$\Delta m=600~\mathrm{keV}$, and $g\simeq0.0515$,
using the Standard Halo Model and finite-momentum
nuclear response.
The spectrum is evaluated at truth level, without
detector efficiency or energy smearing.}
\label{fig:lz}
\end{figure}

For this benchmark, 
the characteristic
xenon recoil scale in Eq.~\eqref{eq:ERstar} is
\begin{equation}
E_R^\star\simeq270\,\keV,
\end{equation}
which lies near the energy of the LZ candidate. Figure~\ref{fig:lz} shows the
full exposure-normalized truth-level recoil energy distribution  (i.e., without taking efficiencies into account ), obtained from
\begin{equation}
\frac{\dd N_{\rm truth}}{\dd E_R}
=
\mathcal E\frac{\dd R}{\dd E_R},
\qquad
\mathcal E=2.84~{\rm tonne\,yr}.
\label{eq:truth-spectrum}
\end{equation}
While the $v_\text{min}=0$ recoil energy is 
$E_R^\star\simeq270~{\rm keV}$, the predicted recoil
spectrum reaches its maximum near $180~{\rm keV}$.
This behavior arises from the interplay between the
halo velocity distribution and the finite-momentum
nuclear response. In particular, the pronounced
diffraction minimum of the adopted xenon response
near $270~{\rm keV}$ suppresses the rate around
$E_R^\star$, shifting the spectral maximum toward
lower recoil energies.

At the reported LZ candidate energy of $248~{\rm keV}$,
the predicted differential event density is
\begin{equation}
\left.
\frac{\dd N_{\rm truth}}{\dd E_R}
\right|_{248\,{\rm keV}}
\simeq
1.8\times10^{-3}~{\rm events/keV},
\label{eq:lz-candidate-density}
\end{equation}
approximately $7\%$ of the maximum spectral density.
Thus, while the mass splitting generates recoils in
the relevant energy range, the nuclear response
favors a substantial population at lower recoil
energies. Integrating the truth-level spectrum over the
recoil-energy range $5.4$--$590~{\rm keV}$ gives
$N_{\rm truth}\simeq2.13$, see Table~\ref{tab:benchmark}.

Integrating over the available digitized efficiency range,
$5.4\leq E_R/{\rm keV}\leq299.9$, we obtain
\begin{equation}
N_\epsilon\simeq1.63.
\label{eq:Nsel-result}
\end{equation}
The integration extends beyond the high-energy
$50\%$ efficiency point at $269.9\,\keV$
\cite{LZ:2026axp}, using the available digitization
up to $299.9\,\keV$. Contributions above this
true-recoil energy are not included.

The recoil-energy distribution provides further
information beyond the total event rate.
Approximately $1.18$ efficiency-weighted events
are expected between $100$ and $200~{\rm keV}$,
compared with about $0.22$ events between
$215$ and $269.9~{\rm keV}$. The latter interval
includes the reported candidate energy and extends
to the high-energy $50\%$ efficiency point.
While a dedicated likelihood analysis for the present
scenario is beyond the scope of this paper, 
the predicted recoil spectrum and event normalization
appear to 
give a reasonable description of the LZ event.

%---------------------------------------------------------
\subsection{Finite-momentum nuclear response}
\label{sec:lz-nuclear}
%---------------------------------------------------------

The predicted recoil spectrum is sensitive to the
finite-momentum scalar nuclear response, particularly
at the high recoil energies relevant to the LZ
candidate. At $E_R=248~{\rm keV}$, the momentum
transfer in xenon is 
$q\simeq 0.25~{\rm GeV}$, and the 
spatial distribution of nucleons is important. 
To examine the dependence on the nuclear response,
we compare our central prediction 
with 
the phenomenological Helm form factor
\cite{Helm:1956zz,Lewin:1995rx}, which increases the
efficiency-weighted event expectation from
$N_\epsilon =1.63$ to $4.70$ events.
The chiral-EFT nuclear structure factors of
Refs.~\cite{Hoferichter:2016nvd,Hoferichter:2018acd}, which also include two-nucleon
contributions, instead yield 1.61
efficiency-weighted events.
These differences reflect
the high sensitivity of the predicted spectrum to the
momentum dependence of the nuclear response,
particularly near its diffraction minima.

Note that the characteristic exothermic recoil scale
depends on the target nucleus through Eq.~(26).
High-recoil searches using xenon, argon, germanium,
and other targets could therefore provide complementary
tests of this scenario~\cite{Su:2026rwz,Baer:2026fpy}.
Other proposed probes of inelastic interpretations
of the LZ event include annual modulation~\cite{McCabe:2026crm},
heavy-element paleodetectors~\cite{Yang:2026wpb,Graham:2026huo},
and searches in large neutrino
detectors~\cite{Graham:2026huo}.
The sensitivity of endothermic interpretations
to the high-velocity tail of the halo, including
perturbations induced by the Large Magellanic Cloud,
has also been studied~\cite{OHare:2026nqi,Bozorgnia:2026emp,Graham:2026huo}.

%

%=========================================================
\section{Correlated Signatures}
\label{sec:indirect}
%=========================================================
%
The interaction in Eq. \eqref{eq:off:portal} governs the thermal relic abundance, 
the scattering on xenon, but also the strength of correlated signatures such as 
the radiative decay of the heavier scalar $\phi_2$
and the annihilation of $\phi_1\phi_2$ into SM particles. Below, we first examine such indirect 
photon signals and then discuss complementary collider searches.

%---------------------------------------------------------
\subsection{Radiative decay of $\phi_2$}
\label{sec:diphoton}
%---------------------------------------------------------

An exothermic scattering signal requires an appreciable
population of the heavier scalar $\phi_2$ to survive
until the present epoch. Although $\phi_2$ is not
absolutely stable, its decay is strongly suppressed
by the small mass splitting.
For
\(\Delta m=600~\keV<2m_e\), the channel
\(\phi_2\to\phi_1e^+e^-\) is kinematically inaccessible.
The leading decay of $\phi_2$
is then $\phi_2\to\phi_1\gamma\gamma$,
which produces a continuous photon spectrum.
Related diphoton
transitions in multicomponent scalar and inelastic
fermionic dark-matter models have been investigated
in Refs.~\cite{Ghosh:2019jzu,Hooper:2025fda}.

After electroweak symmetry breaking, the
interaction in Eq.~\eqref{eq:restrictedportal}
generates the coupling
$-gv_h\phi_1\phi_2h$. The transition to two
photons proceeds through electrically charged
particles coupled to the Higgs sector.
The diphoton invariant mass, 
$ s=(k_1+k_2)^2=2k_1\!\cdot k_2$, with  $k_1$ and $k_2$ the two photon momenta, is limited to the range
$0\leq s\leq(\Delta m)^2$, 
well below the
physical Higgs mass, so that the Higgs can be integrated out. 
The $\phi_2\to\phi_1\gamma\gamma$ transition amplitude 
is~\cite{Marciano:2011gm,Herrero:2020dtv,Feruglio:2025xvc}
\begin{align}
\mathcal M={}&
\epsilon_{1\mu}^{*}\epsilon_{2\nu}^{*}
\frac{4g\,c_\gamma(s)}{m_h^2}
\big[
(k_1\!\cdot k_2)g^{\mu\nu}
-k_2^\mu k_1^\nu
\big],
\label{eq:gamma-amplitude}\\
c_\gamma(s)={}&
\frac{\alpha}{8\pi}
\big[\mathcal P(s)+\mathcal H(s)\big],
\label{eq:gamma-coefficient}
\end{align}
where $m_h$ is the Higgs mass,
$\epsilon_i$ are the photon polarization vectors,
and $\alpha$ the fine-structure
constant.
Here, $\mathcal P(s)$ denotes the perturbative
charged-particle contribution, while
$\mathcal H(s)$ accounts for the low-energy
hadronic contribution to the scalar current.

The perturbative contribution is
\begin{equation}
\mathcal P(s)=
A_1\!\left(\frac{s}{4m_W^2}\right)
+\sum_{f=e,\mu,\tau,c,b,t}
N_c^f Q_f^2
A_{1/2}\!\left(\frac{s}{4m_f^2}\right),
\label{eq:gamma-perturbative}
\end{equation}
where $m_W$ and $m_f$ are the $W$-boson and
fermion masses, $Q_f$ is the electric charge
in units of the positron charge, and
$N_c^f=1$ for leptons and $3$ for quarks. 
For all but the electron, the loop functions can be evaluated 
%the subthreshold kinematics considered here.
in the zero-momentum limit,
$A_1(0)=-7$ and $A_{1/2}(0)=4/3$.
The momentum dependence of the electron loop
must be retained, 
\beq
\begin{split}
A_{1/2}(x)&=
\frac{2[x+(x-1)f(x)]}{x^2},
\\
f(x)&=\arcsin^2\!\sqrt{x},
\label{eq:gamma-loopfunctions}
\end{split}
\eeq
since for the benchmark $s/(4m_e^2)\simeq 0.35$ at the kinematic endpoint.

For the nonperturbative contributions from Higgs coupling to
the light
$u,d,s$ quarks we reuse the LO ChPT results of Ref.~\cite{Feruglio:2025xvc}. Translating to our notation, we get
\begin{equation}
\mathcal H(0)=\frac{23}{27}.
\end{equation}
Including the momentum dependence
of the pion- and kaon-loop contributions changes
the integrated decay width by less than
$10^{-6}$ for $\Delta m=600~\keV$, much less than the size of the neglected NLO ChPT corrections, which are of order $(m_{\pi,K}/4\pi f_{\pi,K})^2\sim {\mathcal O}(20\%)$.

\begin{figure}[t]
\centering
\includegraphics[width=0.48\textwidth]
{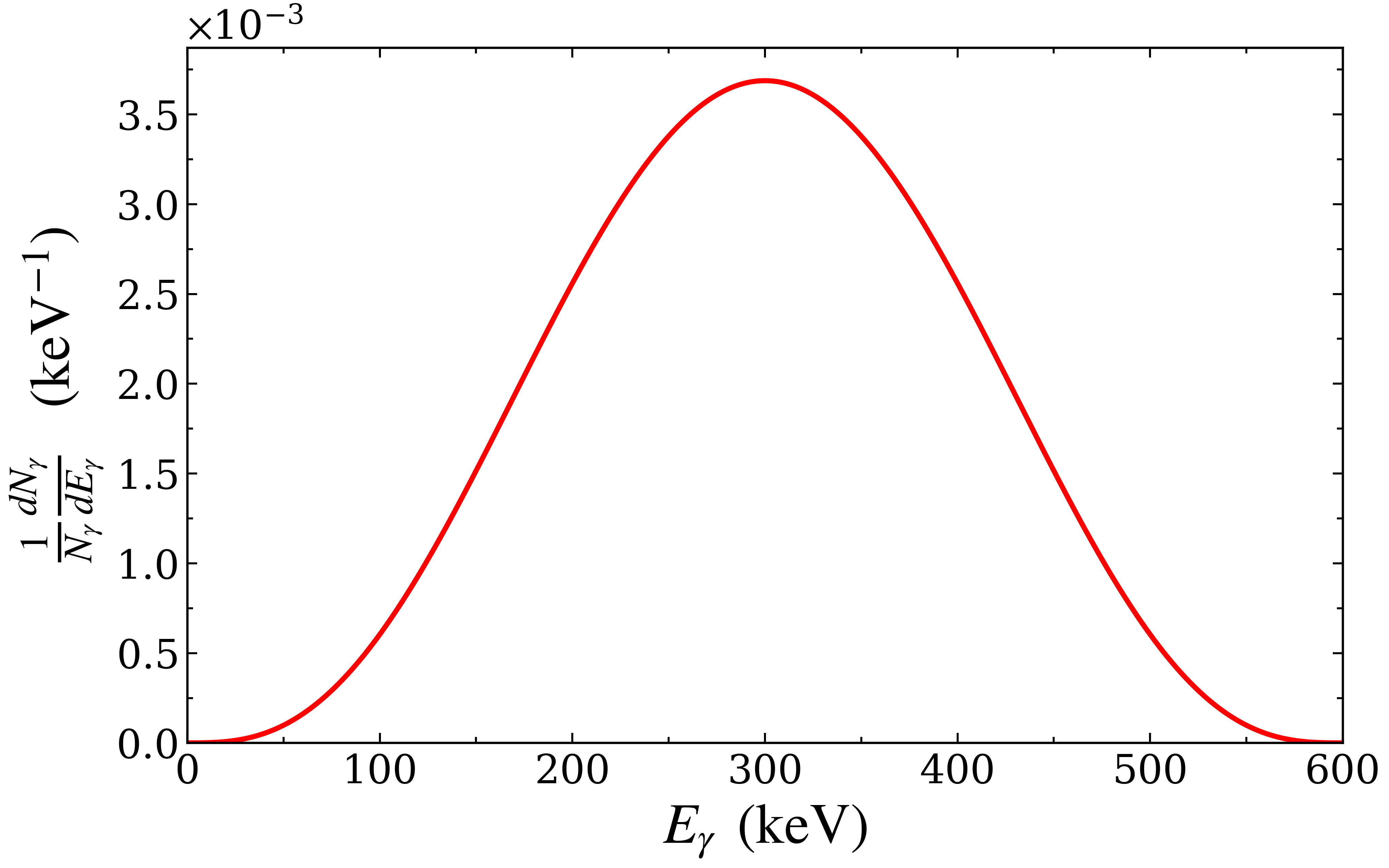}
\caption{
Normalized differential photon-energy distribution,
$(1/N_\gamma)\,dN_\gamma/dE_\gamma$, from
$\phi_2\to\phi_1\gamma\gamma$ in the $\phi_2$
rest frame, for $m_{\phi_1}=100\,\GeV$ and
$\Delta m=600\,\keV$.
The distribution integrates to unity, with
$N_\gamma=2$ photons per decay.
No instrumental response is applied. 
}
\label{fig:photon-yield}
\end{figure}

The differential $\phi_2\to\phi_1\gamma\gamma$ decay width is 
\begin{equation}
\frac{\dd\Gamma_{\gamma\gamma}}{\dd s}
=
\frac{
g^2|c_\gamma(s)|^2s^2
\sqrt{\lambda(m_{\phi_2}^2,m_{\phi_1}^2,s)}
}{
64\pi^3m_{\phi_2}^3(m_h^2-s)^2
},
\label{eq:diphoton-width}
\end{equation}
where
$\lambda(a,b,c)=a^2+b^2+c^2-2ab-2ac-2bc$
is the K\"all\'en function. 
In terms of fractional photon energy, $x=E_\gamma/\Delta m$, the normalized distribution is 
%the normalized spectrum approaches
$(\dd N_\gamma/\dd x )/N_\gamma=140x^3(1-x)^3$~\cite{Krnjaic:2025zjl}, which peaks near
$E_\gamma\simeq\Delta m/2$, see Fig.~\ref{fig:photon-yield}.

In the small-splitting limit, i.e., for approximately constant $c_\gamma(s)\simeq c_\gamma(0)$, the total decay width is
\begin{equation}
\Gamma_{\gamma\gamma}
\simeq
\frac{g^2|c_\gamma(0)|^2}{210\pi^3}
\frac{(\Delta m)^7}
{m_h^4m_{\phi_2}^2}.
\label{eq:diphoton-small-splitting}
\end{equation}
The strong dependence on the mass splitting, 
$\Gamma_{\gamma\gamma}\propto(\Delta m)^7$,
leads to a cosmologically stable
$\phi_2$, despite its large coupling to the Higgs, and the corresponding induced couplings to the
electrically charged SM sector. For the benchmark in Table \ref{tab:benchmark}, 
\begin{equation}
\tau_{\gamma\gamma}^\text{bench}
%=
%\frac{\hbar}{\Gamma_{\gamma\gamma}}
\simeq
4.5\times10^{23}~{\rm s}.
\label{eq:phi2-lifetime}
\end{equation}
For comparison, omitting the hadronic contribution, setting \(\mathcal H=0\), 
while retaining the perturbative loops gives
$\tau_{\gamma\gamma}\simeq1.4\times10^{24}\,\mathrm{s}$.
In either case, the benchmark lifetime is
much longer than the age of the Universe, so
radiative depletion of the excited population is
negligible.

The radiative $\phi_2$ decay signal can be searched for observationally. 
Assuming that $\phi_1$ and $\phi_2$ follow
the same Galactic DM density profile,
the differential photon intensity from
$\phi_2$ decay in the direction $\Omega$
is given by
\begin{equation}
\frac{\dd\Phi_\gamma}
{\dd E_\gamma\,\dd\Omega}
=
\frac{r_2}{4\pi\bar m\,\tau_{\gamma\gamma}}
\frac{\dd N_\gamma}{\dd E_\gamma}
D(\Omega),
\label{eq:decay-flux}
\end{equation}
where 
\begin{equation}
D(\Omega)=
\int_{\rm l.o.s.}\dd\ell\,
\rho_{\rm DM}(\ell,\Omega)
\end{equation}
is the line-of-sight integral of the
Galactic DM density.
We model the Galactic DM distribution
using a Navarro--Frenk--White (NFW) density
profile with scale radius
\(r_s=9.98~{\rm kpc}\).
We take the Sun--Galactic-center distance to be
\(R_{\rm GC}=8.34~{\rm kpc}\) and normalize
the profile to
\(\rho_{\rm DM}(R_{\rm GC})
=\rho_\odot=0.3~{\rm GeV\,cm^{-3}}\),
consistent with the local density adopted
in Sec.~\ref{sec:lz}
\cite{Navarro:1995iw,Navarro:1996gj}.

The predicted photon continuum lies
within the hard X-ray and soft gamma-ray
energy range accessible to the Spectrometer
aboard the INTEGRAL satellite (SPI),
which employs high-resolution germanium
detectors~\cite{2003A&A...411L..63V}.
Analyses of 16 years of SPI observations
have examined diffuse emission from the
inner Galaxy over the energy range
$30~{\rm keV}$--$8~{\rm MeV}$
~\cite{Calore:2022pks}.
The observational analyses employ spatial
templates and models of astrophysical
emission, including diffuse inverse-Compton
radiation, unresolved sources, and
positron-annihilation emission, to constrain
additional photon contributions associated
with Galactic dark-matter decay.
In particular, Ref.~\cite{Krnjaic:2025zjl}
performed a likelihood-based analysis
of related two- and three-photon
continua from inelastic DM,
including the astrophysical background
components in the spectral fit (see also~\cite{Calore:2022pks}).

To constrain the radiative decay of the heavier
scalar, we use the NFW-template spectrum and
energy-redistribution matrix released with the
16-year INTEGRAL/SPI analysis
\cite{Calore:2022pks,Siegert:2022jii}.
The release contains 24 measured-energy bins
spanning \(30~{\rm keV}\)--\(8~{\rm MeV}\)
and a response matrix defined on 377
true-energy bins.
The comparison uses the Galactic region
$|l|,|b|\leq47.5^\circ$, for which the
integrated decay flux is
\begin{equation}
\frac{\dd\Phi_\gamma^{\rm ROI}}
{\dd E_\gamma}
=
\frac{r_2D_{\rm ROI}}
{4\pi\bar m\,\tau_{\gamma\gamma}}
\frac{\dd N_\gamma}{\dd E_\gamma},
\label{eq:spi-flux}
\end{equation}
where
\(D_{\rm ROI}=\int_{\rm ROI}\dd\Omega\,D(\Omega)\).
For the halo parameters specified above,
integration over
\(|l|,|b|\leq47.5^\circ\), with the
line-of-sight integral extended to infinity,
gives
\(D_{\rm ROI}\simeq
8.85\times10^{22}~{\rm GeV\,cm^{-2}}\).
For the numerical evaluation, we use
\(r_2=0.49924\simeq0.50\).
We integrate the Galactic photon flux over the
true-energy bins and apply the released
energy-redistribution matrix to obtain the predicted
channel rates $R_i^{\rm DM}$. These are compared
with the NFW-template rates $R_i$ and their
statistical uncertainties $\sigma_i$. Using a $\chi^2$ constructed from all 
24 measured energy bins gives one-sided $95\%$C.L.
upper bound 
\begin{equation}
\tau_{\rm SPI}^{\rm rec}
\gtrsim
1.1\times10^{22}~{\rm s}.
\label{eq:spi-lifetime}
\end{equation}
The benchmark lifetime in
Eq.~\eqref{eq:phi2-lifetime} is approximately
\(40\times\) longer. Although the reference benchmark lies below the reconstructed INTEGRAL/SPI limit, larger mass splittings can give a substantially enhanced photon signal, since $\Gamma_{\gamma\gamma}\propto g^2(\Delta m)^7/m_{\phi_2}^2$. This is particularly relevant at lower DM masses, where the relic abundance requires a larger portal coupling and an order-one LZ event yield occurs for splittings approaching $1\,\mathrm{MeV}$.
The SPI lifetime constraint therefore excludes part of the LZ favored band in the ($m_\phi, \Delta m$) plane, for higher mass splitting values, shown as white hatched excluded region in  Fig. \ref{fig:lz-thermal-map}.

Note that in obtaining these results,
we neglect inter-bin correlations
and do not propagate systematic uncertainties
from the template and background modeling.
A dedicated analysis of related
inelastic dark-matter photon continua,
including uncertainties in astrophysical background
components, was performed in
Ref.~\cite{Krnjaic:2025zjl}, which also considers
a diphoton continuum with the same leading-order
spectral shape as ours, although a direct comparison
of lifetime limits requires consistent signal
normalization, astrophysical assumptions, and
statistical treatment.

\paragraph{Prospects for COSI.}
The photon continuum from $\phi_2\to\phi_1\gamma\gamma$
falls partly within the $0.2$--$5\,\MeV$ energy range
of the planned COSI mission~\cite{Anthony-Petersen:2026oxk}.
For instance, for the benchmark in Table \ref{tab:benchmark}
about 82.7\% of emitted photons are
above the COSI energy threshold.

Recent forecasts based on all-sky simulations indicate
that two years of COSI observations could strengthen
the lower lifetime limits for
Galactic dark-matter decay lines by more than an order
of magnitude over much of the COSI energy
band~\cite{Anthony-Petersen:2026oxk}. Assuming that comparable sensitivity improvements can
be achieved for continuum decays, COSI is expected
to probe a substantial part of the LZ-favored region
shown in Fig.~\ref{fig:lz-thermal-map}, providing
an important test of this scenario.

%---------------------------------------------------------
\subsection{Annihilation and indirect limits}
\label{sec:annihilation}
%---------------------------------------------------------

The mixed annihilation process $\phi_1\phi_2\to h^*\to {\rm SM}$ that controls the thermal
relic abundance,
also contributes to late-time indirect-detection signals.
This process contains a nonvanishing $s$-wave component,
with the low-velocity cross section
\begin{equation}
\sv_{12}
\simeq 4.52\times10^{-26}
~{\rm cm^3\,s^{-1}}.
\label{eq:sv12-new}
\end{equation}
One can set a constraint on the $\phi_1 \phi_2$ annihilation rate per unit volume for distinguishable initial states, $n_1n_2\sv_{12}$, 
by recasting the bounds on annihilation cross section $\sv_{\rm equiv}$ of a self-conjugate DM, by identifying
\begin{equation}
n_1n_2\sv_{12}
=
\frac12
\left(\frac{\rho_{\rm DM}}{\bar m}\right)^2
\sv_{\rm equiv},
\label{eq:annihilation-normalization}
\end{equation}
which then implies
\begin{equation}
\label{eq:sv:equiv}
\sv_{\rm equiv}
=
2r_1r_2\sv_{12}.
\end{equation}
For 
$r_1\simeq r_2\simeq1/2$ we thus have 
\begin{equation}
\sv_{\rm equiv}
\simeq2.26\times10^{-26}
~{\rm cm^3\,s^{-1}}.
\label{eq:sv-equivalent}
\end{equation}
The factor $2r_1r_2$ in \eqref{eq:sv:equiv} accounts for both the relative
populations and the distinguishable initial states.
At the Table \ref{tab:benchmark} benchmark mass, the  $\phi_1\phi_2\to h^*\to {\rm SM}$ annihilation 
produces predominantly $W^+W^-$ and 
$ZZ$ in roughly $3:1$ proportion, 
with other SM final states contributing below percent level.

We derive the dwarf-galaxy constraint shown in
Fig.~\ref{fig:lz-thermal-map} using the released
energy-bin likelihoods from the 14-year
\emph{Fermi}-LAT analysis~\cite{McDaniel:2023bju}.
For each dark-matter mass, we construct the
predicted gamma-ray spectrum using the
mass-dependent annihilation rates and branching
fractions, including off-shell electroweak
channels below threshold and Higgs-pair
final states where kinematically accessible.
We combine the likelihoods of the 42 dwarf
spheroidal galaxies in the Benchmark sample,
which contains 30 galaxies with measured
$J$ factors and 12 with estimated values.
The $J$-factor uncertainties are included
by profiling the corresponding nuisance
parameters.

We obtain the observed one-sided 95\%
confidence-level upper limit on the
equivalent annihilation cross section by
requiring an increase of 2.71 in the
profiled $-2\ln\mathcal L$ relative to
its minimum, with a nonnegative signal
normalization. Since the predicted spectra
depend only weakly on $\Delta m$, we evaluate
the likelihood as a function of dark-matter
mass at a representative splitting of
$600\,\keV$ and apply the resulting limits
across the scanned splitting range.
For the benchmark in Table~\ref{tab:benchmark},
we obtain
\begin{equation}
\sv_{\rm equiv}
<
3.34\times10^{-26}~{\rm cm^3\,s^{-1}}
\qquad (95\%~{\rm C.L.}).
\end{equation}
The predicted value,
$\sv_{\rm equiv}\simeq
2.26\times10^{-26}~{\rm cm^3\,s^{-1}}$,
lies below this limit. Across the thermal
parameter space, the predicted annihilation
signal exceeds the reconstructed limit
for $m_{\phi_1}\lesssim72\,\GeV$, as shown in Fig.~\ref{fig:lz-thermal-map}.

A combined analysis of dwarf-spheroidal observations by
\emph{Fermi}-LAT, HAWC, H.E.S.S., MAGIC, and VERITAS
provides complementary constraints on annihilation into
electroweak gauge bosons~\cite{Fermi-LAT:2025gei}.
The reported limits depend appreciably on the adopted
dwarf-galaxy $J$ factors. Because the benchmark predicts
a specific mixture of $W^+W^-$ and $ZZ$ final states, an
exact application of this result requires a
likelihood-level evaluation using the corresponding
mixed gamma-ray spectrum.

Annihilation around recombination also deposits energy
into the photon--baryon plasma. The relevant
energy-injection parameter is
\begin{equation}
p_{\rm ann}
=
f_{\rm eff}
\frac{\sv_{\rm equiv}}{\bar m},
\label{eq:pann}
\end{equation}
where $f_{\rm eff}$ is the effective energy-deposition
efficiency for the predicted annihilation products.
We compute the injected photon and
electron--positron spectra per
\(\phi_1\phi_2\) annihilation, including the
branching fractions quoted above, and convolve
them with the deposition-efficiency functions
of Ref.~\cite{Slatyer:2015jla}. This gives
$
f_{\rm eff}\simeq0.179$,
 $p_{\rm ann}
\simeq4.05\times10^{-29}
~{\rm cm^3\,s^{-1}\,GeV^{-1}}$.
This prediction lies about an order of magnitude below the 95\% upper limits
$p_{\rm ann} < 3.2\times10^{-28}~{\rm cm^3\,s^{-1}\,GeV^{-1}}$
from \emph{Planck} (TT,TE,EE+lowE+lensing)~\cite{Planck:2018vyg} and
$p_{\rm ann} < 3.6\times10^{-28}~{\rm cm^3\,s^{-1}\,GeV^{-1}}$
from the combination of \emph{Planck} and ACT DR6 primary CMB,
CMB lensing, and DESI BAO data (P-ACT-LB)~\cite{AtacamaCosmologyTelescope:2025nti}.
For this comparison, we use the state fractions
obtained from the coupled cosmological evolution
at recombination. They differ negligibly from
their present-day values because state-changing
reactions and radiative decay are ineffective
over the intervening epoch.

Although the diagonal Higgs couplings vanish at
tree level, processes with identical incoming
scalars are not strictly absent. In particular,
\(t\)- and \(u\)-channel \(\phi_2\) exchange mediates
\begin{equation}
\phi_1\phi_1\to h h^*\to h f\bar f ,
\end{equation}
where \(f\) denotes a Standard Model fermion. For
the benchmark, its low-velocity cross section is
$
\sv_{11\to h f\bar f}
\simeq
4.55\times10^{-33}
~{\rm cm^3\,s^{-1}}.
$
Including the identical-pair factor \(1/2\) and
the relative abundances, this channel contributes
approximately \(5\times10^{-8}\) of the Galactic
\(\phi_1\phi_2\) annihilation source. The
loop-induced \(\phi_1\phi_1\to W^+W^-\)
contribution is suppressed by a similar factor.
Both are negligible for the dwarf-galaxy and CMB
comparisons above. Their relative importance differs
in the Sun, where capture produces a much larger
population of \(\phi_1\) than of \(\phi_2\); see
Sec.~\ref{sec:solar}.

\subsection{Collider and other probes}
\label{sec:other-tests}
% ---------------------------------------------------------

Collider searches provide an additional probe
of the off-diagonal Higgs portal. If
\(m_{\phi_1}+m_{\phi_2}<m_h\), the mixed pair
can be produced through the on-shell Higgs decay
$h\rightarrow\phi_1\phi_2$.
For particles that escape the detector, the
corresponding partial width is constrained by
Higgs-rate measurements and searches for invisible
Higgs decays
\cite{Curtin:2013fra,Casas:2017jjg,ATLAS:2022yvh,
CMS:2023sdw,ATLAS:2023tkt}. This is not an issue for our benchmark and for masses shown in Fig. \ref{fig:lz-thermal-map}, 
since in all these cases 
\(m_{\phi_1}+m_{\phi_2}>m_h\).
The diagonal channels \(h\to\phi_i\phi_i\)
are also absent at tree level because the
diagonal portal couplings are set to
\(f_1=f_2=0\).

The leading collider production channel of $\phi_1 \phi_2$ is therefore
off-shell Higgs production accompanied by initial-state radiation,
\begin{equation}
pp\rightarrow h^\ast+j\rightarrow\phi_1\phi_2+j,
\end{equation}
resulting in a monojet signature ($\phi_2$ is stable on collider scales)~\cite{Fox:2011pm,ATLAS:2021kxv,CMS:2021far,CMS:2022qva}.
Due to additional 3 body final state phase space suppression, 
existing collider limits on this topology are not expected to compete
with the direct and indirect probes for the moderate portal coupling $g\simeq0.05$ considered here.

The singlet states can also contribute radiatively to off-shell Higgs
processes such as
$gg\rightarrow h^\ast\rightarrow ZZ$, modifying the high-invariant-mass
diboson spectrum
\cite{Goncalves:2017iub,Goncalves:2018pkt}, though for the typical values of $g$ considered here these are again expected to be below experimental sensitivity. 

Finally, the characteristic exothermic recoil scale depends on the
target nucleus, see Eq.~\eqref{eq:ERstar}. High-recoil searches with
xenon, argon, germanium, and other targets could therefore provide
complementary tests of this scenario.

% =========================================================
\section{Solar capture and annihilation}
\label{sec:solar}
% =========================================================

DM entering the Sun can scatter on solar nuclei, $\phi_{1}A\to \phi_{2}A$ and vice versa, and become
gravitationally bound. Such scattering 
changes the
relative populations of \(\phi_1\) and \(\phi_2\), so that these in general no longer equal the
Galactic abundances. Moreover,
because diagonal elastic scattering arises only radiatively, repeated
collisions need not  lead to common thermal distributions. 
We follow the captured particles along their
nonthermal orbits, allowing for subsequent
inelastic transitions and possible ejection,
rather than assuming a thermal spatial distribution.
The numerical solar analysis below focuses on
the benchmark in Table~\ref{tab:benchmark}.

\subsection{Inelastic capture and solar population evolution}
\label{sec:solar-population}

The first of the two possible off-diagonal portal scatterings
\begin{equation}
\phi_2 A\rightarrow\phi_1 A,
\qquad
\phi_1 A\rightarrow\phi_2 A,
\end{equation}
 is exothermic
and provides the dominant capture channel. For a stationary target,
the reverse endothermic transition requires
\begin{equation}
w\geq w_{{\rm thr},A}
=
\sqrt{\frac{2\Delta m}{\mu_{1A}}},
\qquad
\mu_{1A}
=
\frac{m_{\phi_1}m_A}{m_{\phi_1}+m_A},
\label{eq:solar-upthreshold}
\end{equation}
where \(w\) is the local dark-matter--nucleus relative speed. For the
benchmark in Table  \ref{tab:benchmark} and a stationary \(^{56}{\rm Fe}\) nucleus,
\begin{equation}
w_{{\rm thr},{\rm Fe}}
\simeq
1800~{\rm km\,s^{-1}},
\end{equation}
which exceeds the maximum incident speed at the solar center,
\(w_{\max}\simeq1600~{\rm km\,s^{-1}}\), for the adopted halo
distribution and solar potential. 

For stationary iron nuclei, the endothermic
transition is therefore kinematically inaccessible
at the benchmark. In the numerical calculation,
we include thermal nuclear motion in both halo
capture and subsequent transitions along captured
orbits. Thermal motion removes the strict
stationary-target threshold, although the
contribution from iron to endothermic halo capture
remains negligible. For the adopted solar
composition, endothermic transitions are dominated
by heavier trace nuclei. These processes affect
the small captured excited-state population,
but the resulting mixed annihilation rate remains
strongly suppressed, as discussed below.

Exothermic scattering does not necessarily lead to capture, since the
outgoing \(\phi_1\) can remain gravitationally unbound. We define
\(C_{2\to1}\) as the number of incident \(\phi_2\) particles per unit
time that scatter into gravitationally bound \(\phi_1\) orbits. In
terms of the halo speed distribution, this rate can be written as
\begin{align}
\label{eq:solar-capture-integral}
C_{2\to1}
&=
n_{2,\infty}
\sum_A
\int_0^{R_\odot}\!4\pi r^2\,{\rm d}r \\ \nonumber
&~~ \times \int_0^\infty\!{\rm d}u\,
\frac{f_\odot(u)}{u}\,
w_\chi\,
\Omega_{2\to1,A}^{-}(w_\chi,r),
\end{align}
where
\begin{equation}
n_{2,\infty}
=
\frac{r_2\rho_{\rm DM}}{\bar m},
\qquad
w_\chi^2
=
u^2+v_{\rm esc}^2(r).
\end{equation}
Here \(f_\odot(u)\) is the unit-normalized halo speed distribution in
the solar frame. The quantity
\(\Omega_{2\to1,A}^{-}\) is the scattering rate per incident particle
on nuclear species \(A\), averaged over the thermal target
distribution and integrated over final states in which the outgoing
\(\phi_1\) is gravitationally bound. We evaluate
Eq.~\eqref{eq:solar-capture-integral} using the adopted solar density
and composition profiles, the nuclear form factors, and the exact
inelastic two-body kinematics~\cite{Blennow:2018xwu}. For the benchmark,
\begin{equation}
C_{2\to1}
\simeq
9.3\times10^{19}~{\rm s^{-1}}.
\label{eq:C21}
\end{equation}

The phase-space point following the capture collision fixes the
initial bound orbit. We evolve these orbits through subsequent
passages through the solar interior, allowing for further scattering,
conversion, and ejection. For a constant capture rate, the present-day
density is
\begin{equation}
n_i^{\rm cap}(\mathbf r,t_\odot)
=
C_{2\to1}
\int_0^{t_\odot}\!{\rm d}a\,
p_i(\mathbf r,a),
\label{eq:solar-density-construction}
\end{equation}
where \(t_\odot\) is the age of the Sun and
\(p_i(\mathbf r,a)\,{\rm d}^3r\) is the probability that a particle
captured a time \(a\) earlier occupies the volume \({\rm d}^3r\) in
state \(\phi_i\). This probability includes subsequent conversion
and ejection.

 Approximately \(72\%\) of the particles captured as \(\phi_1\)
undergo no further state-changing collision in the resulting orbital
histories. The spatial densities are therefore constructed by
weighting each orbit by the fraction of its period spent in each
radial interval, rather than by imposing a common isothermal profile.
Radiative conversion is likewise negligible on solar timescales.
Even if a bound \(\phi_2\) remained in that state for the full age of
the Sun, $t_\odot$, the decay time is orders of magnitude larger, 
${t_\odot}/{\tau_{\gamma\gamma}}
\simeq
3.2\times10^{-7}$.

\subsection{Solar annihilation and neutrino constraints}
\label{sec:solar-annihilation}

Let \(n_i^{\rm cap}(\mathbf r)\) denote the present-day number density
of gravitationally bound \(\phi_i\) particles. For annihilations
occurring within the solar interior, we define
\begin{subequations}
\begin{align}
I_{11}^{\rm cap}
&=
\int_{r<R_\odot}\!{\rm d}^3r\,
\big[n_1^{\rm cap}(\mathbf r)\big]^2,
\\
I_{12}^{\rm cap}
&=
\int_{r<R_\odot}\!{\rm d}^3r\,
n_1^{\rm cap}(\mathbf r)n_2^{\rm cap}(\mathbf r).
\label{eq:solar-overlaps}
\end{align}
\end{subequations}
The orbital calculation gives
\begin{equation}
I_{11}^{\rm cap}
\simeq
5.3\times10^{41}~{\rm cm^{-3}},
\qquad
I_{12}^{\rm cap}
\simeq
5.8\times10^{30}~{\rm cm^{-3}}.
\label{eq:solar-overlap-values}
\end{equation}
Solar capture is dominated by $\phi_2 A\to\phi_1 A$, leaving most captured dark matter in the lighter state. Subsequent endothermic scattering can regenerate $\phi_2$, but these particles spend relatively little time in the excited state before converting back or being ejected. The resulting small $\phi_2$ population accounts for the hierarchy $I_{12}^{\rm cap}\ll I_{11}^{\rm cap}$.
The rate for
annihilation between the two captured populations is
\begin{equation}
\Gamma_{12}^{\rm cap}
=
I_{12}^{\rm cap}\sv_{12}
\simeq
2.6\times10^5~{\rm s^{-1}},
\label{eq:solar-gamma12}
\end{equation}
where we used $\sv_{12}$ in  \eqref{eq:sv12-new}.
The small value of \(\Gamma_{12}^{\rm cap}\) is caused by the limited
spatial overlap of the captured \(\phi_1\) and \(\phi_2\) populations.

The much larger captured \(\phi_1\) population also permits
annihilation of two \(\phi_1\) particles. For a final state \(X\), the
corresponding event rate is
\begin{equation}
\Gamma_{11}^{\rm cap,X}
=
\frac{1}{2}
I_{11}^{\rm cap}
\sv_{11\to X},
\label{eq:solar-identical-rate}
\end{equation}
where the factor \(1/2\) avoids double counting pairs of identical
incoming particles. At tree level, two off-diagonal portal vertices generate
\begin{equation}
\phi_1\phi_1\rightarrow hh^*\rightarrow h f\bar f
\end{equation}
through intermediate \(\phi_2\) exchange. Although suppressed by the
three-body phase space and the off-shell Higgs propagator, this process
has the low-velocity rate coefficient
\begin{equation}
\sv_{11\to h f\bar f}
\simeq
4.6\times10^{-33}~{\rm cm^3\,s^{-1}},
\label{eq:solar-hff-cross-section}
\end{equation}
and therefore gives
\begin{equation}
\Gamma_{11}^{\rm cap,h f\bar f}
=
\frac{1}{2}
I_{11}^{\rm cap}
\sv_{11\to h f\bar f}
\simeq
1.2\times10^9~{\rm s^{-1}}.
\label{eq:solar-hff-rate}
\end{equation}

The captured \(\phi_1\) population also annihilates through the
loop-induced channel
\begin{equation}
\phi_1\phi_1\rightarrow W^+W^-.
\end{equation}
The renormalized one-loop amplitude includes the induced diagonal
Higgs form factor together with the contact, gauge, and Goldstone
contributions required by electroweak gauge invariance, retaining
their interference. The resulting threshold rate coefficient is
\begin{equation}
\sv_{11\to WW}
\simeq
2.2\times10^{-33}~{\rm cm^3\,s^{-1}}.
\label{eq:solar-ww-cross-section}
\end{equation}
Its velocity dependence is negligible over the range populated by
the captured orbits. Eq.~\eqref{eq:solar-identical-rate} then
gives
\begin{equation}
\Gamma_{11}^{\rm cap,WW}
=
\frac{1}{2}
I_{11}^{\rm cap}
\sv_{11\to WW}
\simeq
5.9\times10^8~{\rm s^{-1}}.
\label{eq:solar-ww-rate}
\end{equation}
The \(h f\bar f\) rate is approximately twice the loop-induced
\(W^+W^-\) rate. Although both \(\phi_1\phi_1\) rate coefficients are
roughly seven orders of magnitude below \(\sv_{12}\), their event
rates are larger than \(\Gamma_{12}^{\rm cap}\) because
\(I_{11}^{\rm cap}\) greatly exceeds \(I_{12}^{\rm cap}\).

The calculated annihilation channels remove particles from the
captured population at a rate 
\begin{equation}
\frac{
2\left(
\Gamma_{12}^{\rm cap}
+
\Gamma_{11}^{\rm cap,h f\bar f}
+
\Gamma_{11}^{\rm cap,WW}
\right)
}{
C_{2\to1}
}
\simeq
3.9\times10^{-11},
\label{eq:solar-depletion}
\end{equation}
where the factor of two counts the two captured particles removed in
each event. Annihilation between captured particles therefore has a
negligible effect on the solar population, and the calculated
captured-pair channels do not establish capture--annihilation
equilibrium.

The preceding rates involve two particles drawn from the captured
solar population. A separate contribution arises when a captured
\(\phi_1\) annihilates with an unbound Galactic \(\phi_2\) passing
through the Sun,
\begin{equation}
\phi_1^{\rm cap}\phi_2^{\rm halo}
\rightarrow
h^*
\rightarrow
W^+W^-,\,ZZ,\ldots .
\end{equation}
Denoting the gravitationally focused density of the incident
population by \(n_2^{\rm halo}(\mathbf r)\), its solar-interior rate is
\begin{equation}
\Gamma_{12}^{\rm cap\text{-}halo}
=
\int_{r<R_\odot}\!{\rm d}^3r\,
n_1^{\rm cap}(\mathbf r)n_2^{\rm halo}(\mathbf r)
\left\langle\sigma v_{\rm rel}\right\rangle_{12,\mathbf r}.
\label{eq:solar-captured-halo}
\end{equation}
The velocity average combines the nonthermal captured distribution
with the incident halo phase-space distribution transported through
the solar potential~\cite{Alenazi:2006wu}. This contribution is not
contained in \(I_{12}^{\rm cap}\) and does not enter
Eq.~\eqref{eq:solar-depletion}; each such event removes only one
captured particle. Its evaluation requires the position-dependent
overlap of the captured orbits with the gravitationally focused halo
distribution and is therefore kept separate from the numerical rates
quoted above.

For comparison with neutrino-telescope searches, we use the calculated
\(\phi_1\phi_1\to W^+W^-\) component. Expressing the released
\(m_\chi=100~{\rm GeV}\), pure-\(W^+W^-\) result from the ten-year
IceCube analysis as an annihilation-rate limit, using the capture
prescription and equilibrium normalization adopted in that analysis,
gives~\cite{IceCube:2025fcu}
\begin{equation}
\Gamma_{\rm IC}^{WW,90}
\simeq
5.4\times10^{21}~{\rm s^{-1}}
\qquad
(90\%~{\rm C.L.}).
\label{eq:icecube-10yr-rate}
\end{equation}
Here the equilibrium relation is used only to translate the
normalization of the IceCube signal template and is not imposed on
the captured population of the present model. The calculated
\(W^+W^-\) component satisfies
\begin{equation}
\frac{
\Gamma_{11}^{\rm cap,WW}
}{
\Gamma_{\rm IC}^{WW,90}
}
\simeq
1.1\times10^{-13}.
\label{eq:icecube-ratio}
\end{equation}

The \(h f\bar f\) event rate lies more than twelve orders of
magnitude below this rate (though the pure-\(W^+W^-\) IceCube
template cannot be applied directly because its neutrino spectrum is
different). The captured-particle--incident-halo contribution produces
a mixture of electroweak final states and has not been evaluated
numerically. A complete solar-neutrino prediction would require its
normalization and spatial distribution, together with the
channel-dependent neutrino yields, propagation through the Sun, and
detector response. The explicitly calculated captured-pair
contributions are nevertheless far below present neutrino-telescope
sensitivity.

\section{Conclusions}
\label{sec:conclusions}

We have studied a minimal thermal realization of exothermic dark
matter in which two nearly degenerate real scalars interact with
the SM predominantly through an off-diagonal Higgs
portal. For a representative benchmark with
$m_{\phi_1}\simeq100\,\mathrm{GeV}$ and
$\Delta m\simeq600\,\mathrm{keV}$, the observed relic abundance
requires a portal coupling $g\simeq0.05$, while the subsequent
conversion history leaves approximately half of the dark matter
in the excited state today. 

In the $\phi_2 A\to\phi_1 A$  scattering the $\phi_2-\phi_1$ mass
splitting can produce nuclear recoils in the few-hundred-keV
range without requiring particles from the extreme high-velocity
tail of the halo. With the interaction strength fixed by the
relic-density requirement, the benchmark gives an order-one
efficiency-weighted event expectation over the available
true-recoil efficiency range for the reported LZ exposure.

The same interaction gives rise to the radiative transition
$\phi_2\to\phi_1\gamma\gamma$, linking the recoil signal to a
broad sub-MeV photon continuum. The predicted lifetime
is sufficiently
long to preserve the excited population and exceeds the lower
bound recast from the released INTEGRAL/SPI spectrum.
Present-day mixed annihilation also lies below the
Fermi-LAT dwarf-galaxy and CMB limits; 
the margin in the dwarf-galaxy comparison is modest and already excludes part of the parameter space 
where $\phi_1$ mass approaches half of the Higgs mass region.

Solar capture produces qualitatively different two-state
dynamics. Exothermic scattering preferentially populates the
ground state, while inefficient regeneration of the excited
state suppresses mixed annihilation between captured particles.
Consequently, the tree-level three-body channel
$\phi_1\phi_1\to h f\bar f$ and the loop-induced channel
$\phi_1\phi_1\to W^+W^-$ exceed the captured mixed-state
annihilation rate despite their much smaller rate coefficients.
The resulting annihilation rates are well below the Ice-Cube bounds.

The central prediction of the above explanation of the LZ event 
is the connection
between a thermally determined two-state population,
target-dependent high-energy nuclear recoils, and a photon
continuum signal in indirect DM searches, such as in COSI.
Higher-statistics
high-recoil measurements in xenon and complementary nuclear
targets, together with sub-MeV gamma-ray observations, could
thus test the proposed scenario. 

{\bf Acknowledgments.} 
SR acknowledges many useful discussions at Fermilab.
The work of SR and JZ at the University of Cincinnati
has been supported by DOE Grants Nos. DE-SC0011784
and DE-SC0026301 and NSF Grant No. OAC-2411215.
JZ also acknowledges support in part from NSF Grants
Nos. OAC-2103889 and OAC-2417682.

\bibliographystyle{apsrev4-2}
\bibliography{references}

\begin{thebibliography}{155}%
\makeatletter
\providecommand \@ifxundefined [1]{%
 \@ifx{#1\undefined}
}%
\providecommand \@ifnum [1]{%
 \ifnum #1\expandafter \@firstoftwo
 \else \expandafter \@secondoftwo
 \fi
}%
\providecommand \@ifx [1]{%
 \ifx #1\expandafter \@firstoftwo
 \else \expandafter \@secondoftwo
 \fi
}%
\providecommand \natexlab [1]{#1}%
\providecommand \enquote  [1]{``#1''}%
\providecommand \bibnamefont  [1]{#1}%
\providecommand \bibfnamefont [1]{#1}%
\providecommand \citenamefont [1]{#1}%
\providecommand \href@noop [0]{\@secondoftwo}%
\providecommand \href [0]{\begingroup \@sanitize@url \@href}%
\providecommand \@href[1]{\@@startlink{#1}\@@href}%
\providecommand \@@href[1]{\endgroup#1\@@endlink}%
\providecommand \@sanitize@url [0]{\catcode `\\12\catcode `\$12\catcode `\&12\catcode `\#12\catcode `\^12\catcode `\_12\catcode `\%12\relax}%
\providecommand \@@startlink[1]{}%
\providecommand \@@endlink[0]{}%
\providecommand \url  [0]{\begingroup\@sanitize@url \@url }%
\providecommand \@url [1]{\endgroup\@href {#1}{\urlprefix }}%
\providecommand \urlprefix  [0]{URL }%
\providecommand \Eprint [0]{\href }%
\providecommand \doibase [0]{https://doi.org/}%
\providecommand \selectlanguage [0]{\@gobble}%
\providecommand \bibinfo  [0]{\@secondoftwo}%
\providecommand \bibfield  [0]{\@secondoftwo}%
\providecommand \translation [1]{[#1]}%
\providecommand \BibitemOpen [0]{}%
\providecommand \bibitemStop [0]{}%
\providecommand \bibitemNoStop [0]{.\EOS\space}%
\providecommand \EOS [0]{\spacefactor3000\relax}%
\providecommand \BibitemShut  [1]{\csname bibitem#1\endcsname}%
\let\auto@bib@innerbib\@empty
%</preamble>
\bibitem [{\citenamefont {Akerib}\ \emph {et~al.}(2026)\citenamefont {Akerib} \emph {et~al.}}]{LZ:2026axp}%
  \BibitemOpen
  \bibfield  {author} {\bibinfo {author} {\bibfnamefont {D.~S.}\ \bibnamefont {Akerib}} \emph {et~al.} (\bibinfo {collaboration} {LZ}),\ }\href {https://doi.org/10.17182/hepdata.182472.v1} {\bibinfo {title} {{Search for dark matter particle interactions in an extended nuclear recoil energy window with the LUX-ZEPLIN (LZ) experiment}}} (\bibinfo {year} {2026}),\ \Eprint {https://arxiv.org/abs/2609.02823} {arXiv:2609.02823 [hep-ex]} \BibitemShut {NoStop}%
\bibitem [{\citenamefont {Tucker-Smith}\ and\ \citenamefont {Weiner}(2001)}]{Tucker-Smith:2001myb}%
  \BibitemOpen
  \bibfield  {author} {\bibinfo {author} {\bibfnamefont {D.}~\bibnamefont {Tucker-Smith}}\ and\ \bibinfo {author} {\bibfnamefont {N.}~\bibnamefont {Weiner}},\ }\href {https://doi.org/10.1103/PhysRevD.64.043502} {\bibfield  {journal} {\bibinfo  {journal} {Phys. Rev. D}\ }\textbf {\bibinfo {volume} {64}},\ \bibinfo {pages} {043502} (\bibinfo {year} {2001})},\ \Eprint {https://arxiv.org/abs/hep-ph/0101138} {arXiv:hep-ph/0101138} \BibitemShut {NoStop}%
\bibitem [{\citenamefont {Fan}\ and\ \citenamefont {Reece}(2026)}]{Fan:2026kxx}%
  \BibitemOpen
  \bibfield  {author} {\bibinfo {author} {\bibfnamefont {J.}~\bibnamefont {Fan}}\ and\ \bibinfo {author} {\bibfnamefont {M.}~\bibnamefont {Reece}},\ }\href@noop {} {\bibinfo {title} {{Higgsino Above the Sea of Fog}}} (\bibinfo {year} {2026}),\ \Eprint {https://arxiv.org/abs/2609.01504} {arXiv:2609.01504 [hep-ph]} \BibitemShut {NoStop}%
\bibitem [{\citenamefont {Freese}\ and\ \citenamefont {Theodosopoulos}(2026)}]{Freese:2026sga}%
  \BibitemOpen
  \bibfield  {author} {\bibinfo {author} {\bibfnamefont {K.}~\bibnamefont {Freese}}\ and\ \bibinfo {author} {\bibfnamefont {D.~P.}\ \bibnamefont {Theodosopoulos}},\ }\href@noop {} {\bibinfo {title} {{Higgsino Dark Matter Interpretation of the LUX-ZEPLIN 248 keV Nuclear-Recoil Event}}} (\bibinfo {year} {2026}),\ \Eprint {https://arxiv.org/abs/2609.01583} {arXiv:2609.01583 [hep-ph]} \BibitemShut {NoStop}%
\bibitem [{\citenamefont {Wu}\ \emph {et~al.}(2026)\citenamefont {Wu}, \citenamefont {Zhang},\ and\ \citenamefont {Zhu}}]{Wu:2026nhi}%
  \BibitemOpen
  \bibfield  {author} {\bibinfo {author} {\bibfnamefont {L.}~\bibnamefont {Wu}}, \bibinfo {author} {\bibfnamefont {Y.}~\bibnamefont {Zhang}},\ and\ \bibinfo {author} {\bibfnamefont {B.}~\bibnamefont {Zhu}},\ }\href@noop {} {\bibinfo {title} {{TeV Higgsino Dark Matter from LZ Nuclear Recoil to Fermi-LAT Gamma Rays}}} (\bibinfo {year} {2026}),\ \Eprint {https://arxiv.org/abs/2609.01590} {arXiv:2609.01590 [hep-ph]} \BibitemShut {NoStop}%
\bibitem [{\citenamefont {Yin}(2026)}]{Yin:2026jnn}%
  \BibitemOpen
  \bibfield  {author} {\bibinfo {author} {\bibfnamefont {W.}~\bibnamefont {Yin}},\ }\href@noop {} {\bibinfo {title} {{A PQ-Symmetric High-Scale SUSY Interpretation of the LZ High-Energy Recoil}}} (\bibinfo {year} {2026}),\ \Eprint {https://arxiv.org/abs/2609.01892} {arXiv:2609.01892 [hep-ph]} \BibitemShut {NoStop}%
\bibitem [{\citenamefont {Nomura}(2026)}]{Nomura:2026qyq}%
  \BibitemOpen
  \bibfield  {author} {\bibinfo {author} {\bibfnamefont {Y.}~\bibnamefont {Nomura}},\ }\href@noop {} {\bibinfo {title} {{Dark Matter as the Z{\_}2 Partner of the Standard Model Higgs Boson}}} (\bibinfo {year} {2026}),\ \Eprint {https://arxiv.org/abs/2609.02505} {arXiv:2609.02505 [hep-ph]} \BibitemShut {NoStop}%
\bibitem [{\citenamefont {Di~Mauro}(2026)}]{DiMauro:2026ldr}%
  \BibitemOpen
  \bibfield  {author} {\bibinfo {author} {\bibfnamefont {M.}~\bibnamefont {Di~Mauro}},\ }\href@noop {} {\bibinfo {title} {{Dark Matter at the Kinematic Edge: Interpreting the 248 keV LZ Nuclear-Recoil Candidate}}} (\bibinfo {year} {2026}),\ \Eprint {https://arxiv.org/abs/2609.02608} {arXiv:2609.02608 [hep-ph]} \BibitemShut {NoStop}%
\bibitem [{\citenamefont {Visinelli}(2026)}]{Visinelli:2026kgt}%
  \BibitemOpen
  \bibfield  {author} {\bibinfo {author} {\bibfnamefont {L.}~\bibnamefont {Visinelli}},\ }\href@noop {} {\bibinfo {title} {{A Peccei-Quinn Origin for Inelastic Electroweak Dark Matter after LUX-ZEPLIN}}} (\bibinfo {year} {2026}),\ \Eprint {https://arxiv.org/abs/2609.02807} {arXiv:2609.02807 [hep-ph]} \BibitemShut {NoStop}%
\bibitem [{\citenamefont {Smirnov}\ \emph {et~al.}(2026)\citenamefont {Smirnov}, \citenamefont {Griffith},\ and\ \citenamefont {Beacom}}]{Smirnov:2026aqk}%
  \BibitemOpen
  \bibfield  {author} {\bibinfo {author} {\bibfnamefont {J.}~\bibnamefont {Smirnov}}, \bibinfo {author} {\bibfnamefont {S.}~\bibnamefont {Griffith}},\ and\ \bibinfo {author} {\bibfnamefont {J.~F.}\ \bibnamefont {Beacom}},\ }\href@noop {} {\bibinfo {title} {{Inelastic Signatures of Electroweak Dark Matter}}} (\bibinfo {year} {2026}),\ \Eprint {https://arxiv.org/abs/2609.04144} {arXiv:2609.04144 [hep-ph]} \BibitemShut {NoStop}%
\bibitem [{\citenamefont {Du}\ and\ \citenamefont {Wang}(2026)}]{Du:2026guj}%
  \BibitemOpen
  \bibfield  {author} {\bibinfo {author} {\bibfnamefont {X.}~\bibnamefont {Du}}\ and\ \bibinfo {author} {\bibfnamefont {F.}~\bibnamefont {Wang}},\ }\href@noop {} {\bibinfo {title} {{TeV Higgsino Interpretation of the LZ High-Recoil Event with Intermediate-Scale Electroweak Gauginos}}} (\bibinfo {year} {2026}),\ \Eprint {https://arxiv.org/abs/2609.04163} {arXiv:2609.04163 [hep-ph]} \BibitemShut {NoStop}%
\bibitem [{\citenamefont {Bisal}\ \emph {et~al.}(2026)\citenamefont {Bisal}, \citenamefont {Cao},\ and\ \citenamefont {Li}}]{Bisal:2026khf}%
  \BibitemOpen
  \bibfield  {author} {\bibinfo {author} {\bibfnamefont {S.}~\bibnamefont {Bisal}}, \bibinfo {author} {\bibfnamefont {J.}~\bibnamefont {Cao}},\ and\ \bibinfo {author} {\bibfnamefont {F.}~\bibnamefont {Li}},\ }\href@noop {} {\bibinfo {title} {{Higgsino Dark Matter Interpretation of the LZ High-Recoil Event in the GNMSSM with TeV-Scale Gauginos}}} (\bibinfo {year} {2026}),\ \Eprint {https://arxiv.org/abs/2609.07811} {arXiv:2609.07811 [hep-ph]} \BibitemShut {NoStop}%
\bibitem [{\citenamefont {Cheung}\ \emph {et~al.}(2026)\citenamefont {Cheung}, \citenamefont {Kang},\ and\ \citenamefont {Kumar}}]{Cheung:2026byg}%
  \BibitemOpen
  \bibfield  {author} {\bibinfo {author} {\bibfnamefont {K.}~\bibnamefont {Cheung}}, \bibinfo {author} {\bibfnamefont {S.~K.}\ \bibnamefont {Kang}},\ and\ \bibinfo {author} {\bibfnamefont {R.}~\bibnamefont {Kumar}},\ }\href@noop {} {\bibinfo {title} {{From LUX-ZEPLIN to Colliders: Probing Higgsino Dark Matter}}} (\bibinfo {year} {2026}),\ \Eprint {https://arxiv.org/abs/2609.08712} {arXiv:2609.08712 [hep-ph]} \BibitemShut {NoStop}%
\bibitem [{\citenamefont {Langhoff}(2026)}]{Langhoff:2026ujr}%
  \BibitemOpen
  \bibfield  {author} {\bibinfo {author} {\bibfnamefont {K.}~\bibnamefont {Langhoff}},\ }\href@noop {} {\bibinfo {title} {{Heavy Higgsino Interpretation of the LZ Event}}} (\bibinfo {year} {2026}),\ \Eprint {https://arxiv.org/abs/2609.09385} {arXiv:2609.09385 [hep-ph]} \BibitemShut {NoStop}%
\bibitem [{\citenamefont {Frolovsky}\ and\ \citenamefont {Ketov}(2026)}]{Frolovsky:2026tvq}%
  \BibitemOpen
  \bibfield  {author} {\bibinfo {author} {\bibfnamefont {D.}~\bibnamefont {Frolovsky}}\ and\ \bibinfo {author} {\bibfnamefont {S.~V.}\ \bibnamefont {Ketov}},\ }\href@noop {} {\bibinfo {title} {{Higgsino dark matter in the Starobinsky supergravity with the MSSM in light of the LUX-ZEPLIN event}}} (\bibinfo {year} {2026}),\ \Eprint {https://arxiv.org/abs/2609.11241} {arXiv:2609.11241 [hep-ph]} \BibitemShut {NoStop}%
\bibitem [{\citenamefont {Paul}\ \emph {et~al.}(2026)\citenamefont {Paul}, \citenamefont {Sahoo}, \citenamefont {Sahu},\ and\ \citenamefont {Sharma}}]{Paul:2026okh}%
  \BibitemOpen
  \bibfield  {author} {\bibinfo {author} {\bibfnamefont {P.~K.}\ \bibnamefont {Paul}}, \bibinfo {author} {\bibfnamefont {S.~K.}\ \bibnamefont {Sahoo}}, \bibinfo {author} {\bibfnamefont {N.}~\bibnamefont {Sahu}},\ and\ \bibinfo {author} {\bibfnamefont {S.}~\bibnamefont {Sharma}},\ }\href@noop {} {\bibinfo {title} {{Resurrecting Electroweak Dark Matter via Type-II Seesaw in light of recent LZ Event}}} (\bibinfo {year} {2026}),\ \Eprint {https://arxiv.org/abs/2609.22063} {arXiv:2609.22063 [hep-ph]} \BibitemShut {NoStop}%
\bibitem [{\citenamefont {Khan}\ \emph {et~al.}(2026{\natexlab{a}})\citenamefont {Khan}, \citenamefont {Muhammad}, \citenamefont {Mustafa}, \citenamefont {Atamurotov}, \citenamefont {Abdujabbarov},\ and\ \citenamefont {Khan}}]{Khan:2026zuj}%
  \BibitemOpen
  \bibfield  {author} {\bibinfo {author} {\bibfnamefont {I.}~\bibnamefont {Khan}}, \bibinfo {author} {\bibfnamefont {A.}~\bibnamefont {Muhammad}}, \bibinfo {author} {\bibfnamefont {G.}~\bibnamefont {Mustafa}}, \bibinfo {author} {\bibfnamefont {F.}~\bibnamefont {Atamurotov}}, \bibinfo {author} {\bibfnamefont {A.}~\bibnamefont {Abdujabbarov}},\ and\ \bibinfo {author} {\bibfnamefont {M.}~\bibnamefont {Khan}},\ }\href@noop {} {\bibinfo {title} {{LZ-Motivated Pseudo-Dirac Higgsinos in the Supersymmetric 331 Model from the Supersymmetric {SU(6)} GUT Model}}} (\bibinfo {year} {2026}{\natexlab{a}}),\ \Eprint {https://arxiv.org/abs/2609.23691} {arXiv:2609.23691 [hep-ph]} \BibitemShut {NoStop}%
\bibitem [{\citenamefont {Unwin}(2026{\natexlab{a}})}]{Unwin:2026qep}%
  \BibitemOpen
  \bibfield  {author} {\bibinfo {author} {\bibfnamefont {J.}~\bibnamefont {Unwin}},\ }\href@noop {} {\bibinfo {title} {{Heavy Higgsino Dark Matter at the Scale of the Vanishing Higgs Quartic}}} (\bibinfo {year} {2026}{\natexlab{a}}),\ \Eprint {https://arxiv.org/abs/2609.28740} {arXiv:2609.28740 [hep-ph]} \BibitemShut {NoStop}%
\bibitem [{\citenamefont {Murayama}\ and\ \citenamefont {Noether}(2026)}]{Murayama:2026apt}%
  \BibitemOpen
  \bibfield  {author} {\bibinfo {author} {\bibfnamefont {H.}~\bibnamefont {Murayama}}\ and\ \bibinfo {author} {\bibfnamefont {B.}~\bibnamefont {Noether}},\ }\href@noop {} {\bibinfo {title} {{Heavy Inert Doublet Reconciles LZ and IceCube}}} (\bibinfo {year} {2026}),\ \Eprint {https://arxiv.org/abs/2609.28819} {arXiv:2609.28819 [hep-ph]} \BibitemShut {NoStop}%
\bibitem [{\citenamefont {Roy}\ \emph {et~al.}(2026)\citenamefont {Roy}, \citenamefont {Schwaller},\ and\ \citenamefont {Wagner}}]{Roy:2026wzb}%
  \BibitemOpen
  \bibfield  {author} {\bibinfo {author} {\bibfnamefont {S.}~\bibnamefont {Roy}}, \bibinfo {author} {\bibfnamefont {P.}~\bibnamefont {Schwaller}},\ and\ \bibinfo {author} {\bibfnamefont {C.~E.~M.}\ \bibnamefont {Wagner}},\ }\href@noop {} {\bibinfo {title} {{Interpreting the High-Recoil LUX-ZEPLIN Event with Bino-/Singlino-like and Higgsino Dark Matter}}} (\bibinfo {year} {2026}),\ \Eprint {https://arxiv.org/abs/2609.40234} {arXiv:2609.40234 [hep-ph]} \BibitemShut {NoStop}%
\bibitem [{\citenamefont {Lee}\ and\ \citenamefont {Youn}(2026{\natexlab{a}})}]{Lee:2026tro}%
  \BibitemOpen
  \bibfield  {author} {\bibinfo {author} {\bibfnamefont {S.~J.}\ \bibnamefont {Lee}}\ and\ \bibinfo {author} {\bibfnamefont {T.}~\bibnamefont {Youn}},\ }\href@noop {} {\bibinfo {title} {{The LZ inelastic splitting as a probe of anomaly mediation}}} (\bibinfo {year} {2026}{\natexlab{a}}),\ \Eprint {https://arxiv.org/abs/2610.13930} {arXiv:2610.13930 [hep-ph]} \BibitemShut {NoStop}%
\bibitem [{\citenamefont {Su}\ \emph {et~al.}(2026)\citenamefont {Su}, \citenamefont {Yang},\ and\ \citenamefont {Yang}}]{Su:2026rwz}%
  \BibitemOpen
  \bibfield  {author} {\bibinfo {author} {\bibfnamefont {L.}~\bibnamefont {Su}}, \bibinfo {author} {\bibfnamefont {J.~M.}\ \bibnamefont {Yang}},\ and\ \bibinfo {author} {\bibfnamefont {W.-N.}\ \bibnamefont {Yang}},\ }\href@noop {} {\bibinfo {title} {{Inelastic Dark Matter Signature at High Recoil Energy in LUX-ZEPLIN and CRESST}}} (\bibinfo {year} {2026}),\ \Eprint {https://arxiv.org/abs/2609.01475} {arXiv:2609.01475 [hep-ph]} \BibitemShut {NoStop}%
\bibitem [{\citenamefont {Yamashita}(2026)}]{Yamashita:2026ump}%
  \BibitemOpen
  \bibfield  {author} {\bibinfo {author} {\bibfnamefont {K.}~\bibnamefont {Yamashita}},\ }\href@noop {} {\bibinfo {title} {{Inelastic Dark Photon Dark Matter for the LUX-ZEPLIN High-Recoil Event and the Galactic Halo Gamma-Ray Excess}}} (\bibinfo {year} {2026}),\ \Eprint {https://arxiv.org/abs/2609.02868} {arXiv:2609.02868 [hep-ph]} \BibitemShut {NoStop}%
\bibitem [{\citenamefont {Lee}(2026)}]{Lee:2026wof}%
  \BibitemOpen
  \bibfield  {author} {\bibinfo {author} {\bibfnamefont {H.~M.}\ \bibnamefont {Lee}},\ }\href@noop {} {\bibinfo {title} {{Inelastic dark matter and baryon flavor symmetry in light of LUX-ZEPLIN (LZ) experiment}}} (\bibinfo {year} {2026}),\ \Eprint {https://arxiv.org/abs/2609.06171} {arXiv:2609.06171 [hep-ph]} \BibitemShut {NoStop}%
\bibitem [{\citenamefont {Wang}\ and\ \citenamefont {Xiao}(2026)}]{Wang:2026ytg}%
  \BibitemOpen
  \bibfield  {author} {\bibinfo {author} {\bibfnamefont {L.}~\bibnamefont {Wang}}\ and\ \bibinfo {author} {\bibfnamefont {Y.}~\bibnamefont {Xiao}},\ }\href@noop {} {\bibinfo {title} {{The Inert Doublet Model of Dark Matter and the LUX-ZEPLIN High-Recoil Event}}} (\bibinfo {year} {2026}),\ \Eprint {https://arxiv.org/abs/2609.06571} {arXiv:2609.06571 [hep-ph]} \BibitemShut {NoStop}%
\bibitem [{\citenamefont {Das}\ \emph {et~al.}(2026)\citenamefont {Das}, \citenamefont {Karmakar}, \citenamefont {Mahapatra},\ and\ \citenamefont {Paul}}]{Das:2026uyy}%
  \BibitemOpen
  \bibfield  {author} {\bibinfo {author} {\bibfnamefont {P.}~\bibnamefont {Das}}, \bibinfo {author} {\bibfnamefont {B.}~\bibnamefont {Karmakar}}, \bibinfo {author} {\bibfnamefont {S.}~\bibnamefont {Mahapatra}},\ and\ \bibinfo {author} {\bibfnamefont {P.~K.}\ \bibnamefont {Paul}},\ }\href@noop {} {\bibinfo {title} {{Inelastic Self-interacting Dark Matter and LUX-ZEPLIN 248 keV Event in a Dirac Modular Inverse Seesaw}}} (\bibinfo {year} {2026}),\ \Eprint {https://arxiv.org/abs/2609.06825} {arXiv:2609.06825 [hep-ph]} \BibitemShut {NoStop}%
\bibitem [{\citenamefont {Okada}\ and\ \citenamefont {Seto}(2026)}]{Okada:2026eol}%
  \BibitemOpen
  \bibfield  {author} {\bibinfo {author} {\bibfnamefont {N.}~\bibnamefont {Okada}}\ and\ \bibinfo {author} {\bibfnamefont {O.}~\bibnamefont {Seto}},\ }\href@noop {} {\bibinfo {title} {{Inelastic $B-L$ scalar dark matter and the LUX-ZEPLIN event}}} (\bibinfo {year} {2026}),\ \Eprint {https://arxiv.org/abs/2609.06909} {arXiv:2609.06909 [hep-ph]} \BibitemShut {NoStop}%
\bibitem [{\citenamefont {Ahmed}\ and\ \citenamefont {Leontaris}(2026{\natexlab{a}})}]{Ahmed:2026qjg}%
  \BibitemOpen
  \bibfield  {author} {\bibinfo {author} {\bibfnamefont {W.}~\bibnamefont {Ahmed}}\ and\ \bibinfo {author} {\bibfnamefont {G.~K.}\ \bibnamefont {Leontaris}},\ }\href@noop {} {\bibinfo {title} {{A Dark-Dimension Origin of Geometric Inelastic Dark Matter: The LUX-ZEPLIN High-Recoil Event and Multi-Target Tests}}} (\bibinfo {year} {2026}{\natexlab{a}}),\ \Eprint {https://arxiv.org/abs/2609.07138} {arXiv:2609.07138 [hep-ph]} \BibitemShut {NoStop}%
\bibitem [{\citenamefont {Du}\ \emph {et~al.}(2026)\citenamefont {Du}, \citenamefont {Huang},\ and\ \citenamefont {Xie}}]{Du:2026lpa}%
  \BibitemOpen
  \bibfield  {author} {\bibinfo {author} {\bibfnamefont {X.-Y.}\ \bibnamefont {Du}}, \bibinfo {author} {\bibfnamefont {W.}~\bibnamefont {Huang}},\ and\ \bibinfo {author} {\bibfnamefont {K.}~\bibnamefont {Xie}},\ }\href@noop {} {\bibinfo {title} {{Pseudo-Dirac Inelastic Dark Matter in the Leptophobic $U(1)_B$ Model: Confronting the LUX-ZEPLIN High-Recoil Event with Collider Searches}}} (\bibinfo {year} {2026}),\ \Eprint {https://arxiv.org/abs/2609.07225} {arXiv:2609.07225 [hep-ph]} \BibitemShut {NoStop}%
\bibitem [{\citenamefont {Bandyopadhyay}\ \emph {et~al.}(2026)\citenamefont {Bandyopadhyay}, \citenamefont {Borah},\ and\ \citenamefont {Borah}}]{Bandyopadhyay:2026gjw}%
  \BibitemOpen
  \bibfield  {author} {\bibinfo {author} {\bibfnamefont {D.}~\bibnamefont {Bandyopadhyay}}, \bibinfo {author} {\bibfnamefont {D.}~\bibnamefont {Borah}},\ and\ \bibinfo {author} {\bibfnamefont {P.}~\bibnamefont {Borah}},\ }\href@noop {} {\bibinfo {title} {{LZ nuclear recoil event from inelastic singlet-doublet scalar dark matter}}} (\bibinfo {year} {2026}),\ \Eprint {https://arxiv.org/abs/2609.07451} {arXiv:2609.07451 [hep-ph]} \BibitemShut {NoStop}%
\bibitem [{\citenamefont {Borah}\ \emph {et~al.}(2026{\natexlab{a}})\citenamefont {Borah}, \citenamefont {Sahoo}, \citenamefont {Sahu},\ and\ \citenamefont {Sharma}}]{Borah:2026zwf}%
  \BibitemOpen
  \bibfield  {author} {\bibinfo {author} {\bibfnamefont {D.}~\bibnamefont {Borah}}, \bibinfo {author} {\bibfnamefont {S.~K.}\ \bibnamefont {Sahoo}}, \bibinfo {author} {\bibfnamefont {N.}~\bibnamefont {Sahu}},\ and\ \bibinfo {author} {\bibfnamefont {S.}~\bibnamefont {Sharma}},\ }\href@noop {} {\bibinfo {title} {{Inelastic Singlet-Doublet Fermion Dark Matter in light of the 248 keV LZ event}}} (\bibinfo {year} {2026}{\natexlab{a}}),\ \Eprint {https://arxiv.org/abs/2609.07800} {arXiv:2609.07800 [hep-ph]} \BibitemShut {NoStop}%
\bibitem [{\citenamefont {Yuan}\ \emph {et~al.}(2026)\citenamefont {Yuan}, \citenamefont {Zhang}, \citenamefont {Cao}, \citenamefont {Feng},\ and\ \citenamefont {Yang}}]{Yuan:2026djt}%
  \BibitemOpen
  \bibfield  {author} {\bibinfo {author} {\bibfnamefont {G.-W.}\ \bibnamefont {Yuan}}, \bibinfo {author} {\bibfnamefont {B.}~\bibnamefont {Zhang}}, \bibinfo {author} {\bibfnamefont {W.-Y.}\ \bibnamefont {Cao}}, \bibinfo {author} {\bibfnamefont {L.}~\bibnamefont {Feng}},\ and\ \bibinfo {author} {\bibfnamefont {R.}~\bibnamefont {Yang}},\ }\href@noop {} {\bibinfo {title} {{ALP-mediated inelastic dark matter and the LUX-ZEPLIN high-recoil candidate event LZ230616}}} (\bibinfo {year} {2026}),\ \Eprint {https://arxiv.org/abs/2609.08893} {arXiv:2609.08893 [hep-ph]} \BibitemShut {NoStop}%
\bibitem [{\citenamefont {Zhu}\ \emph {et~al.}(2026)\citenamefont {Zhu}, \citenamefont {Dalla Valle~Garcia}, \citenamefont {Wang}, \citenamefont {Thomas},\ and\ \citenamefont {White}}]{Zhu:2026dag}%
  \BibitemOpen
  \bibfield  {author} {\bibinfo {author} {\bibfnamefont {P.}~\bibnamefont {Zhu}}, \bibinfo {author} {\bibfnamefont {G.}~\bibnamefont {Dalla Valle~Garcia}}, \bibinfo {author} {\bibfnamefont {X.-G.}\ \bibnamefont {Wang}}, \bibinfo {author} {\bibfnamefont {A.~W.}\ \bibnamefont {Thomas}},\ and\ \bibinfo {author} {\bibfnamefont {M.~J.}\ \bibnamefont {White}},\ }\href@noop {} {\bibinfo {title} {{Endothermic dark matter with a light dark photon and the LUX--ZEPLIN high-energy nuclear-recoil candidate}}} (\bibinfo {year} {2026}),\ \Eprint {https://arxiv.org/abs/2609.09015} {arXiv:2609.09015 [hep-ph]} \BibitemShut {NoStop}%
\bibitem [{\citenamefont {Asadi}\ \emph {et~al.}(2026)\citenamefont {Asadi}, \citenamefont {Batz}, \citenamefont {Fox}, \citenamefont {Homiller},\ and\ \citenamefont {Kribs}}]{Asadi:2026iot}%
  \BibitemOpen
  \bibfield  {author} {\bibinfo {author} {\bibfnamefont {P.}~\bibnamefont {Asadi}}, \bibinfo {author} {\bibfnamefont {A.}~\bibnamefont {Batz}}, \bibinfo {author} {\bibfnamefont {P.~J.}\ \bibnamefont {Fox}}, \bibinfo {author} {\bibfnamefont {S.~D.}\ \bibnamefont {Homiller}},\ and\ \bibinfo {author} {\bibfnamefont {G.~D.}\ \bibnamefont {Kribs}},\ }\href@noop {} {\bibinfo {title} {{For Whom the Xenon Recoils: Magnetic Inelastic Dark Baryons}}} (\bibinfo {year} {2026}),\ \Eprint {https://arxiv.org/abs/2609.09107} {arXiv:2609.09107 [hep-ph]} \BibitemShut {NoStop}%
\bibitem [{\citenamefont {Lee}\ and\ \citenamefont {Youn}(2026{\natexlab{b}})}]{Lee:2026jxl}%
  \BibitemOpen
  \bibfield  {author} {\bibinfo {author} {\bibfnamefont {S.~J.}\ \bibnamefont {Lee}}\ and\ \bibinfo {author} {\bibfnamefont {T.}~\bibnamefont {Youn}},\ }\href@noop {} {\bibinfo {title} {{Mixing-suppressed inelastic dark matter: a minimal model for the LZ 248 keV event}}} (\bibinfo {year} {2026}{\natexlab{b}}),\ \Eprint {https://arxiv.org/abs/2609.09138} {arXiv:2609.09138 [hep-ph]} \BibitemShut {NoStop}%
\bibitem [{\citenamefont {He}(2026)}]{He:2026hqz}%
  \BibitemOpen
  \bibfield  {author} {\bibinfo {author} {\bibfnamefont {Y.}~\bibnamefont {He}},\ }\href@noop {} {\bibinfo {title} {{Transition magnetic-dipole dark matter and the LZ230616 high-recoil candidate}}} (\bibinfo {year} {2026}),\ \Eprint {https://arxiv.org/abs/2609.10453} {arXiv:2609.10453 [hep-ph]} \BibitemShut {NoStop}%
\bibitem [{\citenamefont {Fan}\ \emph {et~al.}(2026)\citenamefont {Fan}, \citenamefont {He}, \citenamefont {Wang},\ and\ \citenamefont {Zhao}}]{Fan:2026hzw}%
  \BibitemOpen
  \bibfield  {author} {\bibinfo {author} {\bibfnamefont {Z.-T.}\ \bibnamefont {Fan}}, \bibinfo {author} {\bibfnamefont {H.-J.}\ \bibnamefont {He}}, \bibinfo {author} {\bibfnamefont {Y.-C.}\ \bibnamefont {Wang}},\ and\ \bibinfo {author} {\bibfnamefont {Y.}~\bibnamefont {Zhao}},\ }\href@noop {} {\bibinfo {title} {{Inelastic Dark Matter and High-Energy Recoil Signatures in LZ}}} (\bibinfo {year} {2026}),\ \Eprint {https://arxiv.org/abs/2609.10491} {arXiv:2609.10491 [hep-ph]} \BibitemShut {NoStop}%
\bibitem [{\citenamefont {Kumar}\ and\ \citenamefont {Prajapati}(2026)}]{Kumar:2026lgi}%
  \BibitemOpen
  \bibfield  {author} {\bibinfo {author} {\bibfnamefont {R.}~\bibnamefont {Kumar}}\ and\ \bibinfo {author} {\bibfnamefont {H.~K.}\ \bibnamefont {Prajapati}},\ }\href@noop {} {\bibinfo {title} {{Generalized Chiral $U(1)_{B-L}$ with Inelastic Scalar Dark Matter for the LZ 248 keV Event}}} (\bibinfo {year} {2026}),\ \Eprint {https://arxiv.org/abs/2609.10827} {arXiv:2609.10827 [hep-ph]} \BibitemShut {NoStop}%
\bibitem [{\citenamefont {Okada}\ \emph {et~al.}(2026)\citenamefont {Okada}, \citenamefont {Shigekami},\ and\ \citenamefont {Wu}}]{Okada:2026upm}%
  \BibitemOpen
  \bibfield  {author} {\bibinfo {author} {\bibfnamefont {H.}~\bibnamefont {Okada}}, \bibinfo {author} {\bibfnamefont {Y.}~\bibnamefont {Shigekami}},\ and\ \bibinfo {author} {\bibfnamefont {J.-J.}\ \bibnamefont {Wu}},\ }\href@noop {} {\bibinfo {title} {{Can a minimal radiative seesaw explain the LZ 248 keV event?}}} (\bibinfo {year} {2026}),\ \Eprint {https://arxiv.org/abs/2609.13038} {arXiv:2609.13038 [hep-ph]} \BibitemShut {NoStop}%
\bibitem [{\citenamefont {Cabo-Almeida}\ \emph {et~al.}(2026)\citenamefont {Cabo-Almeida}, \citenamefont {Costa}, \citenamefont {Feiteira},\ and\ \citenamefont {Oliveira}}]{Cabo-Almeida:2026uqw}%
  \BibitemOpen
  \bibfield  {author} {\bibinfo {author} {\bibfnamefont {D.}~\bibnamefont {Cabo-Almeida}}, \bibinfo {author} {\bibfnamefont {F.}~\bibnamefont {Costa}}, \bibinfo {author} {\bibfnamefont {D.}~\bibnamefont {Feiteira}},\ and\ \bibinfo {author} {\bibfnamefont {V.}~\bibnamefont {Oliveira}},\ }\href@noop {} {\bibinfo {title} {{A Freeze-In Interpretation of the LZ High-Energy Nuclear Recoil Event}}} (\bibinfo {year} {2026}),\ \Eprint {https://arxiv.org/abs/2609.13130} {arXiv:2609.13130 [hep-ph]} \BibitemShut {NoStop}%
\bibitem [{\citenamefont {Borah}\ \emph {et~al.}(2026{\natexlab{b}})\citenamefont {Borah}, \citenamefont {Mahapatra},\ and\ \citenamefont {Nath}}]{Borah:2026ris}%
  \BibitemOpen
  \bibfield  {author} {\bibinfo {author} {\bibfnamefont {P.}~\bibnamefont {Borah}}, \bibinfo {author} {\bibfnamefont {S.}~\bibnamefont {Mahapatra}},\ and\ \bibinfo {author} {\bibfnamefont {N.}~\bibnamefont {Nath}},\ }\href@noop {} {\bibinfo {title} {{Inelastic Dark Matter at LZ from Radiative Dirac Neutrino Mass Paradigm}}} (\bibinfo {year} {2026}{\natexlab{b}}),\ \Eprint {https://arxiv.org/abs/2609.15027} {arXiv:2609.15027 [hep-ph]} \BibitemShut {NoStop}%
\bibitem [{\citenamefont {Das}\ and\ \citenamefont {Nomura}(2026)}]{Das:2026buc}%
  \BibitemOpen
  \bibfield  {author} {\bibinfo {author} {\bibfnamefont {A.}~\bibnamefont {Das}}\ and\ \bibinfo {author} {\bibfnamefont {T.}~\bibnamefont {Nomura}},\ }\href@noop {} {\bibinfo {title} {{Effect of inelastic scalar dark matter in hidden $U(1)$ scenario after the LZ nuclear recoil}}} (\bibinfo {year} {2026}),\ \Eprint {https://arxiv.org/abs/2609.15600} {arXiv:2609.15600 [hep-ph]} \BibitemShut {NoStop}%
\bibitem [{\citenamefont {He}\ \emph {et~al.}(2026)\citenamefont {He}, \citenamefont {Hong},\ and\ \citenamefont {Jeesun}}]{He:2026idw}%
  \BibitemOpen
  \bibfield  {author} {\bibinfo {author} {\bibfnamefont {X.-G.}\ \bibnamefont {He}}, \bibinfo {author} {\bibfnamefont {X.}~\bibnamefont {Hong}},\ and\ \bibinfo {author} {\bibfnamefont {S.}~\bibnamefont {Jeesun}},\ }\href@noop {} {\bibinfo {title} {{Hadrophilic inelastic freeze-in dark matter in $q_1-q_2$ gauge extension and the high energy LZ event}}} (\bibinfo {year} {2026}),\ \Eprint {https://arxiv.org/abs/2609.15714} {arXiv:2609.15714 [hep-ph]} \BibitemShut {NoStop}%
\bibitem [{\citenamefont {Lian}\ and\ \citenamefont {Yang}(2026)}]{Lian:2026hpm}%
  \BibitemOpen
  \bibfield  {author} {\bibinfo {author} {\bibfnamefont {J.}~\bibnamefont {Lian}}\ and\ \bibinfo {author} {\bibfnamefont {J.~M.}\ \bibnamefont {Yang}},\ }\href@noop {} {\bibinfo {title} {{Explain the LZ High-Energy Recoil Event with Inelastic Sneutrino Dark Matter in Supersymmetry}}} (\bibinfo {year} {2026}),\ \Eprint {https://arxiv.org/abs/2609.15742} {arXiv:2609.15742 [hep-ph]} \BibitemShut {NoStop}%
\bibitem [{\citenamefont {An}\ \emph {et~al.}(2026)\citenamefont {An}, \citenamefont {Gao}, \citenamefont {Liu}, \citenamefont {Liu},\ and\ \citenamefont {Xu}}]{An:2026pkc}%
  \BibitemOpen
  \bibfield  {author} {\bibinfo {author} {\bibfnamefont {H.}~\bibnamefont {An}}, \bibinfo {author} {\bibfnamefont {F.}~\bibnamefont {Gao}}, \bibinfo {author} {\bibfnamefont {J.}~\bibnamefont {Liu}}, \bibinfo {author} {\bibfnamefont {M.}~\bibnamefont {Liu}},\ and\ \bibinfo {author} {\bibfnamefont {C.}~\bibnamefont {Xu}},\ }\href@noop {} {\bibinfo {title} {{Cosmological Constrained Axion-Portal Inelastic Dark Matter for the LZ Event}}} (\bibinfo {year} {2026}),\ \Eprint {https://arxiv.org/abs/2609.17412} {arXiv:2609.17412 [hep-ph]} \BibitemShut {NoStop}%
\bibitem [{\citenamefont {Nagata}\ and\ \citenamefont {Yanagida}(2026)}]{Nagata:2026pbj}%
  \BibitemOpen
  \bibfield  {author} {\bibinfo {author} {\bibfnamefont {N.}~\bibnamefont {Nagata}}\ and\ \bibinfo {author} {\bibfnamefont {T.~T.}\ \bibnamefont {Yanagida}},\ }\href@noop {} {\bibinfo {title} {{Asymmetric Inelastic Dark Matter and the LUX-ZEPLIN event}}} (\bibinfo {year} {2026}),\ \Eprint {https://arxiv.org/abs/2609.18564} {arXiv:2609.18564 [hep-ph]} \BibitemShut {NoStop}%
\bibitem [{\citenamefont {Okada}\ and\ \citenamefont {Raut}(2026)}]{Okada:2026fef}%
  \BibitemOpen
  \bibfield  {author} {\bibinfo {author} {\bibfnamefont {N.}~\bibnamefont {Okada}}\ and\ \bibinfo {author} {\bibfnamefont {D.}~\bibnamefont {Raut}},\ }\href@noop {} {\bibinfo {title} {{Endothermic Z'-Portal Dark Matter: LZ-LHC Complementarity}}} (\bibinfo {year} {2026}),\ \Eprint {https://arxiv.org/abs/2609.21011} {arXiv:2609.21011 [hep-ph]} \BibitemShut {NoStop}%
\bibitem [{\citenamefont {Ahmed}\ and\ \citenamefont {Leontaris}(2026{\natexlab{b}})}]{Ahmed:2026com}%
  \BibitemOpen
  \bibfield  {author} {\bibinfo {author} {\bibfnamefont {W.}~\bibnamefont {Ahmed}}\ and\ \bibinfo {author} {\bibfnamefont {G.~K.}\ \bibnamefont {Leontaris}},\ }\href@noop {} {\bibinfo {title} {{Kaluza--Klein Dark-Photon Mediation of Inelastic Dark Matter at LUX-ZEPLIN}}} (\bibinfo {year} {2026}{\natexlab{b}}),\ \Eprint {https://arxiv.org/abs/2609.22739} {arXiv:2609.22739 [hep-ph]} \BibitemShut {NoStop}%
\bibitem [{\citenamefont {Tang}\ and\ \citenamefont {Gu}(2026)}]{Tang:2026lgu}%
  \BibitemOpen
  \bibfield  {author} {\bibinfo {author} {\bibfnamefont {J.}~\bibnamefont {Tang}}\ and\ \bibinfo {author} {\bibfnamefont {P.-H.}\ \bibnamefont {Gu}},\ }\href@noop {} {\bibinfo {title} {{Inelastic dark matter mediated by natural dark photon}}} (\bibinfo {year} {2026}),\ \Eprint {https://arxiv.org/abs/2609.28415} {arXiv:2609.28415 [hep-ph]} \BibitemShut {NoStop}%
\bibitem [{\citenamefont {Langhoff}\ and\ \citenamefont {Xiao}(2026)}]{Langhoff:2026qgo}%
  \BibitemOpen
  \bibfield  {author} {\bibinfo {author} {\bibfnamefont {K.}~\bibnamefont {Langhoff}}\ and\ \bibinfo {author} {\bibfnamefont {H.}~\bibnamefont {Xiao}},\ }\href@noop {} {\bibinfo {title} {{Sneaky Sneutrino Scattering at LZ}}} (\bibinfo {year} {2026}),\ \Eprint {https://arxiv.org/abs/2609.28616} {arXiv:2609.28616 [hep-ph]} \BibitemShut {NoStop}%
\bibitem [{\citenamefont {Choudhury}\ \emph {et~al.}(2026)\citenamefont {Choudhury}, \citenamefont {Das},\ and\ \citenamefont {Sachdeva}}]{Choudhury:2026qjz}%
  \BibitemOpen
  \bibfield  {author} {\bibinfo {author} {\bibfnamefont {D.}~\bibnamefont {Choudhury}}, \bibinfo {author} {\bibfnamefont {J.}~\bibnamefont {Das}},\ and\ \bibinfo {author} {\bibfnamefont {D.}~\bibnamefont {Sachdeva}},\ }\href@noop {} {\bibinfo {title} {{Inelastic Dark Matter from a Higher-Dimensional Operator: The LUX-ZEPLIN High-Recoil Event and Thermal Relic}}} (\bibinfo {year} {2026}),\ \Eprint {https://arxiv.org/abs/2610.00645} {arXiv:2610.00645 [hep-ph]} \BibitemShut {NoStop}%
\bibitem [{\citenamefont {Capolupo}\ \emph {et~al.}(2026)\citenamefont {Capolupo}, \citenamefont {Pisacane},\ and\ \citenamefont {Serao}}]{Capolupo:2026uyi}%
  \BibitemOpen
  \bibfield  {author} {\bibinfo {author} {\bibfnamefont {A.}~\bibnamefont {Capolupo}}, \bibinfo {author} {\bibfnamefont {G.}~\bibnamefont {Pisacane}},\ and\ \bibinfo {author} {\bibfnamefont {R.}~\bibnamefont {Serao}},\ }\href@noop {} {\bibinfo {title} {{Solar capture and annihilation of fermionic and scalar dark matter motivated by the LUX-ZEPLIN high-recoil candidate}}} (\bibinfo {year} {2026}),\ \Eprint {https://arxiv.org/abs/2609.28218} {arXiv:2609.28218 [hep-ph]} \BibitemShut {NoStop}%
\bibitem [{\citenamefont {Unwin}(2026{\natexlab{b}})}]{Unwin:2026rdp}%
  \BibitemOpen
  \bibfield  {author} {\bibinfo {author} {\bibfnamefont {J.}~\bibnamefont {Unwin}},\ }\href@noop {} {\bibinfo {title} {{Axion Portal Dark Matter and the LUX-ZEPLIN High-Recoil Event}}} (\bibinfo {year} {2026}{\natexlab{b}}),\ \Eprint {https://arxiv.org/abs/2609.04186} {arXiv:2609.04186 [hep-ph]} \BibitemShut {NoStop}%
\bibitem [{\citenamefont {Arcadi}\ \emph {et~al.}(2026)\citenamefont {Arcadi}, \citenamefont {di~Mauro}, \citenamefont {Djouadi},\ and\ \citenamefont {Queiroz}}]{Arcadi:2026kev}%
  \BibitemOpen
  \bibfield  {author} {\bibinfo {author} {\bibfnamefont {G.}~\bibnamefont {Arcadi}}, \bibinfo {author} {\bibfnamefont {M.}~\bibnamefont {di~Mauro}}, \bibinfo {author} {\bibfnamefont {A.}~\bibnamefont {Djouadi}},\ and\ \bibinfo {author} {\bibfnamefont {F.}~\bibnamefont {Queiroz}},\ }\href@noop {} {\bibinfo {title} {{A possible interpretation of the LUX-ZEPLIN recoil event in the 2HD+a scenario}}} (\bibinfo {year} {2026}),\ \Eprint {https://arxiv.org/abs/2609.17196} {arXiv:2609.17196 [hep-ph]} \BibitemShut {NoStop}%
\bibitem [{\citenamefont {De}(2026)}]{De:2026win}%
  \BibitemOpen
  \bibfield  {author} {\bibinfo {author} {\bibfnamefont {B.}~\bibnamefont {De}},\ }\href@noop {} {\bibinfo {title} {{The 248 keV LZ Recoil: A Possible Hint of Non-SM-Like Quark Yukawa Couplings with a Scalar-Portal Dark Matter}}} (\bibinfo {year} {2026}),\ \Eprint {https://arxiv.org/abs/2609.23096} {arXiv:2609.23096 [hep-ph]} \BibitemShut {NoStop}%
\bibitem [{\citenamefont {Khan}\ \emph {et~al.}(2026{\natexlab{b}})\citenamefont {Khan}, \citenamefont {Capozziello}, \citenamefont {Mustafa}, \citenamefont {Botirov}, \citenamefont {Abdujabbarov}, \citenamefont {Atamurotov},\ and\ \citenamefont {Channuie}}]{Khan:2026osp}%
  \BibitemOpen
  \bibfield  {author} {\bibinfo {author} {\bibfnamefont {I.}~\bibnamefont {Khan}}, \bibinfo {author} {\bibfnamefont {S.}~\bibnamefont {Capozziello}}, \bibinfo {author} {\bibfnamefont {G.}~\bibnamefont {Mustafa}}, \bibinfo {author} {\bibfnamefont {F.}~\bibnamefont {Botirov}}, \bibinfo {author} {\bibfnamefont {A.}~\bibnamefont {Abdujabbarov}}, \bibinfo {author} {\bibfnamefont {F.}~\bibnamefont {Atamurotov}},\ and\ \bibinfo {author} {\bibfnamefont {P.}~\bibnamefont {Channuie}},\ }\href@noop {} {\bibinfo {title} {{Elastic toroidal vector dark matter through a dark photon in the LUX-ZEPLIN high recoil window}}} (\bibinfo {year} {2026}{\natexlab{b}}),\ \Eprint {https://arxiv.org/abs/2609.25114} {arXiv:2609.25114 [hep-ph]} \BibitemShut {NoStop}%
\bibitem [{\citenamefont {Elahi}\ and\ \citenamefont {Schwaller}(2026)}]{Elahi:2026vlm}%
  \BibitemOpen
  \bibfield  {author} {\bibinfo {author} {\bibfnamefont {F.}~\bibnamefont {Elahi}}\ and\ \bibinfo {author} {\bibfnamefont {P.}~\bibnamefont {Schwaller}},\ }\href@noop {} {\bibinfo {title} {{A Vector-Like Lepton Interpretation of the High-Energy Nuclear Recoil Candidate in LUX-ZEPLIN}}} (\bibinfo {year} {2026}),\ \Eprint {https://arxiv.org/abs/2609.08993} {arXiv:2609.08993 [hep-ph]} \BibitemShut {NoStop}%
\bibitem [{\citenamefont {Liang}\ \emph {et~al.}(2026)\citenamefont {Liang}, \citenamefont {Liu}, \citenamefont {Tran},\ and\ \citenamefont {Xu}}]{Liang:2026coz}%
  \BibitemOpen
  \bibfield  {author} {\bibinfo {author} {\bibfnamefont {J.-H.}\ \bibnamefont {Liang}}, \bibinfo {author} {\bibfnamefont {Z.}~\bibnamefont {Liu}}, \bibinfo {author} {\bibfnamefont {V.~Q.}\ \bibnamefont {Tran}},\ and\ \bibinfo {author} {\bibfnamefont {Y.}~\bibnamefont {Xu}},\ }\href@noop {} {\bibinfo {title} {{LZ Nuclear-Recoil Excess from Boosted Light Magnetic Dipole-dipole Dark Matter}}} (\bibinfo {year} {2026}),\ \Eprint {https://arxiv.org/abs/2609.06756} {arXiv:2609.06756 [hep-ph]} \BibitemShut {NoStop}%
\bibitem [{\citenamefont {Kannike}\ \emph {et~al.}(2026)\citenamefont {Kannike}, \citenamefont {Raidal},\ and\ \citenamefont {Strumia}}]{Kannike:2026qyl}%
  \BibitemOpen
  \bibfield  {author} {\bibinfo {author} {\bibfnamefont {K.}~\bibnamefont {Kannike}}, \bibinfo {author} {\bibfnamefont {M.}~\bibnamefont {Raidal}},\ and\ \bibinfo {author} {\bibfnamefont {A.}~\bibnamefont {Strumia}},\ }\href@noop {} {\bibinfo {title} {{Boosted dark particles and the LZ nuclear recoil event}}} (\bibinfo {year} {2026}),\ \Eprint {https://arxiv.org/abs/2609.07742} {arXiv:2609.07742 [hep-ph]} \BibitemShut {NoStop}%
\bibitem [{\citenamefont {Alhazmi}\ \emph {et~al.}(2026)\citenamefont {Alhazmi}, \citenamefont {Kim}, \citenamefont {Kong}, \citenamefont {Park},\ and\ \citenamefont {Shin}}]{Alhazmi:2026efz}%
  \BibitemOpen
  \bibfield  {author} {\bibinfo {author} {\bibfnamefont {H.}~\bibnamefont {Alhazmi}}, \bibinfo {author} {\bibfnamefont {D.}~\bibnamefont {Kim}}, \bibinfo {author} {\bibfnamefont {K.}~\bibnamefont {Kong}}, \bibinfo {author} {\bibfnamefont {J.-C.}\ \bibnamefont {Park}},\ and\ \bibinfo {author} {\bibfnamefont {S.}~\bibnamefont {Shin}},\ }\href@noop {} {\bibinfo {title} {{High-Energy Nuclear Recoils from Boosted Dark Matter for the LZ 248-keV Event: Beyond the Halo-Dependent High-Velocity Tail}}} (\bibinfo {year} {2026}),\ \Eprint {https://arxiv.org/abs/2609.06890} {arXiv:2609.06890 [hep-ph]} \BibitemShut {NoStop}%
\bibitem [{\citenamefont {Heikinheimo}\ and\ \citenamefont {Zimmermann}(2026)}]{Heikinheimo:2026kwp}%
  \BibitemOpen
  \bibfield  {author} {\bibinfo {author} {\bibfnamefont {M.}~\bibnamefont {Heikinheimo}}\ and\ \bibinfo {author} {\bibfnamefont {N.}~\bibnamefont {Zimmermann}},\ }\href@noop {} {\bibinfo {title} {{Cosmic ray boosted dark matter with momentum dependent interactions can explain the LZ 248 keV event}}} (\bibinfo {year} {2026}),\ \Eprint {https://arxiv.org/abs/2609.11600} {arXiv:2609.11600 [hep-ph]} \BibitemShut {NoStop}%
\bibitem [{\citenamefont {Mahapatra}\ and\ \citenamefont {Paul}(2026)}]{Mahapatra:2026glu}%
  \BibitemOpen
  \bibfield  {author} {\bibinfo {author} {\bibfnamefont {S.}~\bibnamefont {Mahapatra}}\ and\ \bibinfo {author} {\bibfnamefont {P.~K.}\ \bibnamefont {Paul}},\ }\href@noop {} {\bibinfo {title} {{Boosted or Inelastic? Discriminating Interpretations of the LZ 248 keV Event}}} (\bibinfo {year} {2026}),\ \Eprint {https://arxiv.org/abs/2609.14799} {arXiv:2609.14799 [hep-ph]} \BibitemShut {NoStop}%
\bibitem [{\citenamefont {Chauhan}\ \emph {et~al.}(2026)\citenamefont {Chauhan}, \citenamefont {Sahasrabuddhe},\ and\ \citenamefont {Sen}}]{Chauhan:2026udz}%
  \BibitemOpen
  \bibfield  {author} {\bibinfo {author} {\bibfnamefont {B.}~\bibnamefont {Chauhan}}, \bibinfo {author} {\bibfnamefont {S.}~\bibnamefont {Sahasrabuddhe}},\ and\ \bibinfo {author} {\bibfnamefont {M.}~\bibnamefont {Sen}},\ }\href@noop {} {\bibinfo {title} {{A closer look at the LZ 248 keV event through the lens of cosmic-ray boosted dark matter}}} (\bibinfo {year} {2026}),\ \Eprint {https://arxiv.org/abs/2609.24982} {arXiv:2609.24982 [hep-ph]} \BibitemShut {NoStop}%
\bibitem [{\citenamefont {McCabe}(2026)}]{McCabe:2026crm}%
  \BibitemOpen
  \bibfield  {author} {\bibinfo {author} {\bibfnamefont {C.}~\bibnamefont {McCabe}},\ }\href@noop {} {\bibinfo {title} {{Seasonal dark matter from the LUX-ZEPLIN high-energy event}}} (\bibinfo {year} {2026}),\ \Eprint {https://arxiv.org/abs/2609.04181} {arXiv:2609.04181 [hep-ph]} \BibitemShut {NoStop}%
\bibitem [{\citenamefont {O'Hare}(2026)}]{OHare:2026nqi}%
  \BibitemOpen
  \bibfield  {author} {\bibinfo {author} {\bibfnamefont {C.~A.~J.}\ \bibnamefont {O'Hare}},\ }\href@noop {} {\bibinfo {title} {{The high-velocity dark matter halo of the Milky Way in light of the LZ 248 keV event}}} (\bibinfo {year} {2026}),\ \Eprint {https://arxiv.org/abs/2609.21444} {arXiv:2609.21444 [hep-ph]} \BibitemShut {NoStop}%
\bibitem [{\citenamefont {Lou}\ and\ \citenamefont {Lu}(2026)}]{Lou:2026idn}%
  \BibitemOpen
  \bibfield  {author} {\bibinfo {author} {\bibfnamefont {Y.}~\bibnamefont {Lou}}\ and\ \bibinfo {author} {\bibfnamefont {C.-T.}\ \bibnamefont {Lu}},\ }\href@noop {} {\bibinfo {title} {{Fermionic Dark Matter Absorption and the High-Energy Event in LUX-ZEPLIN}}} (\bibinfo {year} {2026}),\ \Eprint {https://arxiv.org/abs/2609.01592} {arXiv:2609.01592 [hep-ph]} \BibitemShut {NoStop}%
\bibitem [{\citenamefont {Aghaie}\ and\ \citenamefont {Strumia}(2026)}]{Aghaie:2026vsu}%
  \BibitemOpen
  \bibfield  {author} {\bibinfo {author} {\bibfnamefont {M.}~\bibnamefont {Aghaie}}\ and\ \bibinfo {author} {\bibfnamefont {A.}~\bibnamefont {Strumia}},\ }\href@noop {} {\bibinfo {title} {{Neutron disappearance and the LZ nuclear recoil event}}} (\bibinfo {year} {2026}),\ \Eprint {https://arxiv.org/abs/2609.09037} {arXiv:2609.09037 [hep-ph]} \BibitemShut {NoStop}%
\bibitem [{\citenamefont {Lee}\ \emph {et~al.}(2026)\citenamefont {Lee}, \citenamefont {Takahashi},\ and\ \citenamefont {Tsai}}]{Lee:2026zbr}%
  \BibitemOpen
  \bibfield  {author} {\bibinfo {author} {\bibfnamefont {J.}~\bibnamefont {Lee}}, \bibinfo {author} {\bibfnamefont {F.}~\bibnamefont {Takahashi}},\ and\ \bibinfo {author} {\bibfnamefont {Y.-D.}\ \bibnamefont {Tsai}},\ }\href@noop {} {\bibinfo {title} {{Nuclear Recoils from Invisible Neutron-Pair Annihilation and the LZ event}}} (\bibinfo {year} {2026}),\ \Eprint {https://arxiv.org/abs/2609.12045} {arXiv:2609.12045 [hep-ph]} \BibitemShut {NoStop}%
\bibitem [{\citenamefont {Uttayarat}\ \emph {et~al.}(2026)\citenamefont {Uttayarat}, \citenamefont {Julio},\ and\ \citenamefont {Primulando}}]{Uttayarat:2026isp}%
  \BibitemOpen
  \bibfield  {author} {\bibinfo {author} {\bibfnamefont {P.}~\bibnamefont {Uttayarat}}, \bibinfo {author} {\bibfnamefont {J.}~\bibnamefont {Julio}},\ and\ \bibinfo {author} {\bibfnamefont {R.}~\bibnamefont {Primulando}},\ }\href@noop {} {\bibinfo {title} {{DM induced neutron disappearance as the origin of the LZ nuclear recoil event}}} (\bibinfo {year} {2026}),\ \Eprint {https://arxiv.org/abs/2609.15933} {arXiv:2609.15933 [hep-ph]} \BibitemShut {NoStop}%
\bibitem [{\citenamefont {Jeesun}\ and\ \citenamefont {Majumdar}(2026)}]{Jeesun:2026vzo}%
  \BibitemOpen
  \bibfield  {author} {\bibinfo {author} {\bibfnamefont {S.}~\bibnamefont {Jeesun}}\ and\ \bibinfo {author} {\bibfnamefont {A.}~\bibnamefont {Majumdar}},\ }\href@noop {} {\bibinfo {title} {{Atmospheric neutrino up-scattering explanation of LZ 2026 excess}}} (\bibinfo {year} {2026}),\ \Eprint {https://arxiv.org/abs/2609.04185} {arXiv:2609.04185 [hep-ph]} \BibitemShut {NoStop}%
\bibitem [{\citenamefont {Chattaraj}\ \emph {et~al.}(2026)\citenamefont {Chattaraj}, \citenamefont {Majumdar}, \citenamefont {Papoulias},\ and\ \citenamefont {Srivastava}}]{Chattaraj:2026fxn}%
  \BibitemOpen
  \bibfield  {author} {\bibinfo {author} {\bibfnamefont {A.}~\bibnamefont {Chattaraj}}, \bibinfo {author} {\bibfnamefont {A.}~\bibnamefont {Majumdar}}, \bibinfo {author} {\bibfnamefont {D.~K.}\ \bibnamefont {Papoulias}},\ and\ \bibinfo {author} {\bibfnamefont {R.}~\bibnamefont {Srivastava}},\ }\href@noop {} {\bibinfo {title} {{Can Elastic Neutrino Scattering Account for the LZ230616 Event?}}} (\bibinfo {year} {2026}),\ \Eprint {https://arxiv.org/abs/2609.10504} {arXiv:2609.10504 [hep-ph]} \BibitemShut {NoStop}%
\bibitem [{\citenamefont {Brdar}\ and\ \citenamefont {Chattopadhyay}(2026)}]{Brdar:2026ukx}%
  \BibitemOpen
  \bibfield  {author} {\bibinfo {author} {\bibfnamefont {V.}~\bibnamefont {Brdar}}\ and\ \bibinfo {author} {\bibfnamefont {D.~S.}\ \bibnamefont {Chattopadhyay}},\ }\href@noop {} {\bibinfo {title} {{A Narrow Neutrino Window for the LZ Event}}} (\bibinfo {year} {2026}),\ \Eprint {https://arxiv.org/abs/2609.30255} {arXiv:2609.30255 [hep-ph]} \BibitemShut {NoStop}%
\bibitem [{\citenamefont {Sheng}\ and\ \citenamefont {Zhang}(2026)}]{Sheng:2026tqt}%
  \BibitemOpen
  \bibfield  {author} {\bibinfo {author} {\bibfnamefont {J.}~\bibnamefont {Sheng}}\ and\ \bibinfo {author} {\bibfnamefont {K.}~\bibnamefont {Zhang}},\ }\href@noop {} {\bibinfo {title} {{The LUX-ZEPLIN Event as Hyperfine Spectroscopy of Composite Dark Matter}}} (\bibinfo {year} {2026}),\ \Eprint {https://arxiv.org/abs/2609.23477} {arXiv:2609.23477 [hep-ph]} \BibitemShut {NoStop}%
\bibitem [{\citenamefont {Jung}\ and\ \citenamefont {Park}(2026)}]{Jung:2026otm}%
  \BibitemOpen
  \bibfield  {author} {\bibinfo {author} {\bibfnamefont {H.}~\bibnamefont {Jung}}\ and\ \bibinfo {author} {\bibfnamefont {S.~C.}\ \bibnamefont {Park}},\ }\href@noop {} {\bibinfo {title} {{Dark Diffraction at LZ from a Screened Neutral Composite Baryon}}} (\bibinfo {year} {2026}),\ \Eprint {https://arxiv.org/abs/2609.25723} {arXiv:2609.25723 [hep-ph]} \BibitemShut {NoStop}%
\bibitem [{\citenamefont {Sannino}\ and\ \citenamefont {Turner}(2026)}]{Sannino:2026hkc}%
  \BibitemOpen
  \bibfield  {author} {\bibinfo {author} {\bibfnamefont {F.}~\bibnamefont {Sannino}}\ and\ \bibinfo {author} {\bibfnamefont {J.}~\bibnamefont {Turner}},\ }\href@noop {} {\bibinfo {title} {{Interpreting the LZ 248 keV Event using Dark QCD}}} (\bibinfo {year} {2026}),\ \Eprint {https://arxiv.org/abs/2609.24988} {arXiv:2609.24988 [hep-ph]} \BibitemShut {NoStop}%
\bibitem [{\citenamefont {Lee}\ and\ \citenamefont {Randall}(2026)}]{Lee:2026xxh}%
  \BibitemOpen
  \bibfield  {author} {\bibinfo {author} {\bibfnamefont {V.~S.~H.}\ \bibnamefont {Lee}}\ and\ \bibinfo {author} {\bibfnamefont {L.}~\bibnamefont {Randall}},\ }\href@noop {} {\bibinfo {title} {{A Warped Extra Dimensional Candidate for the LZ 248 keV Event}}} (\bibinfo {year} {2026}),\ \Eprint {https://arxiv.org/abs/2609.09136} {arXiv:2609.09136 [hep-ph]} \BibitemShut {NoStop}%
\bibitem [{\citenamefont {Batell}\ \emph {et~al.}(2009)\citenamefont {Batell}, \citenamefont {Pospelov},\ and\ \citenamefont {Ritz}}]{Batell:2009vb}%
  \BibitemOpen
  \bibfield  {author} {\bibinfo {author} {\bibfnamefont {B.}~\bibnamefont {Batell}}, \bibinfo {author} {\bibfnamefont {M.}~\bibnamefont {Pospelov}},\ and\ \bibinfo {author} {\bibfnamefont {A.}~\bibnamefont {Ritz}},\ }\href {https://doi.org/10.1103/PhysRevD.79.115019} {\bibfield  {journal} {\bibinfo  {journal} {Phys. Rev. D}\ }\textbf {\bibinfo {volume} {79}},\ \bibinfo {pages} {115019} (\bibinfo {year} {2009})},\ \Eprint {https://arxiv.org/abs/0903.3396} {arXiv:0903.3396 [hep-ph]} \BibitemShut {NoStop}%
\bibitem [{\citenamefont {Finkbeiner}\ \emph {et~al.}(2009)\citenamefont {Finkbeiner}, \citenamefont {Slatyer}, \citenamefont {Weiner},\ and\ \citenamefont {Yavin}}]{Finkbeiner:2009mi}%
  \BibitemOpen
  \bibfield  {author} {\bibinfo {author} {\bibfnamefont {D.~P.}\ \bibnamefont {Finkbeiner}}, \bibinfo {author} {\bibfnamefont {T.~R.}\ \bibnamefont {Slatyer}}, \bibinfo {author} {\bibfnamefont {N.}~\bibnamefont {Weiner}},\ and\ \bibinfo {author} {\bibfnamefont {I.}~\bibnamefont {Yavin}},\ }\href {https://doi.org/10.1088/1475-7516/2009/09/037} {\bibfield  {journal} {\bibinfo  {journal} {JCAP}\ }\textbf {\bibinfo {volume} {09}},\ \bibinfo {pages} {037}},\ \Eprint {https://arxiv.org/abs/0903.1037} {arXiv:0903.1037 [hep-ph]} \BibitemShut {NoStop}%
\bibitem [{\citenamefont {Graham}\ \emph {et~al.}(2010)\citenamefont {Graham}, \citenamefont {Harnik}, \citenamefont {Rajendran},\ and\ \citenamefont {Saraswat}}]{Graham:2010ca}%
  \BibitemOpen
  \bibfield  {author} {\bibinfo {author} {\bibfnamefont {P.~W.}\ \bibnamefont {Graham}}, \bibinfo {author} {\bibfnamefont {R.}~\bibnamefont {Harnik}}, \bibinfo {author} {\bibfnamefont {S.}~\bibnamefont {Rajendran}},\ and\ \bibinfo {author} {\bibfnamefont {P.}~\bibnamefont {Saraswat}},\ }\href {https://doi.org/10.1103/PhysRevD.82.063512} {\bibfield  {journal} {\bibinfo  {journal} {Phys. Rev. D}\ }\textbf {\bibinfo {volume} {82}},\ \bibinfo {pages} {063512} (\bibinfo {year} {2010})},\ \Eprint {https://arxiv.org/abs/1004.0937} {arXiv:1004.0937 [hep-ph]} \BibitemShut {NoStop}%
\bibitem [{\citenamefont {Dent}\ and\ \citenamefont {Newstead}(2026)}]{Dent:2026bji}%
  \BibitemOpen
  \bibfield  {author} {\bibinfo {author} {\bibfnamefont {J.~B.}\ \bibnamefont {Dent}}\ and\ \bibinfo {author} {\bibfnamefont {J.~L.}\ \bibnamefont {Newstead}},\ }\href@noop {} {\bibinfo {title} {{Exothermic and Endothermic Inelastic Dark Matter Interpretations at LZ: Sideband Constraints and Future Prospects}}} (\bibinfo {year} {2026}),\ \Eprint {https://arxiv.org/abs/2609.04673} {arXiv:2609.04673 [hep-ph]} \BibitemShut {NoStop}%
\bibitem [{\citenamefont {de~Lima}(2026)}]{deLima:2026shq}%
  \BibitemOpen
  \bibfield  {author} {\bibinfo {author} {\bibfnamefont {C.~H.}\ \bibnamefont {de~Lima}},\ }\href@noop {} {\bibinfo {title} {{Exothermic Dark Matter at LZ}}} (\bibinfo {year} {2026}),\ \Eprint {https://arxiv.org/abs/2609.05204} {arXiv:2609.05204 [hep-ph]} \BibitemShut {NoStop}%
\bibitem [{\citenamefont {Baer}\ and\ \citenamefont {Barger}(2026)}]{Baer:2026fpy}%
  \BibitemOpen
  \bibfield  {author} {\bibinfo {author} {\bibfnamefont {H.}~\bibnamefont {Baer}}\ and\ \bibinfo {author} {\bibfnamefont {V.}~\bibnamefont {Barger}},\ }\href@noop {} {\bibinfo {title} {{Exothermic dark matter and the 248 keV nuclear recoil in LUX-ZEPLIN}}} (\bibinfo {year} {2026}),\ \Eprint {https://arxiv.org/abs/2609.06153} {arXiv:2609.06153 [hep-ph]} \BibitemShut {NoStop}%
\bibitem [{\citenamefont {Yang}\ \emph {et~al.}(2026)\citenamefont {Yang}, \citenamefont {Wu}, \citenamefont {Duan}, \citenamefont {Xia}, \citenamefont {Tsai},\ and\ \citenamefont {Fan}}]{Yang:2026wpb}%
  \BibitemOpen
  \bibfield  {author} {\bibinfo {author} {\bibfnamefont {M.}~\bibnamefont {Yang}}, \bibinfo {author} {\bibfnamefont {Q.-f.}\ \bibnamefont {Wu}}, \bibinfo {author} {\bibfnamefont {K.-K.}\ \bibnamefont {Duan}}, \bibinfo {author} {\bibfnamefont {Z.-Q.}\ \bibnamefont {Xia}}, \bibinfo {author} {\bibfnamefont {Y.-L.~S.}\ \bibnamefont {Tsai}},\ and\ \bibinfo {author} {\bibfnamefont {Y.-Z.}\ \bibnamefont {Fan}},\ }\href@noop {} {\bibinfo {title} {{A Common $\sim$55 GeV Inelastic Dark Matter Origin of the LZ, Fermi-LAT, and AMS-02 Excesses and Paleo-Detector Prospects}}} (\bibinfo {year} {2026}),\ \Eprint {https://arxiv.org/abs/2609.06640} {arXiv:2609.06640 [hep-ph]} \BibitemShut {NoStop}%
\bibitem [{\citenamefont {Xing}(2026)}]{Xing:2026civ}%
  \BibitemOpen
  \bibfield  {author} {\bibinfo {author} {\bibfnamefont {C.-Y.}\ \bibnamefont {Xing}},\ }\href@noop {} {\bibinfo {title} {{Galactic Endothermic Production and Exothermic Detection of Excited Dark Matter: Implications for LUX-ZEPLIN}}} (\bibinfo {year} {2026}),\ \Eprint {https://arxiv.org/abs/2609.17935} {arXiv:2609.17935 [hep-ph]} \BibitemShut {NoStop}%
\bibitem [{\citenamefont {Gemmell}\ \emph {et~al.}(2026)\citenamefont {Gemmell}, \citenamefont {Hooper},\ and\ \citenamefont {Krnjaic}}]{Gemmell:2026yaw}%
  \BibitemOpen
  \bibfield  {author} {\bibinfo {author} {\bibfnamefont {C.}~\bibnamefont {Gemmell}}, \bibinfo {author} {\bibfnamefont {D.}~\bibnamefont {Hooper}},\ and\ \bibinfo {author} {\bibfnamefont {G.}~\bibnamefont {Krnjaic}},\ }\href@noop {} {\bibinfo {title} {{A Simple Dark Matter Model to Explain the LZ Event and Galactic Center Excess}}} (\bibinfo {year} {2026}),\ \Eprint {https://arxiv.org/abs/2609.26570} {arXiv:2609.26570 [hep-ph]} \BibitemShut {NoStop}%
\bibitem [{\citenamefont {Oliveira}(2026)}]{Oliveira:2026jtn}%
  \BibitemOpen
  \bibfield  {author} {\bibinfo {author} {\bibfnamefont {V.}~\bibnamefont {Oliveira}},\ }\href@noop {} {\bibinfo {title} {{The LZ event with a two-state dark matter halo}}} (\bibinfo {year} {2026}),\ \Eprint {https://arxiv.org/abs/2609.35650} {arXiv:2609.35650 [hep-ph]} \BibitemShut {NoStop}%
\bibitem [{\citenamefont {Jho}\ \emph {et~al.}(2026)\citenamefont {Jho}, \citenamefont {Kim},\ and\ \citenamefont {Park}}]{Jho:2026zzj}%
  \BibitemOpen
  \bibfield  {author} {\bibinfo {author} {\bibfnamefont {Y.}~\bibnamefont {Jho}}, \bibinfo {author} {\bibfnamefont {S.}~\bibnamefont {Kim}},\ and\ \bibinfo {author} {\bibfnamefont {S.~C.}\ \bibnamefont {Park}},\ }\href@noop {} {\bibinfo {title} {{Decaying Dark Matter for the Exothermic Recoil at LZ}}} (\bibinfo {year} {2026}),\ \Eprint {https://arxiv.org/abs/2609.37561} {arXiv:2609.37561 [hep-ph]} \BibitemShut {NoStop}%
\bibitem [{\citenamefont {Sheikh}\ \emph {et~al.}(2026)\citenamefont {Sheikh}, \citenamefont {Toma},\ and\ \citenamefont {Tsumura}}]{Sheikh:2026lxx}%
  \BibitemOpen
  \bibfield  {author} {\bibinfo {author} {\bibfnamefont {R.}~\bibnamefont {Sheikh}}, \bibinfo {author} {\bibfnamefont {T.}~\bibnamefont {Toma}},\ and\ \bibinfo {author} {\bibfnamefont {K.}~\bibnamefont {Tsumura}},\ }\href@noop {} {\bibinfo {title} {{Exothermic nuclear recoils from thermal quasi-pNGB dark matter}}} (\bibinfo {year} {2026}),\ \Eprint {https://arxiv.org/abs/2610.09516} {arXiv:2610.09516 [hep-ph]} \BibitemShut {NoStop}%
\bibitem [{\citenamefont {Deka}(2026)}]{Deka:2026zeu}%
  \BibitemOpen
  \bibfield  {author} {\bibinfo {author} {\bibfnamefont {K.}~\bibnamefont {Deka}},\ }\href@noop {} {\bibinfo {title} {{Exothermic dark-photon dark matter and the LZ high-recoil event: finite-reheating constraints}}} (\bibinfo {year} {2026}),\ \Eprint {https://arxiv.org/abs/2610.18974} {arXiv:2610.18974 [hep-ph]} \BibitemShut {NoStop}%
\bibitem [{\citenamefont {Kim}\ \emph {et~al.}(2026)\citenamefont {Kim}, \citenamefont {Kim},\ and\ \citenamefont {Ko}}]{Kim:2026euj}%
  \BibitemOpen
  \bibfield  {author} {\bibinfo {author} {\bibfnamefont {J.}~\bibnamefont {Kim}}, \bibinfo {author} {\bibfnamefont {J.}~\bibnamefont {Kim}},\ and\ \bibinfo {author} {\bibfnamefont {P.}~\bibnamefont {Ko}},\ }\href@noop {} {\bibinfo {title} {{Exothermic Dark Matter with a Dark Higgs: The LZ 248 keV Event and the Galactic Center Excess}}} (\bibinfo {year} {2026}),\ \Eprint {https://arxiv.org/abs/2610.25423} {arXiv:2610.25423 [hep-ph]} \BibitemShut {NoStop}%
\bibitem [{\citenamefont {Silveira}\ and\ \citenamefont {Zee}(1985)}]{Silveira:1985rk}%
  \BibitemOpen
  \bibfield  {author} {\bibinfo {author} {\bibfnamefont {V.}~\bibnamefont {Silveira}}\ and\ \bibinfo {author} {\bibfnamefont {A.}~\bibnamefont {Zee}},\ }\href {https://doi.org/10.1016/0370-2693(85)90624-0} {\bibfield  {journal} {\bibinfo  {journal} {Phys. Lett. B}\ }\textbf {\bibinfo {volume} {161}},\ \bibinfo {pages} {136} (\bibinfo {year} {1985})}\BibitemShut {NoStop}%
\bibitem [{\citenamefont {McDonald}(1994)}]{McDonald:1993ex}%
  \BibitemOpen
  \bibfield  {author} {\bibinfo {author} {\bibfnamefont {J.}~\bibnamefont {McDonald}},\ }\href {https://doi.org/10.1103/PhysRevD.50.3637} {\bibfield  {journal} {\bibinfo  {journal} {Phys. Rev. D}\ }\textbf {\bibinfo {volume} {50}},\ \bibinfo {pages} {3637} (\bibinfo {year} {1994})},\ \Eprint {https://arxiv.org/abs/hep-ph/0702143} {arXiv:hep-ph/0702143} \BibitemShut {NoStop}%
\bibitem [{\citenamefont {Burgess}\ \emph {et~al.}(2001)\citenamefont {Burgess}, \citenamefont {Pospelov},\ and\ \citenamefont {ter Veldhuis}}]{Burgess:2000yq}%
  \BibitemOpen
  \bibfield  {author} {\bibinfo {author} {\bibfnamefont {C.~P.}\ \bibnamefont {Burgess}}, \bibinfo {author} {\bibfnamefont {M.}~\bibnamefont {Pospelov}},\ and\ \bibinfo {author} {\bibfnamefont {T.}~\bibnamefont {ter Veldhuis}},\ }\href {https://doi.org/10.1016/S0550-3213(01)00513-2} {\bibfield  {journal} {\bibinfo  {journal} {Nucl. Phys. B}\ }\textbf {\bibinfo {volume} {619}},\ \bibinfo {pages} {709} (\bibinfo {year} {2001})},\ \Eprint {https://arxiv.org/abs/hep-ph/0011335} {arXiv:hep-ph/0011335} \BibitemShut {NoStop}%
\bibitem [{\citenamefont {Ghorbani}\ and\ \citenamefont {Ghorbani}(2016)}]{Ghorbani:2014gka}%
  \BibitemOpen
  \bibfield  {author} {\bibinfo {author} {\bibfnamefont {K.}~\bibnamefont {Ghorbani}}\ and\ \bibinfo {author} {\bibfnamefont {H.}~\bibnamefont {Ghorbani}},\ }\href {https://doi.org/10.1103/PhysRevD.93.055012} {\bibfield  {journal} {\bibinfo  {journal} {Phys. Rev. D}\ }\textbf {\bibinfo {volume} {93}},\ \bibinfo {pages} {055012} (\bibinfo {year} {2016})},\ \Eprint {https://arxiv.org/abs/1501.00206} {arXiv:1501.00206 [hep-ph]} \BibitemShut {NoStop}%
\bibitem [{\citenamefont {Casas}\ \emph {et~al.}(2017)\citenamefont {Casas}, \citenamefont {Cerde{\~n}o}, \citenamefont {Moreno},\ and\ \citenamefont {Quilis}}]{Casas:2017jjg}%
  \BibitemOpen
  \bibfield  {author} {\bibinfo {author} {\bibfnamefont {J.~A.}\ \bibnamefont {Casas}}, \bibinfo {author} {\bibfnamefont {D.~G.}\ \bibnamefont {Cerde{\~n}o}}, \bibinfo {author} {\bibfnamefont {J.~M.}\ \bibnamefont {Moreno}},\ and\ \bibinfo {author} {\bibfnamefont {J.}~\bibnamefont {Quilis}},\ }\href {https://doi.org/10.1007/JHEP05(2017)036} {\bibfield  {journal} {\bibinfo  {journal} {JHEP}\ }\textbf {\bibinfo {volume} {05}},\ \bibinfo {pages} {036}},\ \Eprint {https://arxiv.org/abs/1701.08134} {arXiv:1701.08134 [hep-ph]} \BibitemShut {NoStop}%
\bibitem [{\citenamefont {D{\'\i}az~S{\'a}ez}\ \emph {et~al.}(2024)\citenamefont {D{\'\i}az~S{\'a}ez}, \citenamefont {Lahiri},\ and\ \citenamefont {M{\"o}hling}}]{DiazSaez:2024nrq}%
  \BibitemOpen
  \bibfield  {author} {\bibinfo {author} {\bibfnamefont {B.}~\bibnamefont {D{\'\i}az~S{\'a}ez}}, \bibinfo {author} {\bibfnamefont {J.}~\bibnamefont {Lahiri}},\ and\ \bibinfo {author} {\bibfnamefont {K.}~\bibnamefont {M{\"o}hling}},\ }\href {https://doi.org/10.1088/1475-7516/2024/10/001} {\bibfield  {journal} {\bibinfo  {journal} {JCAP}\ }\textbf {\bibinfo {volume} {10}},\ \bibinfo {pages} {001}},\ \Eprint {https://arxiv.org/abs/2404.19057} {arXiv:2404.19057 [hep-ph]} \BibitemShut {NoStop}%
\bibitem [{\citenamefont {Casas}\ \emph {et~al.}(2018)\citenamefont {Casas}, \citenamefont {G{\'o}mez~Vargas}, \citenamefont {Moreno}, \citenamefont {Quilis},\ and\ \citenamefont {Ruiz~de Austri}}]{Casas:2017rww}%
  \BibitemOpen
  \bibfield  {author} {\bibinfo {author} {\bibfnamefont {J.~A.}\ \bibnamefont {Casas}}, \bibinfo {author} {\bibfnamefont {G.~A.}\ \bibnamefont {G{\'o}mez~Vargas}}, \bibinfo {author} {\bibfnamefont {J.~M.}\ \bibnamefont {Moreno}}, \bibinfo {author} {\bibfnamefont {J.}~\bibnamefont {Quilis}},\ and\ \bibinfo {author} {\bibfnamefont {R.}~\bibnamefont {Ruiz~de Austri}},\ }\href {https://doi.org/10.1088/1475-7516/2018/06/031} {\bibfield  {journal} {\bibinfo  {journal} {JCAP}\ }\textbf {\bibinfo {volume} {06}},\ \bibinfo {pages} {031}},\ \Eprint {https://arxiv.org/abs/1711.10957} {arXiv:1711.10957 [hep-ph]} \BibitemShut {NoStop}%
\bibitem [{\citenamefont {Guo}\ \emph {et~al.}(2022)\citenamefont {Guo}, \citenamefont {He}, \citenamefont {Liu},\ and\ \citenamefont {Wang}}]{Guo:2021vpb}%
  \BibitemOpen
  \bibfield  {author} {\bibinfo {author} {\bibfnamefont {J.}~\bibnamefont {Guo}}, \bibinfo {author} {\bibfnamefont {Y.}~\bibnamefont {He}}, \bibinfo {author} {\bibfnamefont {J.}~\bibnamefont {Liu}},\ and\ \bibinfo {author} {\bibfnamefont {X.-P.}\ \bibnamefont {Wang}},\ }\href {https://doi.org/10.1007/JHEP04(2022)024} {\bibfield  {journal} {\bibinfo  {journal} {JHEP}\ }\textbf {\bibinfo {volume} {04}},\ \bibinfo {pages} {024}},\ \Eprint {https://arxiv.org/abs/2111.01164} {arXiv:2111.01164 [hep-ph]} \BibitemShut {NoStop}%
\bibitem [{\citenamefont {Hooper}\ \emph {et~al.}(2026)\citenamefont {Hooper}, \citenamefont {Krnjaic}, \citenamefont {Rocha},\ and\ \citenamefont {Roy}}]{Hooper:2025fda}%
  \BibitemOpen
  \bibfield  {author} {\bibinfo {author} {\bibfnamefont {D.}~\bibnamefont {Hooper}}, \bibinfo {author} {\bibfnamefont {G.}~\bibnamefont {Krnjaic}}, \bibinfo {author} {\bibfnamefont {D.}~\bibnamefont {Rocha}},\ and\ \bibinfo {author} {\bibfnamefont {S.}~\bibnamefont {Roy}},\ }\href {https://doi.org/10.1103/4vqv-vmxz} {\bibfield  {journal} {\bibinfo  {journal} {Phys. Rev. D}\ }\textbf {\bibinfo {volume} {113}},\ \bibinfo {pages} {115009} (\bibinfo {year} {2026})},\ \Eprint {https://arxiv.org/abs/2507.22975} {arXiv:2507.22975 [hep-ph]} \BibitemShut {NoStop}%
\bibitem [{\citenamefont {Roy}(2026)}]{Roy:2025zvo}%
  \BibitemOpen
  \bibfield  {author} {\bibinfo {author} {\bibfnamefont {S.}~\bibnamefont {Roy}},\ }\href {https://doi.org/10.1103/4mgy-mp7p} {\bibfield  {journal} {\bibinfo  {journal} {Phys. Rev. D}\ }\textbf {\bibinfo {volume} {113}},\ \bibinfo {pages} {055040} (\bibinfo {year} {2026})},\ \Eprint {https://arxiv.org/abs/2509.19982} {arXiv:2509.19982 [hep-ph]} \BibitemShut {NoStop}%
\bibitem [{\citenamefont {Dutra}\ \emph {et~al.}(2026)\citenamefont {Dutra}, \citenamefont {Neto},\ and\ \citenamefont {Siqueira}}]{Dutra:2026fng}%
  \BibitemOpen
  \bibfield  {author} {\bibinfo {author} {\bibfnamefont {M.}~\bibnamefont {Dutra}}, \bibinfo {author} {\bibfnamefont {J.~P.}\ \bibnamefont {Neto}},\ and\ \bibinfo {author} {\bibfnamefont {C.}~\bibnamefont {Siqueira}},\ }\href@noop {} {\bibinfo {title} {{Sub-MeV gamma-ray lines from an endothermic interpretation of the LZ recoil candidate}}} (\bibinfo {year} {2026}),\ \Eprint {https://arxiv.org/abs/2609.33952} {arXiv:2609.33952 [hep-ph]} \BibitemShut {NoStop}%
\bibitem [{\citenamefont {Roques}\ \emph {et~al.}(2003)\citenamefont {Roques} \emph {et~al.}}]{Roques:2003xg}%
  \BibitemOpen
  \bibfield  {author} {\bibinfo {author} {\bibfnamefont {J.~P.}\ \bibnamefont {Roques}} \emph {et~al.},\ }\href {https://doi.org/10.1051/0004-6361:20031501} {\bibfield  {journal} {\bibinfo  {journal} {Astron. Astrophys.}\ }\textbf {\bibinfo {volume} {411}},\ \bibinfo {pages} {L91} (\bibinfo {year} {2003})},\ \Eprint {https://arxiv.org/abs/astro-ph/0310793} {arXiv:astro-ph/0310793} \BibitemShut {NoStop}%
\bibitem [{\citenamefont {Tomsick}\ \emph {et~al.}(2023)\citenamefont {Tomsick} \emph {et~al.}}]{Tomsick:2023aue}%
  \BibitemOpen
  \bibfield  {author} {\bibinfo {author} {\bibfnamefont {J.~A.}\ \bibnamefont {Tomsick}} \emph {et~al.},\ }\href {https://doi.org/10.22323/1.444.0745} {\bibfield  {journal} {\bibinfo  {journal} {PoS}\ }\textbf {\bibinfo {volume} {ICRC2023}},\ \bibinfo {pages} {745} (\bibinfo {year} {2023})},\ \Eprint {https://arxiv.org/abs/2308.12362} {arXiv:2308.12362 [astro-ph.HE]} \BibitemShut {NoStop}%
\bibitem [{\citenamefont {Pospelov}\ and\ \citenamefont {Ramani}(2026)}]{Pospelov:2026ewn}%
  \BibitemOpen
  \bibfield  {author} {\bibinfo {author} {\bibfnamefont {M.}~\bibnamefont {Pospelov}}\ and\ \bibinfo {author} {\bibfnamefont {H.}~\bibnamefont {Ramani}},\ }\href@noop {} {\bibinfo {title} {{Strong Constraints on Higgsino Dark Matter from Solar Capture}}} (\bibinfo {year} {2026}),\ \Eprint {https://arxiv.org/abs/2609.02775} {arXiv:2609.02775 [hep-ph]} \BibitemShut {NoStop}%
\bibitem [{\citenamefont {Bose}\ \emph {et~al.}(2026)\citenamefont {Bose} \emph {et~al.}}]{Bose:2026ndd}%
  \BibitemOpen
  \bibfield  {author} {\bibinfo {author} {\bibfnamefont {D.}~\bibnamefont {Bose}} \emph {et~al.},\ }\href@noop {} {\bibinfo {title} {{Not so good $\nu$s for Higgsino dark matter as LZ excess: stringent limits from Super-Kamiokande and IceCube}}} (\bibinfo {year} {2026}),\ \Eprint {https://arxiv.org/abs/2609.07807} {arXiv:2609.07807 [hep-ph]} \BibitemShut {NoStop}%
\bibitem [{\citenamefont {Di~Mauro}\ and\ \citenamefont {Shaikh}(2026)}]{DiMauro:2026dqp}%
  \BibitemOpen
  \bibfield  {author} {\bibinfo {author} {\bibfnamefont {M.}~\bibnamefont {Di~Mauro}}\ and\ \bibinfo {author} {\bibfnamefont {H.}~\bibnamefont {Shaikh}},\ }\href@noop {} {\bibinfo {title} {{Solar Capture Tests of Inelastic Dark Matter after the LZ High-Recoil Event}}} (\bibinfo {year} {2026}),\ \Eprint {https://arxiv.org/abs/2609.06760} {arXiv:2609.06760 [hep-ph]} \BibitemShut {NoStop}%
\bibitem [{\citenamefont {Nguyen}\ \emph {et~al.}(2026)\citenamefont {Nguyen}, \citenamefont {Linden},\ and\ \citenamefont {Hooper}}]{Nguyen:2026lui}%
  \BibitemOpen
  \bibfield  {author} {\bibinfo {author} {\bibfnamefont {T.~T.~Q.}\ \bibnamefont {Nguyen}}, \bibinfo {author} {\bibfnamefont {T.}~\bibnamefont {Linden}},\ and\ \bibinfo {author} {\bibfnamefont {D.}~\bibnamefont {Hooper}},\ }\href@noop {} {\bibinfo {title} {{Solar Neutrino Constraints on Inelastic Dark Matter Scattering in Light of Recent LUX-ZEPLIN Observations}}} (\bibinfo {year} {2026}),\ \Eprint {https://arxiv.org/abs/2609.11833} {arXiv:2609.11833 [hep-ph]} \BibitemShut {NoStop}%
\bibitem [{\citenamefont {Ghosh}\ \emph {et~al.}(2026)\citenamefont {Ghosh}, \citenamefont {Chavez},\ and\ \citenamefont {Kelso}}]{Ghosh:2026txe}%
  \BibitemOpen
  \bibfield  {author} {\bibinfo {author} {\bibfnamefont {A.}~\bibnamefont {Ghosh}}, \bibinfo {author} {\bibfnamefont {I.}~\bibnamefont {Chavez}},\ and\ \bibinfo {author} {\bibfnamefont {C.}~\bibnamefont {Kelso}},\ }\href@noop {} {\bibinfo {title} {{Confronting the Higgsino Interpretation of the LZ Event with Astrophysical Uncertainties and the Solar Capture Constraints}}} (\bibinfo {year} {2026}),\ \Eprint {https://arxiv.org/abs/2609.15321} {arXiv:2609.15321 [hep-ph]} \BibitemShut {NoStop}%
\bibitem [{\citenamefont {Luengas}\ \emph {et~al.}(2026)\citenamefont {Luengas}, \citenamefont {Pospelov},\ and\ \citenamefont {Ramani}}]{Luengas:2026tlf}%
  \BibitemOpen
  \bibfield  {author} {\bibinfo {author} {\bibfnamefont {J.~P.}\ \bibnamefont {Luengas}}, \bibinfo {author} {\bibfnamefont {M.}~\bibnamefont {Pospelov}},\ and\ \bibinfo {author} {\bibfnamefont {H.}~\bibnamefont {Ramani}},\ }\href@noop {} {\bibinfo {title} {{Constraints on Complex Electroweak WIMPs from Solar Capture}}} (\bibinfo {year} {2026}),\ \Eprint {https://arxiv.org/abs/2609.38316} {arXiv:2609.38316 [hep-ph]} \BibitemShut {NoStop}%
\bibitem [{\citenamefont {Cline}\ \emph {et~al.}(2013)\citenamefont {Cline}, \citenamefont {Kainulainen}, \citenamefont {Scott},\ and\ \citenamefont {Weniger}}]{Cline:2013gha}%
  \BibitemOpen
  \bibfield  {author} {\bibinfo {author} {\bibfnamefont {J.~M.}\ \bibnamefont {Cline}}, \bibinfo {author} {\bibfnamefont {K.}~\bibnamefont {Kainulainen}}, \bibinfo {author} {\bibfnamefont {P.}~\bibnamefont {Scott}},\ and\ \bibinfo {author} {\bibfnamefont {C.}~\bibnamefont {Weniger}},\ }\href {https://doi.org/10.1103/PhysRevD.88.055025} {\bibfield  {journal} {\bibinfo  {journal} {Phys. Rev. D}\ }\textbf {\bibinfo {volume} {88}},\ \bibinfo {pages} {055025} (\bibinfo {year} {2013})},\ \bibinfo {note} {[Erratum: Phys.Rev.D 92, 039906 (2015)]},\ \Eprint {https://arxiv.org/abs/1306.4710} {arXiv:1306.4710 [hep-ph]} \BibitemShut {NoStop}%
\bibitem [{\citenamefont {Arcadi}\ \emph {et~al.}(2018)\citenamefont {Arcadi}, \citenamefont {Dutra}, \citenamefont {Ghosh}, \citenamefont {Lindner}, \citenamefont {Mambrini}, \citenamefont {Pierre}, \citenamefont {Profumo},\ and\ \citenamefont {Queiroz}}]{Arcadi:2017kky}%
  \BibitemOpen
  \bibfield  {author} {\bibinfo {author} {\bibfnamefont {G.}~\bibnamefont {Arcadi}}, \bibinfo {author} {\bibfnamefont {M.}~\bibnamefont {Dutra}}, \bibinfo {author} {\bibfnamefont {P.}~\bibnamefont {Ghosh}}, \bibinfo {author} {\bibfnamefont {M.}~\bibnamefont {Lindner}}, \bibinfo {author} {\bibfnamefont {Y.}~\bibnamefont {Mambrini}}, \bibinfo {author} {\bibfnamefont {M.}~\bibnamefont {Pierre}}, \bibinfo {author} {\bibfnamefont {S.}~\bibnamefont {Profumo}},\ and\ \bibinfo {author} {\bibfnamefont {F.~S.}\ \bibnamefont {Queiroz}},\ }\href {https://doi.org/10.1140/epjc/s10052-018-5662-y} {\bibfield  {journal} {\bibinfo  {journal} {Eur. Phys. J. C}\ }\textbf {\bibinfo {volume} {78}},\ \bibinfo {pages} {203} (\bibinfo {year} {2018})},\ \Eprint {https://arxiv.org/abs/1703.07364} {arXiv:1703.07364 [hep-ph]} \BibitemShut {NoStop}%
\bibitem [{\citenamefont {Ghosh}\ \emph {et~al.}(2023)\citenamefont {Ghosh}, \citenamefont {Ghosh},\ and\ \citenamefont {Roy}}]{Ghosh:2022fzp}%
  \BibitemOpen
  \bibfield  {author} {\bibinfo {author} {\bibfnamefont {P.}~\bibnamefont {Ghosh}}, \bibinfo {author} {\bibfnamefont {T.}~\bibnamefont {Ghosh}},\ and\ \bibinfo {author} {\bibfnamefont {S.}~\bibnamefont {Roy}},\ }\href {https://doi.org/10.1007/JHEP10(2023)057} {\bibfield  {journal} {\bibinfo  {journal} {JHEP}\ }\textbf {\bibinfo {volume} {10}},\ \bibinfo {pages} {057}},\ \Eprint {https://arxiv.org/abs/2211.15640} {arXiv:2211.15640 [hep-ph]} \BibitemShut {NoStop}%
\bibitem [{\citenamefont {Roy}(2025)}]{Roy:2022gop}%
  \BibitemOpen
  \bibfield  {author} {\bibinfo {author} {\bibfnamefont {S.}~\bibnamefont {Roy}},\ }\href {https://doi.org/10.1103/PhysRevD.111.015037} {\bibfield  {journal} {\bibinfo  {journal} {Phys. Rev. D}\ }\textbf {\bibinfo {volume} {111}},\ \bibinfo {pages} {015037} (\bibinfo {year} {2025})},\ \Eprint {https://arxiv.org/abs/2212.11230} {arXiv:2212.11230 [hep-ph]} \BibitemShut {NoStop}%
\bibitem [{\citenamefont {Griest}\ and\ \citenamefont {Seckel}(1991)}]{Griest:1990kh}%
  \BibitemOpen
  \bibfield  {author} {\bibinfo {author} {\bibfnamefont {K.}~\bibnamefont {Griest}}\ and\ \bibinfo {author} {\bibfnamefont {D.}~\bibnamefont {Seckel}},\ }\href {https://doi.org/10.1103/PhysRevD.43.3191} {\bibfield  {journal} {\bibinfo  {journal} {Phys. Rev. D}\ }\textbf {\bibinfo {volume} {43}},\ \bibinfo {pages} {3191} (\bibinfo {year} {1991})}\BibitemShut {NoStop}%
\bibitem [{\citenamefont {Fitzpatrick}\ \emph {et~al.}(2013)\citenamefont {Fitzpatrick}, \citenamefont {Haxton}, \citenamefont {Katz}, \citenamefont {Lubbers},\ and\ \citenamefont {Xu}}]{Fitzpatrick:2012ix}%
  \BibitemOpen
  \bibfield  {author} {\bibinfo {author} {\bibfnamefont {A.~L.}\ \bibnamefont {Fitzpatrick}}, \bibinfo {author} {\bibfnamefont {W.}~\bibnamefont {Haxton}}, \bibinfo {author} {\bibfnamefont {E.}~\bibnamefont {Katz}}, \bibinfo {author} {\bibfnamefont {N.}~\bibnamefont {Lubbers}},\ and\ \bibinfo {author} {\bibfnamefont {Y.}~\bibnamefont {Xu}},\ }\href {https://doi.org/10.1088/1475-7516/2013/02/004} {\bibfield  {journal} {\bibinfo  {journal} {JCAP}\ }\textbf {\bibinfo {volume} {02}},\ \bibinfo {pages} {004}},\ \Eprint {https://arxiv.org/abs/1203.3542} {arXiv:1203.3542 [hep-ph]} \BibitemShut {NoStop}%
\bibitem [{\citenamefont {Anand}\ \emph {et~al.}(2014)\citenamefont {Anand}, \citenamefont {Fitzpatrick},\ and\ \citenamefont {Haxton}}]{Anand:2013yka}%
  \BibitemOpen
  \bibfield  {author} {\bibinfo {author} {\bibfnamefont {N.}~\bibnamefont {Anand}}, \bibinfo {author} {\bibfnamefont {A.~L.}\ \bibnamefont {Fitzpatrick}},\ and\ \bibinfo {author} {\bibfnamefont {W.~C.}\ \bibnamefont {Haxton}},\ }\href {https://doi.org/10.1103/PhysRevC.89.065501} {\bibfield  {journal} {\bibinfo  {journal} {Phys. Rev. C}\ }\textbf {\bibinfo {volume} {89}},\ \bibinfo {pages} {065501} (\bibinfo {year} {2014})},\ \Eprint {https://arxiv.org/abs/1308.6288} {arXiv:1308.6288 [hep-ph]} \BibitemShut {NoStop}%
\bibitem [{\citenamefont {Dent}\ \emph {et~al.}(2015)\citenamefont {Dent}, \citenamefont {Krauss}, \citenamefont {Newstead},\ and\ \citenamefont {Sabharwal}}]{Dent:2015zpa}%
  \BibitemOpen
  \bibfield  {author} {\bibinfo {author} {\bibfnamefont {J.~B.}\ \bibnamefont {Dent}}, \bibinfo {author} {\bibfnamefont {L.~M.}\ \bibnamefont {Krauss}}, \bibinfo {author} {\bibfnamefont {J.~L.}\ \bibnamefont {Newstead}},\ and\ \bibinfo {author} {\bibfnamefont {S.}~\bibnamefont {Sabharwal}},\ }\href {https://doi.org/10.1103/PhysRevD.92.063515} {\bibfield  {journal} {\bibinfo  {journal} {Phys. Rev. D}\ }\textbf {\bibinfo {volume} {92}},\ \bibinfo {pages} {063515} (\bibinfo {year} {2015})},\ \Eprint {https://arxiv.org/abs/1505.03117} {arXiv:1505.03117 [hep-ph]} \BibitemShut {NoStop}%
\bibitem [{\citenamefont {Catena}\ \emph {et~al.}(2019)\citenamefont {Catena}, \citenamefont {Fridell},\ and\ \citenamefont {Krauss}}]{Catena:2019hzw}%
  \BibitemOpen
  \bibfield  {author} {\bibinfo {author} {\bibfnamefont {R.}~\bibnamefont {Catena}}, \bibinfo {author} {\bibfnamefont {K.}~\bibnamefont {Fridell}},\ and\ \bibinfo {author} {\bibfnamefont {M.~B.}\ \bibnamefont {Krauss}},\ }\href {https://doi.org/10.1007/JHEP08(2019)030} {\bibfield  {journal} {\bibinfo  {journal} {JHEP}\ }\textbf {\bibinfo {volume} {08}},\ \bibinfo {pages} {030}},\ \Eprint {https://arxiv.org/abs/1907.02910} {arXiv:1907.02910 [hep-ph]} \BibitemShut {NoStop}%
\bibitem [{\citenamefont {Kavanagh}\ and\ \citenamefont {Edwards}(2024)}]{WIMpyNREFT2024}%
  \BibitemOpen
  \bibfield  {author} {\bibinfo {author} {\bibfnamefont {B.~J.}\ \bibnamefont {Kavanagh}}\ and\ \bibinfo {author} {\bibfnamefont {T.~D.~P.}\ \bibnamefont {Edwards}},\ }\href {https://doi.org/10.5281/zenodo.1230503} {\bibinfo {title} {{WIMpy\_NREFT} v1.2 [computer software]}} (\bibinfo {year} {2024})\BibitemShut {NoStop}%
\bibitem [{\citenamefont {McCabe}(2014)}]{McCabe:2013kea}%
  \BibitemOpen
  \bibfield  {author} {\bibinfo {author} {\bibfnamefont {C.}~\bibnamefont {McCabe}},\ }\href {https://doi.org/10.1088/1475-7516/2014/02/027} {\bibfield  {journal} {\bibinfo  {journal} {JCAP}\ }\textbf {\bibinfo {volume} {02}},\ \bibinfo {pages} {027}},\ \Eprint {https://arxiv.org/abs/1312.1355} {arXiv:1312.1355 [astro-ph.CO]} \BibitemShut {NoStop}%
\bibitem [{\citenamefont {Calore}\ \emph {et~al.}(2023)\citenamefont {Calore}, \citenamefont {Dekker}, \citenamefont {Serpico},\ and\ \citenamefont {Siegert}}]{Calore:2022pks}%
  \BibitemOpen
  \bibfield  {author} {\bibinfo {author} {\bibfnamefont {F.}~\bibnamefont {Calore}}, \bibinfo {author} {\bibfnamefont {A.}~\bibnamefont {Dekker}}, \bibinfo {author} {\bibfnamefont {P.~D.}\ \bibnamefont {Serpico}},\ and\ \bibinfo {author} {\bibfnamefont {T.}~\bibnamefont {Siegert}},\ }\href {https://doi.org/10.1093/mnras/stad457} {\bibinfo {title} {{Constraints on light decaying dark matter candidates from 16~yr of INTEGRAL/SPI observations}}} (\bibinfo {year} {2023}),\ \bibinfo {note} {[Erratum: Mon.Not.Roy.Astron.Soc. 538, (2025), Erratum: Mon.Not.Roy.Astron.Soc. 547, stag045 (2026)]},\ \Eprint {https://arxiv.org/abs/2209.06299} {arXiv:2209.06299 [hep-ph]} \BibitemShut {NoStop}%
\bibitem [{\citenamefont {Siegert}\ \emph {et~al.}(2022)\citenamefont {Siegert}, \citenamefont {Berteaud}, \citenamefont {Calore}, \citenamefont {Serpico},\ and\ \citenamefont {Weinberger}}]{Siegert:2022jii}%
  \BibitemOpen
  \bibfield  {author} {\bibinfo {author} {\bibfnamefont {T.}~\bibnamefont {Siegert}}, \bibinfo {author} {\bibfnamefont {J.}~\bibnamefont {Berteaud}}, \bibinfo {author} {\bibfnamefont {F.}~\bibnamefont {Calore}}, \bibinfo {author} {\bibfnamefont {P.~D.}\ \bibnamefont {Serpico}},\ and\ \bibinfo {author} {\bibfnamefont {C.}~\bibnamefont {Weinberger}},\ }\href {https://doi.org/10.1051/0004-6361/202142639} {\bibfield  {journal} {\bibinfo  {journal} {Astron. Astrophys.}\ }\textbf {\bibinfo {volume} {660}},\ \bibinfo {pages} {A130} (\bibinfo {year} {2022})},\ \Eprint {https://arxiv.org/abs/2202.04574} {arXiv:2202.04574 [astro-ph.HE]} \BibitemShut {NoStop}%
\bibitem [{\citenamefont {McDaniel}\ \emph {et~al.}(2024)\citenamefont {McDaniel}, \citenamefont {Ajello}, \citenamefont {Karwin}, \citenamefont {Di~Mauro}, \citenamefont {Drlica-Wagner},\ and\ \citenamefont {S{\'a}nchez-Conde}}]{McDaniel:2023bju}%
  \BibitemOpen
  \bibfield  {author} {\bibinfo {author} {\bibfnamefont {A.}~\bibnamefont {McDaniel}}, \bibinfo {author} {\bibfnamefont {M.}~\bibnamefont {Ajello}}, \bibinfo {author} {\bibfnamefont {C.~M.}\ \bibnamefont {Karwin}}, \bibinfo {author} {\bibfnamefont {M.}~\bibnamefont {Di~Mauro}}, \bibinfo {author} {\bibfnamefont {A.}~\bibnamefont {Drlica-Wagner}},\ and\ \bibinfo {author} {\bibfnamefont {M.~A.}\ \bibnamefont {S{\'a}nchez-Conde}},\ }\href {https://doi.org/10.1103/PhysRevD.109.063024} {\bibfield  {journal} {\bibinfo  {journal} {Phys. Rev. D}\ }\textbf {\bibinfo {volume} {109}},\ \bibinfo {pages} {063024} (\bibinfo {year} {2024})},\ \Eprint {https://arxiv.org/abs/2311.04982} {arXiv:2311.04982 [astro-ph.HE]} \BibitemShut {NoStop}%
\bibitem [{\citenamefont {Helm}(1956)}]{Helm:1956zz}%
  \BibitemOpen
  \bibfield  {author} {\bibinfo {author} {\bibfnamefont {R.~H.}\ \bibnamefont {Helm}},\ }\href {https://doi.org/10.1103/PhysRev.104.1466} {\bibfield  {journal} {\bibinfo  {journal} {Phys. Rev.}\ }\textbf {\bibinfo {volume} {104}},\ \bibinfo {pages} {1466} (\bibinfo {year} {1956})}\BibitemShut {NoStop}%
\bibitem [{\citenamefont {Lewin}\ and\ \citenamefont {Smith}(1996)}]{Lewin:1995rx}%
  \BibitemOpen
  \bibfield  {author} {\bibinfo {author} {\bibfnamefont {J.~D.}\ \bibnamefont {Lewin}}\ and\ \bibinfo {author} {\bibfnamefont {P.~F.}\ \bibnamefont {Smith}},\ }\href {https://doi.org/10.1016/S0927-6505(96)00047-3} {\bibfield  {journal} {\bibinfo  {journal} {Astropart. Phys.}\ }\textbf {\bibinfo {volume} {6}},\ \bibinfo {pages} {87} (\bibinfo {year} {1996})}\BibitemShut {NoStop}%
\bibitem [{\citenamefont {Hoferichter}\ \emph {et~al.}(2016)\citenamefont {Hoferichter}, \citenamefont {Klos}, \citenamefont {Men{\'e}ndez},\ and\ \citenamefont {Schwenk}}]{Hoferichter:2016nvd}%
  \BibitemOpen
  \bibfield  {author} {\bibinfo {author} {\bibfnamefont {M.}~\bibnamefont {Hoferichter}}, \bibinfo {author} {\bibfnamefont {P.}~\bibnamefont {Klos}}, \bibinfo {author} {\bibfnamefont {J.}~\bibnamefont {Men{\'e}ndez}},\ and\ \bibinfo {author} {\bibfnamefont {A.}~\bibnamefont {Schwenk}},\ }\href {https://doi.org/10.1103/PhysRevD.94.063505} {\bibfield  {journal} {\bibinfo  {journal} {Phys. Rev. D}\ }\textbf {\bibinfo {volume} {94}},\ \bibinfo {pages} {063505} (\bibinfo {year} {2016})},\ \Eprint {https://arxiv.org/abs/1605.08043} {arXiv:1605.08043 [hep-ph]} \BibitemShut {NoStop}%
\bibitem [{\citenamefont {Hoferichter}\ \emph {et~al.}(2019)\citenamefont {Hoferichter}, \citenamefont {Klos}, \citenamefont {Men{\'e}ndez},\ and\ \citenamefont {Schwenk}}]{Hoferichter:2018acd}%
  \BibitemOpen
  \bibfield  {author} {\bibinfo {author} {\bibfnamefont {M.}~\bibnamefont {Hoferichter}}, \bibinfo {author} {\bibfnamefont {P.}~\bibnamefont {Klos}}, \bibinfo {author} {\bibfnamefont {J.}~\bibnamefont {Men{\'e}ndez}},\ and\ \bibinfo {author} {\bibfnamefont {A.}~\bibnamefont {Schwenk}},\ }\href {https://doi.org/10.1103/PhysRevD.99.055031} {\bibfield  {journal} {\bibinfo  {journal} {Phys. Rev. D}\ }\textbf {\bibinfo {volume} {99}},\ \bibinfo {pages} {055031} (\bibinfo {year} {2019})},\ \Eprint {https://arxiv.org/abs/1812.05617} {arXiv:1812.05617 [hep-ph]} \BibitemShut {NoStop}%
\bibitem [{\citenamefont {Graham}\ \emph {et~al.}(2026)\citenamefont {Graham}, \citenamefont {Luengas}, \citenamefont {Ramani},\ and\ \citenamefont {Wong}}]{Graham:2026huo}%
  \BibitemOpen
  \bibfield  {author} {\bibinfo {author} {\bibfnamefont {P.~W.}\ \bibnamefont {Graham}}, \bibinfo {author} {\bibfnamefont {J.~P.}\ \bibnamefont {Luengas}}, \bibinfo {author} {\bibfnamefont {H.}~\bibnamefont {Ramani}},\ and\ \bibinfo {author} {\bibfnamefont {S.~S.~Y.}\ \bibnamefont {Wong}},\ }\href@noop {} {\bibinfo {title} {{Testing the inelastic dark matter interpretation of the LZ event with heavy-element experiments}}} (\bibinfo {year} {2026}),\ \Eprint {https://arxiv.org/abs/2610.21041} {arXiv:2610.21041 [hep-ph]} \BibitemShut {NoStop}%
\bibitem [{\citenamefont {Bozorgnia}\ \emph {et~al.}(2026)\citenamefont {Bozorgnia}, \citenamefont {Contreras}, \citenamefont {Gelmini}, \citenamefont {Kamaha}, \citenamefont {Reynoso-Cordova},\ and\ \citenamefont {Xu}}]{Bozorgnia:2026emp}%
  \BibitemOpen
  \bibfield  {author} {\bibinfo {author} {\bibfnamefont {N.}~\bibnamefont {Bozorgnia}}, \bibinfo {author} {\bibfnamefont {S.}~\bibnamefont {Contreras}}, \bibinfo {author} {\bibfnamefont {G.~B.}\ \bibnamefont {Gelmini}}, \bibinfo {author} {\bibfnamefont {A.~C.}\ \bibnamefont {Kamaha}}, \bibinfo {author} {\bibfnamefont {J.}~\bibnamefont {Reynoso-Cordova}},\ and\ \bibinfo {author} {\bibfnamefont {Y.}~\bibnamefont {Xu}},\ }\href@noop {} {\bibinfo {title} {{LMC-Perturbed LZ Dark Matter Landscape}}} (\bibinfo {year} {2026}),\ \Eprint {https://arxiv.org/abs/2609.36184} {arXiv:2609.36184 [hep-ph]} \BibitemShut {NoStop}%
\bibitem [{\citenamefont {Ghosh}\ \emph {et~al.}(2020)\citenamefont {Ghosh}, \citenamefont {Ibarra}, \citenamefont {Mondal},\ and\ \citenamefont {Mukhopadhyaya}}]{Ghosh:2019jzu}%
  \BibitemOpen
  \bibfield  {author} {\bibinfo {author} {\bibfnamefont {A.}~\bibnamefont {Ghosh}}, \bibinfo {author} {\bibfnamefont {A.}~\bibnamefont {Ibarra}}, \bibinfo {author} {\bibfnamefont {T.}~\bibnamefont {Mondal}},\ and\ \bibinfo {author} {\bibfnamefont {B.}~\bibnamefont {Mukhopadhyaya}},\ }\href {https://doi.org/10.1088/1475-7516/2020/01/011} {\bibfield  {journal} {\bibinfo  {journal} {JCAP}\ }\textbf {\bibinfo {volume} {01}},\ \bibinfo {pages} {011}},\ \Eprint {https://arxiv.org/abs/1909.13292} {arXiv:1909.13292 [hep-ph]} \BibitemShut {NoStop}%
\bibitem [{\citenamefont {Marciano}\ \emph {et~al.}(2012)\citenamefont {Marciano}, \citenamefont {Zhang},\ and\ \citenamefont {Willenbrock}}]{Marciano:2011gm}%
  \BibitemOpen
  \bibfield  {author} {\bibinfo {author} {\bibfnamefont {W.~J.}\ \bibnamefont {Marciano}}, \bibinfo {author} {\bibfnamefont {C.}~\bibnamefont {Zhang}},\ and\ \bibinfo {author} {\bibfnamefont {S.}~\bibnamefont {Willenbrock}},\ }\href {https://doi.org/10.1103/PhysRevD.85.013002} {\bibfield  {journal} {\bibinfo  {journal} {Phys. Rev. D}\ }\textbf {\bibinfo {volume} {85}},\ \bibinfo {pages} {013002} (\bibinfo {year} {2012})},\ \Eprint {https://arxiv.org/abs/1109.5304} {arXiv:1109.5304 [hep-ph]} \BibitemShut {NoStop}%
\bibitem [{\citenamefont {Herrero}\ and\ \citenamefont {Morales}(2020)}]{Herrero:2020dtv}%
  \BibitemOpen
  \bibfield  {author} {\bibinfo {author} {\bibfnamefont {M.}~\bibnamefont {Herrero}}\ and\ \bibinfo {author} {\bibfnamefont {R.~A.}\ \bibnamefont {Morales}},\ }\href {https://doi.org/10.1103/PhysRevD.102.075040} {\bibfield  {journal} {\bibinfo  {journal} {Phys. Rev. D}\ }\textbf {\bibinfo {volume} {102}},\ \bibinfo {pages} {075040} (\bibinfo {year} {2020})},\ \Eprint {https://arxiv.org/abs/2005.03537} {arXiv:2005.03537 [hep-ph]} \BibitemShut {NoStop}%
\bibitem [{\citenamefont {Feruglio}\ \emph {et~al.}(2026)\citenamefont {Feruglio}, \citenamefont {Levati},\ and\ \citenamefont {Ziegler}}]{Feruglio:2025xvc}%
  \BibitemOpen
  \bibfield  {author} {\bibinfo {author} {\bibfnamefont {F.}~\bibnamefont {Feruglio}}, \bibinfo {author} {\bibfnamefont {G.}~\bibnamefont {Levati}},\ and\ \bibinfo {author} {\bibfnamefont {R.}~\bibnamefont {Ziegler}},\ }\href {https://doi.org/10.1007/JHEP05(2026)307} {\bibfield  {journal} {\bibinfo  {journal} {JHEP}\ }\textbf {\bibinfo {volume} {05}},\ \bibinfo {pages} {307}},\ \Eprint {https://arxiv.org/abs/2512.20695} {arXiv:2512.20695 [hep-ph]} \BibitemShut {NoStop}%
\bibitem [{\citenamefont {Krnjaic}\ \emph {et~al.}(2025)\citenamefont {Krnjaic}, \citenamefont {McKeen}, \citenamefont {Mizuta}, \citenamefont {Mohlabeng}, \citenamefont {Morrissey},\ and\ \citenamefont {Tuckler}}]{Krnjaic:2025zjl}%
  \BibitemOpen
  \bibfield  {author} {\bibinfo {author} {\bibfnamefont {G.}~\bibnamefont {Krnjaic}}, \bibinfo {author} {\bibfnamefont {D.}~\bibnamefont {McKeen}}, \bibinfo {author} {\bibfnamefont {R.}~\bibnamefont {Mizuta}}, \bibinfo {author} {\bibfnamefont {G.}~\bibnamefont {Mohlabeng}}, \bibinfo {author} {\bibfnamefont {D.~E.}\ \bibnamefont {Morrissey}},\ and\ \bibinfo {author} {\bibfnamefont {D.}~\bibnamefont {Tuckler}},\ }\href {https://doi.org/10.1103/99z7-kz4s} {\bibfield  {journal} {\bibinfo  {journal} {Phys. Rev. D}\ }\textbf {\bibinfo {volume} {112}},\ \bibinfo {pages} {115039} (\bibinfo {year} {2025})},\ \Eprint {https://arxiv.org/abs/2509.19428} {arXiv:2509.19428 [hep-ph]} \BibitemShut {NoStop}%
\bibitem [{\citenamefont {Navarro}\ \emph {et~al.}(1996)\citenamefont {Navarro}, \citenamefont {Frenk},\ and\ \citenamefont {White}}]{Navarro:1995iw}%
  \BibitemOpen
  \bibfield  {author} {\bibinfo {author} {\bibfnamefont {J.~F.}\ \bibnamefont {Navarro}}, \bibinfo {author} {\bibfnamefont {C.~S.}\ \bibnamefont {Frenk}},\ and\ \bibinfo {author} {\bibfnamefont {S.~D.~M.}\ \bibnamefont {White}},\ }\href {https://doi.org/10.1086/177173} {\bibfield  {journal} {\bibinfo  {journal} {Astrophys. J.}\ }\textbf {\bibinfo {volume} {462}},\ \bibinfo {pages} {563} (\bibinfo {year} {1996})},\ \Eprint {https://arxiv.org/abs/astro-ph/9508025} {arXiv:astro-ph/9508025} \BibitemShut {NoStop}%
\bibitem [{\citenamefont {Navarro}\ \emph {et~al.}(1997)\citenamefont {Navarro}, \citenamefont {Frenk},\ and\ \citenamefont {White}}]{Navarro:1996gj}%
  \BibitemOpen
  \bibfield  {author} {\bibinfo {author} {\bibfnamefont {J.~F.}\ \bibnamefont {Navarro}}, \bibinfo {author} {\bibfnamefont {C.~S.}\ \bibnamefont {Frenk}},\ and\ \bibinfo {author} {\bibfnamefont {S.~D.~M.}\ \bibnamefont {White}},\ }\href {https://doi.org/10.1086/304888} {\bibfield  {journal} {\bibinfo  {journal} {Astrophys. J.}\ }\textbf {\bibinfo {volume} {490}},\ \bibinfo {pages} {493} (\bibinfo {year} {1997})},\ \Eprint {https://arxiv.org/abs/astro-ph/9611107} {arXiv:astro-ph/9611107} \BibitemShut {NoStop}%
\bibitem [{\citenamefont {{Vedrenne}}\ \emph {et~al.}(2003)\citenamefont {{Vedrenne}}, \citenamefont {{Roques}}, \citenamefont {{Sch{\"o}nfelder}}, \citenamefont {{Mandrou}}, \citenamefont {{Lichti}}, \citenamefont {{von Kienlin}}, \citenamefont {{Cordier}}, \citenamefont {{Schanne}}, \citenamefont {{Kn{\"o}dlseder}}, \citenamefont {{Skinner}}, \citenamefont {{Jean}}, \citenamefont {{Sanchez}}, \citenamefont {{Caraveo}}, \citenamefont {{Teegarden}}, \citenamefont {{von Ballmoos}}, \citenamefont {{Bouchet}}, \citenamefont {{Paul}}, \citenamefont {{Matteson}}, \citenamefont {{Boggs}}, \citenamefont {{Wunderer}}, \citenamefont {{Leleux}}, \citenamefont {{Weidenspointner}}, \citenamefont {{Durouchoux}}, \citenamefont {{Diehl}}, \citenamefont {{Strong}}, \citenamefont {{Cass{\'e}}}, \citenamefont {{Clair}},\ and\ \citenamefont {{Andr{\'e}}}}]{2003A&A...411L..63V}%
  \BibitemOpen
  \bibfield  {author} {\bibinfo {author} {\bibfnamefont {G.}~\bibnamefont {{Vedrenne}}}, \bibinfo {author} {\bibfnamefont {J.-P.}\ \bibnamefont {{Roques}}}, \bibinfo {author} {\bibfnamefont {V.}~\bibnamefont {{Sch{\"o}nfelder}}}, \bibinfo {author} {\bibfnamefont {P.}~\bibnamefont {{Mandrou}}}, \bibinfo {author} {\bibfnamefont {G.~G.}\ \bibnamefont {{Lichti}}}, \bibinfo {author} {\bibfnamefont {A.}~\bibnamefont {{von Kienlin}}}, \bibinfo {author} {\bibfnamefont {B.}~\bibnamefont {{Cordier}}}, \bibinfo {author} {\bibfnamefont {S.}~\bibnamefont {{Schanne}}}, \bibinfo {author} {\bibfnamefont {J.}~\bibnamefont {{Kn{\"o}dlseder}}}, \bibinfo {author} {\bibfnamefont {G.}~\bibnamefont {{Skinner}}}, \bibinfo {author} {\bibfnamefont {P.}~\bibnamefont {{Jean}}}, \bibinfo {author} {\bibfnamefont {F.}~\bibnamefont {{Sanchez}}}, \bibinfo {author} {\bibfnamefont {P.}~\bibnamefont {{Caraveo}}}, \bibinfo {author} {\bibfnamefont {B.}~\bibnamefont {{Teegarden}}}, \bibinfo {author} {\bibfnamefont {P.}~\bibnamefont {{von
  Ballmoos}}}, \bibinfo {author} {\bibfnamefont {L.}~\bibnamefont {{Bouchet}}}, \bibinfo {author} {\bibfnamefont {P.}~\bibnamefont {{Paul}}}, \bibinfo {author} {\bibfnamefont {J.}~\bibnamefont {{Matteson}}}, \bibinfo {author} {\bibfnamefont {S.}~\bibnamefont {{Boggs}}}, \bibinfo {author} {\bibfnamefont {C.}~\bibnamefont {{Wunderer}}}, \bibinfo {author} {\bibfnamefont {P.}~\bibnamefont {{Leleux}}}, \bibinfo {author} {\bibfnamefont {G.}~\bibnamefont {{Weidenspointner}}}, \bibinfo {author} {\bibfnamefont {P.}~\bibnamefont {{Durouchoux}}}, \bibinfo {author} {\bibfnamefont {R.}~\bibnamefont {{Diehl}}}, \bibinfo {author} {\bibfnamefont {A.}~\bibnamefont {{Strong}}}, \bibinfo {author} {\bibfnamefont {M.}~\bibnamefont {{Cass{\'e}}}}, \bibinfo {author} {\bibfnamefont {M.~A.}\ \bibnamefont {{Clair}}},\ and\ \bibinfo {author} {\bibfnamefont {Y.}~\bibnamefont {{Andr{\'e}}}},\ }\href {https://doi.org/10.1051/0004-6361:20031482} {\bibfield  {journal} {\bibinfo  {journal} {\aap}\ }\textbf {\bibinfo {volume} {411}},\
  \bibinfo {pages} {L63} (\bibinfo {year} {2003})}\BibitemShut {NoStop}%
\bibitem [{\citenamefont {Anthony-Petersen}\ \emph {et~al.}(2026)\citenamefont {Anthony-Petersen} \emph {et~al.}}]{Anthony-Petersen:2026oxk}%
  \BibitemOpen
  \bibfield  {author} {\bibinfo {author} {\bibfnamefont {R.}~\bibnamefont {Anthony-Petersen}} \emph {et~al.},\ }\href@noop {} {\bibinfo {title} {{MeV dark matter decay forecasts for the COSI mission}}} (\bibinfo {year} {2026}),\ \Eprint {https://arxiv.org/abs/2610.17217} {arXiv:2610.17217 [astro-ph.CO]} \BibitemShut {NoStop}%
\bibitem [{\citenamefont {Abdollahi}\ \emph {et~al.}(2026)\citenamefont {Abdollahi} \emph {et~al.}}]{Fermi-LAT:2025gei}%
  \BibitemOpen
  \bibfield  {author} {\bibinfo {author} {\bibfnamefont {S.}~\bibnamefont {Abdollahi}} \emph {et~al.} (\bibinfo {collaboration} {Fermi-LAT, HAWC, H.E.S.S., MAGIC, VERITAS}),\ }\href {https://doi.org/10.1088/1475-7516/2026/03/035} {\bibfield  {journal} {\bibinfo  {journal} {JCAP}\ }\textbf {\bibinfo {volume} {03}},\ \bibinfo {pages} {035}},\ \Eprint {https://arxiv.org/abs/2508.20229} {arXiv:2508.20229 [astro-ph.HE]} \BibitemShut {NoStop}%
\bibitem [{\citenamefont {Slatyer}(2016)}]{Slatyer:2015jla}%
  \BibitemOpen
  \bibfield  {author} {\bibinfo {author} {\bibfnamefont {T.~R.}\ \bibnamefont {Slatyer}},\ }\href {https://doi.org/10.1103/PhysRevD.93.023527} {\bibfield  {journal} {\bibinfo  {journal} {Phys. Rev. D}\ }\textbf {\bibinfo {volume} {93}},\ \bibinfo {pages} {023527} (\bibinfo {year} {2016})},\ \Eprint {https://arxiv.org/abs/1506.03811} {arXiv:1506.03811 [hep-ph]} \BibitemShut {NoStop}%
\bibitem [{\citenamefont {Aghanim}\ \emph {et~al.}(2020)\citenamefont {Aghanim} \emph {et~al.}}]{Planck:2018vyg}%
  \BibitemOpen
  \bibfield  {author} {\bibinfo {author} {\bibfnamefont {N.}~\bibnamefont {Aghanim}} \emph {et~al.} (\bibinfo {collaboration} {Planck}),\ }\href {https://doi.org/10.1051/0004-6361/201833910} {\bibfield  {journal} {\bibinfo  {journal} {Astron. Astrophys.}\ }\textbf {\bibinfo {volume} {641}},\ \bibinfo {pages} {A6} (\bibinfo {year} {2020})},\ \bibinfo {note} {[Erratum: Astron.Astrophys. 652, C4 (2021)]},\ \Eprint {https://arxiv.org/abs/1807.06209} {arXiv:1807.06209 [astro-ph.CO]} \BibitemShut {NoStop}%
\bibitem [{\citenamefont {Calabrese}\ \emph {et~al.}(2025)\citenamefont {Calabrese} \emph {et~al.}}]{AtacamaCosmologyTelescope:2025nti}%
  \BibitemOpen
  \bibfield  {author} {\bibinfo {author} {\bibfnamefont {E.}~\bibnamefont {Calabrese}} \emph {et~al.} (\bibinfo {collaboration} {Atacama Cosmology Telescope}),\ }\href {https://doi.org/10.1088/1475-7516/2025/11/063} {\bibfield  {journal} {\bibinfo  {journal} {JCAP}\ }\textbf {\bibinfo {volume} {11}},\ \bibinfo {pages} {063}},\ \Eprint {https://arxiv.org/abs/2503.14454} {arXiv:2503.14454 [astro-ph.CO]} \BibitemShut {NoStop}%
\bibitem [{\citenamefont {Curtin}\ \emph {et~al.}(2014)\citenamefont {Curtin} \emph {et~al.}}]{Curtin:2013fra}%
  \BibitemOpen
  \bibfield  {author} {\bibinfo {author} {\bibfnamefont {D.}~\bibnamefont {Curtin}} \emph {et~al.},\ }\href {https://doi.org/10.1103/PhysRevD.90.075004} {\bibfield  {journal} {\bibinfo  {journal} {Phys. Rev. D}\ }\textbf {\bibinfo {volume} {90}},\ \bibinfo {pages} {075004} (\bibinfo {year} {2014})},\ \Eprint {https://arxiv.org/abs/1312.4992} {arXiv:1312.4992 [hep-ph]} \BibitemShut {NoStop}%
\bibitem [{\citenamefont {Aad}\ \emph {et~al.}(2022)\citenamefont {Aad} \emph {et~al.}}]{ATLAS:2022yvh}%
  \BibitemOpen
  \bibfield  {author} {\bibinfo {author} {\bibfnamefont {G.}~\bibnamefont {Aad}} \emph {et~al.} (\bibinfo {collaboration} {ATLAS}),\ }\href {https://doi.org/10.1007/JHEP08(2022)104} {\bibfield  {journal} {\bibinfo  {journal} {JHEP}\ }\textbf {\bibinfo {volume} {08}},\ \bibinfo {pages} {104}},\ \Eprint {https://arxiv.org/abs/2202.07953} {arXiv:2202.07953 [hep-ex]} \BibitemShut {NoStop}%
\bibitem [{\citenamefont {Tumasyan}\ \emph {et~al.}(2023)\citenamefont {Tumasyan} \emph {et~al.}}]{CMS:2023sdw}%
  \BibitemOpen
  \bibfield  {author} {\bibinfo {author} {\bibfnamefont {A.}~\bibnamefont {Tumasyan}} \emph {et~al.} (\bibinfo {collaboration} {CMS}),\ }\href {https://doi.org/10.1140/epjc/s10052-023-11952-7} {\bibfield  {journal} {\bibinfo  {journal} {Eur. Phys. J. C}\ }\textbf {\bibinfo {volume} {83}},\ \bibinfo {pages} {933} (\bibinfo {year} {2023})},\ \Eprint {https://arxiv.org/abs/2303.01214} {arXiv:2303.01214 [hep-ex]} \BibitemShut {NoStop}%
\bibitem [{\citenamefont {Aad}\ \emph {et~al.}(2023)\citenamefont {Aad} \emph {et~al.}}]{ATLAS:2023tkt}%
  \BibitemOpen
  \bibfield  {author} {\bibinfo {author} {\bibfnamefont {G.}~\bibnamefont {Aad}} \emph {et~al.} (\bibinfo {collaboration} {ATLAS}),\ }\href {https://doi.org/10.1016/j.physletb.2023.137963} {\bibfield  {journal} {\bibinfo  {journal} {Phys. Lett. B}\ }\textbf {\bibinfo {volume} {842}},\ \bibinfo {pages} {137963} (\bibinfo {year} {2023})},\ \Eprint {https://arxiv.org/abs/2301.10731} {arXiv:2301.10731 [hep-ex]} \BibitemShut {NoStop}%
\bibitem [{\citenamefont {Fox}\ \emph {et~al.}(2012)\citenamefont {Fox}, \citenamefont {Harnik}, \citenamefont {Kopp},\ and\ \citenamefont {Tsai}}]{Fox:2011pm}%
  \BibitemOpen
  \bibfield  {author} {\bibinfo {author} {\bibfnamefont {P.~J.}\ \bibnamefont {Fox}}, \bibinfo {author} {\bibfnamefont {R.}~\bibnamefont {Harnik}}, \bibinfo {author} {\bibfnamefont {J.}~\bibnamefont {Kopp}},\ and\ \bibinfo {author} {\bibfnamefont {Y.}~\bibnamefont {Tsai}},\ }\href {https://doi.org/10.1103/PhysRevD.85.056011} {\bibfield  {journal} {\bibinfo  {journal} {Phys. Rev. D}\ }\textbf {\bibinfo {volume} {85}},\ \bibinfo {pages} {056011} (\bibinfo {year} {2012})},\ \Eprint {https://arxiv.org/abs/1109.4398} {arXiv:1109.4398 [hep-ph]} \BibitemShut {NoStop}%
\bibitem [{\citenamefont {Aad}\ \emph {et~al.}(2021)\citenamefont {Aad} \emph {et~al.}}]{ATLAS:2021kxv}%
  \BibitemOpen
  \bibfield  {author} {\bibinfo {author} {\bibfnamefont {G.}~\bibnamefont {Aad}} \emph {et~al.} (\bibinfo {collaboration} {ATLAS}),\ }\href {https://doi.org/10.1103/PhysRevD.103.112006} {\bibfield  {journal} {\bibinfo  {journal} {Phys. Rev. D}\ }\textbf {\bibinfo {volume} {103}},\ \bibinfo {pages} {112006} (\bibinfo {year} {2021})},\ \Eprint {https://arxiv.org/abs/2102.10874} {arXiv:2102.10874 [hep-ex]} \BibitemShut {NoStop}%
\bibitem [{\citenamefont {Tumasyan}\ \emph {et~al.}(2021)\citenamefont {Tumasyan} \emph {et~al.}}]{CMS:2021far}%
  \BibitemOpen
  \bibfield  {author} {\bibinfo {author} {\bibfnamefont {A.}~\bibnamefont {Tumasyan}} \emph {et~al.} (\bibinfo {collaboration} {CMS}),\ }\href {https://doi.org/10.1007/JHEP11(2021)153} {\bibfield  {journal} {\bibinfo  {journal} {JHEP}\ }\textbf {\bibinfo {volume} {11}},\ \bibinfo {pages} {153}},\ \Eprint {https://arxiv.org/abs/2107.13021} {arXiv:2107.13021 [hep-ex]} \BibitemShut {NoStop}%
\bibitem [{\citenamefont {Tumasyan}\ \emph {et~al.}(2022)\citenamefont {Tumasyan} \emph {et~al.}}]{CMS:2022qva}%
  \BibitemOpen
  \bibfield  {author} {\bibinfo {author} {\bibfnamefont {A.}~\bibnamefont {Tumasyan}} \emph {et~al.} (\bibinfo {collaboration} {CMS}),\ }\href {https://doi.org/10.1103/PhysRevD.105.092007} {\bibfield  {journal} {\bibinfo  {journal} {Phys. Rev. D}\ }\textbf {\bibinfo {volume} {105}},\ \bibinfo {pages} {092007} (\bibinfo {year} {2022})},\ \Eprint {https://arxiv.org/abs/2201.11585} {arXiv:2201.11585 [hep-ex]} \BibitemShut {NoStop}%
\bibitem [{\citenamefont {Goncalves}\ \emph {et~al.}(2018)\citenamefont {Goncalves}, \citenamefont {Han},\ and\ \citenamefont {Mukhopadhyay}}]{Goncalves:2017iub}%
  \BibitemOpen
  \bibfield  {author} {\bibinfo {author} {\bibfnamefont {D.}~\bibnamefont {Goncalves}}, \bibinfo {author} {\bibfnamefont {T.}~\bibnamefont {Han}},\ and\ \bibinfo {author} {\bibfnamefont {S.}~\bibnamefont {Mukhopadhyay}},\ }\href {https://doi.org/10.1103/PhysRevLett.120.111801} {\bibfield  {journal} {\bibinfo  {journal} {Phys. Rev. Lett.}\ }\textbf {\bibinfo {volume} {120}},\ \bibinfo {pages} {111801} (\bibinfo {year} {2018})},\ \bibinfo {note} {[Erratum: Phys.Rev.Lett. 121, 079902 (2018)]},\ \Eprint {https://arxiv.org/abs/1710.02149} {arXiv:1710.02149 [hep-ph]} \BibitemShut {NoStop}%
\bibitem [{\citenamefont {Gon{\c{c}}alves}\ \emph {et~al.}(2018)\citenamefont {Gon{\c{c}}alves}, \citenamefont {Han},\ and\ \citenamefont {Mukhopadhyay}}]{Goncalves:2018pkt}%
  \BibitemOpen
  \bibfield  {author} {\bibinfo {author} {\bibfnamefont {D.}~\bibnamefont {Gon{\c{c}}alves}}, \bibinfo {author} {\bibfnamefont {T.}~\bibnamefont {Han}},\ and\ \bibinfo {author} {\bibfnamefont {S.}~\bibnamefont {Mukhopadhyay}},\ }\href {https://doi.org/10.1103/PhysRevD.98.015023} {\bibfield  {journal} {\bibinfo  {journal} {Phys. Rev. D}\ }\textbf {\bibinfo {volume} {98}},\ \bibinfo {pages} {015023} (\bibinfo {year} {2018})},\ \Eprint {https://arxiv.org/abs/1803.09751} {arXiv:1803.09751 [hep-ph]} \BibitemShut {NoStop}%
\bibitem [{\citenamefont {Blennow}\ \emph {et~al.}(2018)\citenamefont {Blennow}, \citenamefont {Clementz},\ and\ \citenamefont {Herrero-Garcia}}]{Blennow:2018xwu}%
  \BibitemOpen
  \bibfield  {author} {\bibinfo {author} {\bibfnamefont {M.}~\bibnamefont {Blennow}}, \bibinfo {author} {\bibfnamefont {S.}~\bibnamefont {Clementz}},\ and\ \bibinfo {author} {\bibfnamefont {J.}~\bibnamefont {Herrero-Garcia}},\ }\href {https://doi.org/10.1140/epjc/s10052-018-5863-4} {\bibfield  {journal} {\bibinfo  {journal} {Eur. Phys. J. C}\ }\textbf {\bibinfo {volume} {78}},\ \bibinfo {pages} {386} (\bibinfo {year} {2018})},\ \bibinfo {note} {[Erratum: Eur.Phys.J.C 79, 407 (2019)]},\ \Eprint {https://arxiv.org/abs/1802.06880} {arXiv:1802.06880 [hep-ph]} \BibitemShut {NoStop}%
\bibitem [{\citenamefont {Alenazi}\ and\ \citenamefont {Gondolo}(2006)}]{Alenazi:2006wu}%
  \BibitemOpen
  \bibfield  {author} {\bibinfo {author} {\bibfnamefont {M.~S.}\ \bibnamefont {Alenazi}}\ and\ \bibinfo {author} {\bibfnamefont {P.}~\bibnamefont {Gondolo}},\ }\href {https://doi.org/10.1103/PhysRevD.74.083518} {\bibfield  {journal} {\bibinfo  {journal} {Phys. Rev. D}\ }\textbf {\bibinfo {volume} {74}},\ \bibinfo {pages} {083518} (\bibinfo {year} {2006})},\ \Eprint {https://arxiv.org/abs/astro-ph/0608390} {arXiv:astro-ph/0608390} \BibitemShut {NoStop}%
\bibitem [{\citenamefont {Abbasi}\ \emph {et~al.}(2025)\citenamefont {Abbasi} \emph {et~al.}}]{IceCube:2025fcu}%
  \BibitemOpen
  \bibfield  {author} {\bibinfo {author} {\bibfnamefont {R.}~\bibnamefont {Abbasi}} \emph {et~al.} (\bibinfo {collaboration} {IceCube}),\ }\href@noop {} {\bibinfo {title} {{Search for High-Energy Neutrinos From the Sun Using Ten Years of IceCube Data}}} (\bibinfo {year} {2025}),\ \Eprint {https://arxiv.org/abs/2507.08457} {arXiv:2507.08457 [hep-ex]} \BibitemShut {NoStop}%
\end{thebibliography}%
\end{document}